\documentclass[a4paper,12pt]{article}
\usepackage[english]{babel}
\usepackage{authblk}
\usepackage[utf8]{inputenc}
\emergencystretch=\maxdimen
\usepackage{hyperref}
\usepackage{fullpage}
\usepackage[pdftex]{graphicx}
\usepackage{csquotes}
\usepackage{cite}
\usepackage{amsmath}
\usepackage{amssymb}
\usepackage{amsfonts}
\usepackage{graphicx}
\usepackage{epstopdf}
\usepackage{multirow}
\usepackage{multimedia}
\usepackage{mathptmx}
\usepackage{listings}
\usepackage{color}
\usepackage{courier}
\usepackage{wrapfig}
\usepackage{tikz}
\usepackage{framed}
\usepackage{xcolor}
\usepackage{fix-cm}
\usepackage{times}
\usepackage{verbatim}
\usepackage{caption}
\usepackage[mathscr]{euscript}
\usepackage{mathtools,xparse}
\usepackage{fixltx2e}
\usetikzlibrary{arrows,shapes}
\usepackage{booktabs}
\usepackage[font=scriptsize,labelfont=bf]{caption}
\usepackage[bbgreekl]{mathbbol}
\DeclareSymbolFontAlphabet{\mathbb}{AMSb}
\DeclareSymbolFontAlphabet{\mathbbl}{bbold}
\usepackage{siunitx}
\DeclareSIUnit{\molar}{M}
\DeclareSIUnit{\cells}{cells}
\usepackage{subcaption}
\usepackage{enumitem}
\providecommand{\keywords}[1]
{
	\small	
	\textbf{\textit{Keywords:}} #1
}
\numberwithin{equation}{section}

\newcommand{\eps}{\varepsilon}
\newcommand{\R}{\mathbb R}
\newcommand{\Sbb}{\mathbb S^{N-1}}
\newcommand{\om}{\omega}
\newcommand{\Om}{\Omega }

\newcommand{\xbf}{{\bf x}}

\newcommand{\Ubf}{{\bf U}}
\newcommand{\Sbf}{{\bf S}}
\newcommand{\vbf}{{\bf v}}

\newcommand{\thetabf}{\mbox{\boldmath$\theta$}}

\newcommand{\vthet}{\mbox{\boldmath$\vartheta$}}
\newcommand{\phibf}{\mbox{\boldmath$\phi$}}
\newcommand{\Eti}{\tilde {\mathbb E}}
\newcommand{\Vti}{\tilde {\mathbb V}}

\newcounter{expcounter}			

\usepackage{titlesec}
\titleformat{\section}
{\normalfont\fontfamily{phv}\fontsize{14}{17}\bfseries}{\thesection}{0.75em}{}
\titleformat{\subsection}
{\normalfont\fontfamily{phv}\fontsize{12}{17}\selectfont}{\thesubsection}{0.75em}{}
\titleformat{\subsubsection}
{\normalfont\fontfamily{phv}\fontsize{13}{17}\selectfont}{\thesubsubsection}{0.75em}{}
\title{A flux-limited model for glioma patterning with hypoxia-induced angiogenesis}

\author{Pawan Kumar and Christina Surulescu\\ TU Kaiserslautern,  Felix-Klein-Zentrum f{\"u}r Mathematik,\\
	Paul-Ehrlich-Str. 31, 67663 Kaiserslautern, Germany\\
{\small (kumar, surulescu@mathematik.uni-kl.de)}}

\begin{document}
	\maketitle
	
\begin{abstract}
	\noindent

We propose a model for glioma patterns in a microlocal tumor environment under the influence of acidity, angiogenesis, and tissue anisotropy. The bottom-up model deduction 
eventually leads to a system of reaction-diffusion-taxis equations for glioma and endothelial cell population densities, of which the former infers flux limitation both in the self-diffusion and taxis terms. The model extends a recently introduced \cite{kumar2020multiscale} description of glioma pseudopalisade formation, with the aim of studying the effect of hypoxia-induced tumor vascularization on the establishment and maintenance of these histological patterns which are typical for high grade brain cancer. Numerical simulations of the population level dynamics are performed to investigate several model scenarios containing this and further effects.
\end{abstract}

\medskip
\section{Introduction}\label{sec:intro}
\noindent
Classification of glioma, the most common type of primary brain tumors, typically comprises four grades, according to the degree of malignancy \cite{kleih,Perry2016}. The highest grade IV (including the most aggressive type glioblastoma multiforme, shortly GBM) corresponds to characteristic histological patterns called pseudopalisades; they exhibit garland-like hypercellular structures surrounding necrotic regions usually centered around one or several sites of vasoocclusion \cite{brat2004pseudopalisades,brat2004vaso,rong2006pseudopalisading,wippold2006neuropathology}, and are associated with poor patient survival prognosis \cite{kleih}.\\[-2ex]  


\noindent
More specifically, vascular occlusion and thrombosis \cite{rong2006pseudopalisading, brat2004pseudopalisades} result into hypoxia, which promotes migration of glioma cells away from the highly acidic area, as well as tissue necrotization therein. Hypoxia around the pseudopalisading cells triggers the process of capillary formation, which involves a cascade of coordinated steps \cite{tate2009biology, onishi2011angiogenesis}. It typically starts with tumor cells overexpressing hypoxia-inducible regulators of angiogenesis, including vascular endothelial growth factor (VEGF) \cite{brat2004vaso}. The latter acts as a chemoattractant and proliferation promotor for endothelial cells (ECs) lining the surrounding blood vessels. The chemotactic migration of ECs towards VEGF gradients is accompanied by formation of new capillaries off existing ones, thus leading to microvascular hyperplasia which is common in GBM \cite{brat2002genetic,brat2004vaso} and which facilitates the growth of the neoplasm \cite{brat2004vaso}. Understanding the mechanisms of tumor malignancy can contribute towards developing and tuning  therapeutic strategies, e.g. by targeting angiogenesis \cite{Batchelor2014} or by tumor alkalization \cite{Zhang2019}, used in a (neo)adjuvant way to other approaches (such as surgery and radiotherapy).\\[-2ex]  

\noindent
Among the few mathematical models explicitly addressing glioma pseudopalisade formation (we refer to \cite{cai2016mathematical,caiazzo} for agent-based approaches and to \cite{alfonso2016one,martinez2012hypoxic,kumar2020multiscale} for continuous descriptions), the latter proposed a multiscale approach to deducing a reaction-(myopic)diffusion equation with repellent pH-taxis for  glioma cell density coupled with a reaction-diffusion PDE for the acidity (expressed as proton concentration) produced by tumor cells. The derivation of this system started with describing the evolution of the glioma distribution function on the mesoscopic scale, in the KTAP (kinetic theory of active particles) framework \cite{bellom3}; thereby, the activity variable represented the amount of glioma transmembrane units occupied with protons. A parabolic scaling led to the announced re\-ac\-tion-dif\-fu\-sion-ta\-xis (RDT) equation on the so-called macroscopic level of the glioma population\footnote{We will use in the sequel the term 'macroscopic' to designate quantities only depending on time and space, not to be confounded with the true biological size scale on which pseudopalisades are observed and which is -in that acception- microscopic. Thus, in the following 'microscopic' means single-cell level, while 'mesoscopic' refers to distributions of cells depending on time, space, and further kinetic and/or activity variables.}, including in its terms influences of acidity and tissue coming from the lower modeling scales. The two equations for glioma and acidity were capable to qualitatively reproduce pseudopalisade-like patterns. In this work we aim to use a related approach and to extend the model in \cite{kumar2020multiscale} such as to account for hypoxia-triggered angiogenesis and its effects on the pseudopalisade patterns. We thereby continue in the line of previous works \cite{chauviere2007modeling,conte2020mathematical,corbin2020modeling,Corbin2018,DKSS20,engwer2015glioma,engwer2016effective,hillen2013transport} dealing with cancer cell migration in heterogeneous tissues.\\[-2ex] 

\noindent
The rest of the paper is structured as follows: Section \ref{sec:modeling} is dedicated to setting up the micro-meso formulation in the KTAP framework and to deduce a macroscopic system of differential equations characterizing the dynamics of glioma density (RDT) with flux-limited diffusion and repellent pH-taxis, EC density (RDT) with linear diffusion and chemotaxis towards VEGF gradients, concentrations of acidity and VEGF (both with linear diffusion and tumor- and EC-dependent source terms). 
In Section \ref{sec:numerics} we perform numerical simulations for various scenarios of the macroscopic model obtained in Section \ref{sec:modeling} and interpret the results. Finally, Section \ref{sec:discussions} provides a discussion and an outlook on further problems of interest in this context.

\section{Modeling}\label{sec:modeling}
We begin by introducing some notations for the various variables and functions involved in our modeling process:
\begin{itemize}
	\item variables: time $t\ge 0$, position $\xbf \in \R^N$, velocity of glioma cells $\vbf \in V=s\Sbb \subset \R^N$, velocity of ECs $\vthet\in \Theta =\sigma \Sbb \subset \R^N$; we assume the cell speeds $s,\sigma >0$ to be constant, $\Sbb$ denotes the unit sphere in $\R^N$;
	\item $\hat \vbf =\frac{\vbf }{|\vbf |}$,  $\hat \vthet =\frac{\vthet }{|\vthet |}$ are unit vectors denoting the directions of vectors $\vbf \in V$ and $\vthet \in \Theta$, respectively;
	\item $p(t, \xbf, \vbf )$: (mesoscopic) density function of glioma cells and $M(t,\xbf )=\int _Vp(t, \xbf, \vbf )d\vbf $: macroscopic density of glioma;
	\item $w(t,\mathbf{x},\vthet)$: (mesoscopic) density function of ECs and $W(t,\xbf )=\int _\Theta w(t,\mathbf{x},\vthet)d\vthet$: macroscopic density of ECs.
	\item $q(\xbf ,\thetabf )$: (known) directional distribution function of tissue fibers with orientation $\theta \in \Sbb$. It holds that $\int _{\Sbb }q(\xbf ,\thetabf )d\thetabf =1$; for $\om =\int _V q(\xbf ,\hat \vbf )d\vbf =s^{N-1}$ it holds then that $\int _V \frac{q(\xbf ,\hat \vbf)}{\om }d\vbf =1$;
	\item $\mathbb E_q(\xbf):=\int _{\Sbb}\thetabf q(\xbf,\thetabf )d\thetabf$: mean fiber orientation. We also denote $\tilde {\mathbb E}_q(\xbf):=\int _V\vbf \frac{q(\xbf ,\hat \vbf)}{\om } d\vbf$ and call it mean fiber direction; 
	\item $\mathbb V_q(\xbf ):=\int _{\Sbb}(\thetabf -\mathbb E_q(\xbf ))\otimes (\thetabf -\mathbb E_q(\xbf ))q(\xbf,\thetabf )d\thetabf $: variance-covariance matrix for orientation distribution of tissue fibers, and, correspondingly, $\tilde {\mathbb V}_q(\xbf )=\int _V(\vbf -\tilde {\mathbb E}_q(\xbf))\otimes (\vbf -\tilde {\mathbb E}_q(\xbf)) \frac{q(\xbf ,\hat \vbf)}{\om } d\vbf$;
	\item $h(t,\mathbf{x})$: concentration of protons (acidity), a macroscopic quantity;
	\item $g(t,\mathbf{x})$: concentration of VEGF, also a macroscopic quantity;
	\item $\lambda >0$: constant glioma turning rate;
	\item $\eta (\vthet,g(t,\xbf ))>0$: EC turning rate.
\end{itemize}
\subsection{Glioma cells}\label{subsec:glioma}

 In the spirit of \cite{corbin2020modeling} we consider single cell (microscale) dynamics in the form 
\begin{align}\label{eq:micro}
\frac{d\xbf }{dt}=\vbf ,\quad \frac{d\mathbf{v}}{dt} = -\alpha \Sbf (\mathbf{v},h,M),
\end{align}
with 
\begin{align}
\Sbf (\mathbf{v},h,M)&= \mathbb{B}(\mathbf{v}) \left(\gamma_1 \frac{\nabla h}{\sqrt{K_h^2/X^2 + |\nabla h |^2}}  + \gamma_2 \frac{\nabla M}{\sqrt{K_M^2/X^2 + |\nabla M |^2}}  \right)\notag \\
&=:\mathbb{B}(\mathbf{v})\phibf (h,M),\label{eq:bigS}
\end{align}
where $\gamma_1,\gamma_2\ge 0$ are some dimensionless weighting constants,  $K_h,K_M,X>0$ are further, dimensional constants (to be explained below\footnote{with $X$ to be chosen in Paragraph \ref{par:nondim}}), and the tensor 
$\mathbb{B}(\mathbf{v}) := |\mathbf{v}|^2 \mathbb{I}_N - \mathbf{v} \otimes \mathbf{v}$ models biomechanical cell stress; thereby  $\mathbf{v} \otimes \mathbf{v}$ represents active cell stress, while $-|\mathbf{v}|^2 \mathbb{I}_N $ is the isotropic part. Hence, similarly to  \cite{chauviere2007modeling,corbin2020modeling,DKSS20}, the velocity (and in particular the direction) of glioma cells is also influenced by a weighted sum of gradients (each inferring some limitation): increasing gradients of proton concentration have a repelling effect, and the cells also try to avoid too crowded regions. The constant $\alpha >0$ in \eqref{eq:micro} is a scaling parameter influenced by the gradients of $h$ and $M$; for further details we refer to \cite{corbin2020modeling}. Notice that \eqref{eq:micro} actually implies $\frac{d|\mathbf{v}|^2}{dt} = 0$, which is in line with our assumption of $s=|\vbf |$ being constant.\\[-2ex] 

\noindent
Correspondingly, on the mesoscale we have the following kinetic transport equation (KTE) for the evolution of glioma cell distribution function $p(t,\mathbf{x,v})$:
\begin{align}\label{eq_tumor_KTE}
\nabla _{(t,\xbf ,\vbf )}\cdot ((1,\vbf ,-\alpha \Sbf (\vbf ,h,M))p)=p_t + \nabla_\xbf \cdot (\mathbf{v} p) - \alpha \nabla_{\mathbf{v}} \cdot \left( \Sbf (\mathbf{v},h,M)p  \right) = \mathcal {L}[\lambda]p + \mathscr{P}(M,h,W),
\end{align}
where $\mathcal L[\lambda ]p(t,\xbf ,\vbf)=-\lambda p(t,\xbf ,\vbf)+\lambda \int _VK(\vbf ,\vbf ')p(t,\xbf ,\vbf ')d\vbf '$ is a turning operator with constant turning rate $\lambda >0$ and turning kernel $K$. As in \cite{hillen2006m} and many subsequent works  \cite{conte2020mathematical,corbin2020modeling,Corbin2018,engwer2015glioma,engwer2016effective,kumar2020multiscale} we choose $K(\mathbf{v},\mathbf{v}^{\prime}):=\frac{q(\mathbf{x}, \theta)}{\om}$, meaning that the reorientation of cells is due to contact guidance by tissue, with its orientation distribution $q$. Hence, $\mathcal L[\lambda ]p(t,\xbf ,\vbf)=\lambda (M\frac{q(\xbf ,\hat \vbf )}{\om}-p(t,\xbf ,\vbf ))$. Moreover, as e.g., in \cite{corbin2020modeling} we assume $p$ to be compactly supported in the velocity space. \\[-2ex] 

\noindent
The term $\mathscr{P}(M,h,W)$ models glioma cell proliferation, which is supposed to depend on the amount of cells already present (irrespective of their orientation) and on favorable (nutrients provided by ECs of density $W$) and unfavorable (acidity $h$) factors in the environment. More detailed descriptions account for proliferative interactions mediated by kernels depending on kinetic or activity variables, as in \cite{corbin2020modeling,engwer2016effective} or even differentiate between moving and proliferating cells \cite{EKS,HS}. Here we would rather focus on the macrolevel influences and choose a uniform velocity kernel to characterize such interactions, i.e.
\begin{align}\label{eq_proli_operator}
\mathscr{P}(M,h,W) &= \left( 1 - \frac{M}{K_M}\right)\int _V\left( \mu_1\left(  1 - \frac{h}{K_h}\right)+\mu_2\frac{W/K_W}{1+W/K_W}\right ) \frac{1}{|V|}p(t,\xbf ,\vbf ')d\vbf '\notag \\
&=\frac{\mu_1}{|V|} M \left( 1 - \frac{M}{K_M}\right) \left(  1 - \frac{h}{K_h}\right)+ \frac{\mu_2}{|V|} M\frac{W/K_W}{1+W/K_W} ,
\end{align}
where $\mu_1,\mu_2>0$ are rates related to the cell growth or decay due to acidity and ECs. The positive constants $K_M$ and $K_W$ represent carrying capacities of glioma and endothelial cells, respectively, while $K_h>0$ denotes a threshold of proton concentration, which, when exceeded, leads to glioma cell death. Similarly to \cite{kumar2020multiscale} we consider a logistic type growth (as long as the environment is not too acidic), supplemented by a  limited growth due to the interaction with ECs here representing the nutrient-supplying vasculature.\\[-2ex] 

\noindent
To deduce a macroscopic equation for the dynamics of macroscopic glioma density $M$ from the KTE formulation, in \cite{kumar2020multiscale} was used a  parabolic upscaling\footnote{with the observation that a hyperbolic one leads to a PDE which is too much drift-dominated and not able to reproduce pseudopalisade patterns}; the microscale dynamics in that setting focused on the evolution of an activity variable quantifying the amount of transmembrane units occupied by protons. In \cite{corbin2020modeling}, where velocity dynamics akin to \eqref{eq:micro} was considered (however without limitations as in the denominators of $\Sbf$), it was yet a parabolic scaling leading to the reaction-diffusion-taxis PDE for glioma evolution of the macroscale. Instead,  we will consider here a moment closure approach similar to that in \cite{chauviere2007modeling,hillen2013transport}, ensuring the closure by way of an equilibrium distribution.\\[-2ex] 

\noindent
For that purpose we introduce the following quantities:
\begin{align}
(M\Ubf )(t,\xbf ) &:= \int_V \vbf p(t,\mathbf{x},\vbf) d \vbf, \\
\mathbb{P}(t,\mathbf{x}) &:= \int_V \left( \vbf - \Ubf(t,\xbf) \right) \otimes \left( \vbf -\Ubf (t,\xbf )\right)p(t,\xbf ,\vbf ) d \vbf,\label{eq_pressure_tensor}
\end{align}
commonly designated as momentum and pressure tensor, respectively. Thereby,  $\Ubf $ represents the so-called ensemble (macroscopic) cell velocity.\\[-2ex]

\noindent
Integrating \eqref{eq_tumor_KTE} w.r.t. $\mathbf{v}$ we get
\begin{equation}\label{eq_tumor_ineg_v}
\partial_t M + \nabla_\xbf \cdot (M\Ubf ) ={  \mu_1M \left(1-\frac{M}{K_M}\right)\left(  1 - \frac{h}{K_h}\right)+\mu_2M\frac{W/K_W}{1+W/K_W}}
\end{equation}
Multiplying \eqref{eq_tumor_KTE} by $\mathbf{v}$ and integrating again w.r.t. $\mathbf{v}$ we get
\begin{align}\label{eq_integ_v_times_KTE}
\partial_t(M\Ubf) + \nabla_\xbf  \cdot \int_{V} \mathbf{v} \otimes \mathbf{v} pd\mathbf{v} +\alpha \int_{V} \Sbf (\mathbf{v},h,M)pd\mathbf{v} = \lambda M(\Eti_q - \Ubf) 
\end{align}
From (\ref{eq_pressure_tensor}),
\begin{equation*}
\mathbb P = \int_{V} \mathbf{v} \otimes \mathbf{v}\ p d \mathbf{v} - M \Ubf \otimes \Ubf,
\end{equation*}
hence due to \eqref{eq:bigS} we can rewrite \eqref{eq_integ_v_times_KTE} as
\begin{equation}\label{eq_tumor_with_Eq}
\partial_t(M\Ubf) + \nabla_x \cdot (\mathbb P + M \Ubf \otimes \Ubf) - \alpha (\mathbb P + M \Ubf \otimes \Ubf - s^2M \mathbb{I}_N)\phibf(h,M) = \lambda M(\Eti_q -\Ubf).
\end{equation}
System \eqref{eq_integ_v_times_KTE}, \eqref{eq_tumor_with_Eq} is not closed, since the pressure tensor $\mathbb P$ involves a second order moment of $p$ and is therefore unknown. In order to close the system we proceed as e.g., in \cite{hillen2013transport} and assume that it is close to equilibrium, which also dominates the second order moments. If we also assume that at equilibrium we can neglect cell proliferation (as it indeed happens much slower than the motility-related processes), then in virtue of \eqref{eq_tumor_KTE} we can write 
\begin{align}
p_{\text{equil}} = \frac{q}{\om} M,
\end{align} 
which also leads to 
\begin{align}
\Ubf_{\text{equil}} &= \frac{1}{M} \int_{V}\mathbf{v} p_{\text{equil}} d \mathbf{v} =\Eti _q
\end{align}
and 
\begin{align}
\mathbb P_{\text{equil}} = \int_{V} (\mathbf{v} - \tilde{\mathbb{E}}_q) \otimes (\mathbf{v} - \tilde{\mathbb{E}}_q) \frac{q}{ \om } M d\mathbf{v}= M \Vti_q,
\end{align}
hence at equilibrium the momentum of glioma population is driven by the cell ensemble aligning their motion to the average fiber orientation $\Eti_q$, and the 'population pressure' is generated by the variance-covariance matrix of the orientation distribution $q/\om$ of tissue surrounding the cells. \\[-2ex]

\noindent
Thus, we get from 
\eqref{eq_tumor_with_Eq} 
\begin{align}
\partial_t(M\Ubf) + \nabla_\xbf \cdot (M \Ubf \otimes \Ubf) = -\nabla_\xbf \cdot \mathbb P_{\text{equil}} + \alpha \left(   \mathbb P_{\text{equil}} + M \Ubf \otimes \Ubf - s^2 M \mathbb{I}_N\right)\phibf(h,M) + \lambda (M\Eti_q-M\Ubf). \label{eq_tumor_momentum_final}
\end{align}
The middle term on the right hand side of \eqref{eq_tumor_momentum_final} stems from the description of forces acting on the cells; we suppose that the tensor $M\Ubf \otimes \Ubf$ therein is itself relating to the equilibrium situation, thus replace it by $M\Eti_q\otimes \Eti_q$. On the other hand, the left side in  \eqref{eq_tumor_momentum_final} can be interpreted as directional  derivative of the momentum $M\Ubf $ in the direction $\Ubf $ of the 'cell flow'. As in \cite{hillen2013transport} we consider that the cell flux relaxes quickly to its equilibrium, so that
\begin{align}
M\Ubf =M\Eti _q-\frac{1}{\lambda }\nabla_\xbf \cdot \mathbb P_{\text{equil}}+\frac{\alpha }{\lambda }\left(   \mathbb P_{\text{equil}} + M \Eti_q \otimes \Eti_q - s^2 M \mathbb{I}_N\right)\phibf(h,M).
\end{align} 
\noindent
If, furthermore, we assume that the tissue is undirected\footnote{as suggested by the simulation results in \cite{kumar2020multiscale}}, this translates into $\Eti _q={\bf 0}$, which simplifies the above expression of the momentum:
\begin{align}\label{eq:MU}
M\Ubf =-\nabla_\xbf \cdot (\mathbb D_T(\xbf )M)+\alpha (\mathbb D_T(\xbf )-\frac{s^2}{\lambda }\mathbb{I}_N)M\phibf (h,M),
\end{align}
with $\mathbb D_T(\xbf ):=\frac{1}{\lambda }\int _V\vbf \otimes \vbf \frac{q(\xbf ,\hat \vbf)}{\om }d\vbf $ denoting as in previous works (see e.g. \cite{engwer2015glioma,kumar2020multiscale}) the so-called tumor diffusion tensor.\\[-2ex]

\noindent
Plugging \eqref{eq:MU} into \eqref{eq_tumor_ineg_v} then leads to 
the macroscopic PDE for glioma evolution
\begin{equation}\label{eq_tumor_macro_1}
\partial_t M =  \nabla \nabla : (\mathbb{D}_T M) + \alpha \nabla \cdot \left( \left( \frac{s^2}{\lambda}\mathbb{I}_N- \mathbb{D}_T\right) \phibf(h,M)\ M \right) + { \mu_1M\left(1-\frac{M}{K_M} \right) \left(  1 - \frac{h}{K_h}\right)+\mu_2\frac{MW/K_W}{1+W/K_W}},
\end{equation}
which is of course merely an approximation, in view of the many assumptions made above. \\[-2ex]

\noindent
The myopic diffusion term in \eqref{eq_tumor_macro_1} is made of a drift and a 'classical' diffusion term:
$$\nabla \nabla : (\mathbb{D}_T M)=\nabla \cdot (\nabla\cdot \mathbb{D}_T\ M)+\nabla \cdot (\mathbb{D}_T\nabla M),$$
hence using the expression of $\phibf (h,M)$ we can rewrite \eqref{eq_tumor_macro_1} in the form of a PDE in which both self-diffusion (first term in \eqref{eq:eq_tumor_macro_FL}) and repellent pH-taxis (third term in \eqref{eq:eq_tumor_macro_FL}) are (at least partially) flux-limited:
\begin{align}\label{eq:eq_tumor_macro_FL}
\partial_t M =&\nabla \cdot \left (\left (\mathbb{D}_T+\alpha \gamma _2\left( \frac{s^2}{\lambda}\mathbb{I}_N- \mathbb{D}_T\right)\frac{M}{\sqrt{K_M^2/X^2+|\nabla M|^2}}\right )\nabla M\right )+\nabla \cdot (\nabla \cdot \mathbb{D}_T\ M)\notag \\
&+\nabla \cdot \left(\alpha \gamma_1\left( \frac{s^2}{\lambda}\mathbb{I}_N- \mathbb{D}_T\right)M\frac{\nabla h}{\sqrt{K_h^2/X^2+|\nabla h|^2}}\right )+ \mu_1M\left(1-\frac{M}{K_M} \right) \left(  1 - \frac{h}{K_h}\right)+\mu_2\frac{MW/K_W}{1+W/K_W}.
\end{align}
Notice that in order to have a genuine repellent pH-taxis the tensor $\frac{s^2}{\lambda}\mathbb{I}_N- \mathbb{D}_T$ should be positive definite, which also ensures a nondegenerating, proper diffusion\footnote{without being, however, a necessary condition therefor}.  Under the hypothesis of undirected tissue, i.e. for $\mathbb E_q={\bf 0}$ this amounts to $\mathbb{I}_N-\mathbb V_q=\mathbb{I}_N-\int _{\Sbb}\thetabf \otimes \thetabf q(\thetabf) d\thetabf$ being positive definite. This is, for instance, the case if $q$ is the so-called peanut distribution \cite{Painter2013}: $$q(\xbf ,\thetabf)=\frac{N}{|\Sbb|\text{ trace }\mathbb D_{water}(\xbf )}\thetabf ^T\mathbb D_{water}(\xbf )\thetabf,$$ where $\mathbb D_{water}(\xbf )$ denotes the (positive definite and usually diagonalized) water diffusion tensor. Same applies to our choice of $q$ in Subsection \ref{subsec:tissue}. 

\noindent
A much simplified situation is encountered when the stress tensor $\mathbb B(\vbf )$ has only an isotropic part, while the active cell stress is neglected. The above deduction then directly yields instead of \eqref{eq:eq_tumor_macro_FL}:
\begin{align*}
\partial_t M =&\nabla \cdot \left (\left (\mathbb{D}_T+\alpha \gamma _2\frac{s^2}{\lambda}\mathbb{I}_N\frac{M}{\sqrt{K_M^2/X^2+|\nabla M|^2}}\right )\nabla M\right )+\nabla \cdot (\nabla \cdot \mathbb{D}_T\ M)\notag \\
&+\nabla \cdot \left(\alpha \gamma_1 \frac{s^2}{\lambda}M\frac{\nabla h}{\sqrt{K_h^2/X^2+|\nabla h|^2}}\right )+ \mu_1M\left(1-\frac{M}{K_M} \right) \left(  1 - \frac{h}{K_h}\right)+\mu_2\frac{MW/K_W}{1+W/K_W}.
\end{align*}

\noindent
For yet another choice of the right hand side in \eqref{eq:micro} leading in a different, but related way to a flux-limited reaction-diffusion-taxis PDE for glioma evolving in a tissue (also under the influence of acidity and vascularization) we refer to \cite{DKSS20}. 

\subsection{Endothelial cells}\label{subsec:ECs}

We turn now to obtaining a macroscopic description for 
the evolution of ECs. For this purpose we can reproduce the derivation made in \cite{conte2020mathematical}, the only difference being that ECs are supposed to respond here to (gradients of) VEGF concentration rather than proliferating glioma cells.\footnote{Indeed, the latter assumption would be inappropriate in the present context of glioma patterns, where most of the cells are migrating to avoid hypoxic regions, only starting to proliferate when there is enough vascularization.} Therefore, we will only sketch the main steps and refer for details to \cite{conte2020mathematical}.\\[-2ex]

\noindent
We start with the KTE formulation for mesoscopic dynamics of EC distribution function $w(t,\xbf ,\vthet)$: 
\begin{equation}{\label{eq_KTE_EC}}
w_t + \nabla_\xbf  \cdot (\vthet w) = \mathscr{L}_w[\eta]w + \mu_w(W,g)w,
\end{equation}
with turning operator 
\begin{align}
\mathscr{L}_w[\eta] = -\eta(\vthet,g )w(t,\mathbf{x},\vthet) + \int_{\Theta} \frac{1}{|\Theta|} \eta (\vthet^{\prime},g)w(t,\xbf,\vthet^{\prime})d \vthet^{\prime},
\end{align}
involving a uniform turning kernel and the turning rate
\begin{align}
\eta(\vthet,g) &= \eta_0 e^{-a(g)D_t g} \label{eq_EC_turning_rate}
\end{align}
depending upon the pathwise gradient $D_t g = g_t + \vthet \cdot \nabla g$ of VEGF concentration. The coefficient function $a(g)$ takes into account the way in which ECs perceive VEGF and respond to it, correspondingly adapting their turning. Following \cite{conte2020mathematical} we choose $a(g)=\chi_a\frac{K_R}{(K_R+g/K_g)^2}$, with $\chi_a, K_R >0$ constants. This corresponds to the rate of change of the expression $y^*(g)=\frac{g/K_g}{g/K_g+K_R}$ representing the equilibrium of EC receptor binding dynamics to $g$; the constant $\chi_a$ scales changes in the turning rate per unit of change in $dy^*/dt$.\\[-2ex]

\noindent  
The coefficient $\mu_w(W,g)$ in the proliferation term is chosen as 
\begin{equation}
\mu_w(W,g)=\mu_W \left(1-\frac{W}{K_W}\right)\frac{g}{K_g},
\end{equation}
with $\mu_W,K_W,K_g>0$ constants representing EC growth rate, EC carrying capacity, and maximum amount of VEGF, respectively.

\noindent
A parabolic scaling $t\leadsto \eps^2t$, $\xbf \leadsto \eps \xbf$ performed in the usual way in combination with a Hilbert expansion $ w = w_0 + \epsilon w_1 + \epsilon^2 w_2 + ... $, a linearization of $\eta $, and the assumption that EC proliferation is much slower than migration then leads to the reaction-diffusion-chemotaxis PDE for the leading term in the (macroscopic) Hilbert expansion  $W = W_0 + \epsilon W_1 + \epsilon^2 W_2 + ...$ (for details compare \cite{conte2020mathematical}):

\begin{equation}\label{eq_macro_EC_1}
\partial_t W_0 - \nabla  \cdot \left(\mathbb{D}_{\mathbb{EC}} \nabla W_0 \right) + \nabla  \cdot \left( \bbchi_a(g)W_0 \nabla  g\right) = \mu_w (W_0,g)W_0,
\end{equation}
where $\mathbb{D}_{\mathbb{EC}}(\mathbf{x}):= \frac{\sigma^2}{N \eta_0}\mathbb{I}_N$ and $\bbchi_a(g):= \frac{\sigma^2 a(g)}{N}\mathbb{I}_N = \eta_0 a(g)\mathbb{D}_{\mathbb{EC}}(\mathbf{x})$. \\[-2ex]

\noindent
In the sequel we will (formally) approximate the evolution of $W$ by that of its leading term $W_0$ and will use the notation $W$ instead of $W_0$.

\subsection{Full macroscopic system}

Equations  \eqref{eq:eq_tumor_macro_FL} and \eqref{eq_macro_EC_1} are coupled with the following equations set directly on the macroscopic scale and describing the dynamics of VEGF and acidity:
\begin{align}
g_t &= D_g  \Delta g + \gamma_g \frac{M \frac{h}{K_h}}{M+ K_M} - \zeta_g g\frac{W}{K_W},\\
h_t &= D_h \Delta h + \gamma_h \frac{M}{K_M + M} - \zeta_h h \frac{W}{K_W} \label{eq_acidity_macro_dim},
\end{align}
where, $D_i$, $\gamma_i$ and $ \zeta_i$ , for $i = h,g$, are positive constants representing diffusion coefficients, production, and uptake rates for the involved quantities.\\[-2ex]

\noindent
The above PDE system needs to be supplied with initial conditions; they  will be addressed in Section \ref{sec:numerics}. Furthermore, since we are interested in the solution behavior inside a bounded region (representing e.g., a slice of a histologic sample), we restrict our equations (hitherto written for $\xbf \in \R^N$) to a bounded domain $\Om \subset \R^N$ and consider no-flux boundary conditions for $M,W,g,h$. They are naturally assumed for $g,h$ and can be obtained (at least for $W$) during the upscaling process, as detailed in \cite{corbin2020modeling}. 

%
%

\section{Numerical simulations}\label{sec:numerics}

This section is dedicated to numerical simulations of the macroscopic system   obtained in Section \ref{sec:modeling} for the quantities $M,W,g,h$.  \\[-2ex]

\subsection{Nondimensionalization and choice of parameters}\label{par:nondim}

We consider the following rescalings of the hitherto dimensional quantities:
\begin{align*}
& \tilde{M}:= \frac{M}{K_M}, \quad \tilde{h}:= \frac{h}{K_h}, \quad \tilde{W}:= \frac{W}{K_W}, \quad \tilde{g}:= \frac{g}{K_g}, \quad 
\tilde{t} := \mu_1 t, \quad  \tilde{x} := x\sqrt{\frac{\mu_1}{D_h}},\\
&\tilde{\mathbb{D}}_T := \frac{\mathbb{D}_T}{D_h}, \quad \tilde{\mathbb{D}}_{EC}: = \frac{\mathbb{D}_{EC}}{D_h}, \quad\tilde{D}_g:= \frac{D_g}{D_h},\quad \tilde D_s=\frac{s^2}{\lambda D_h},\quad  \tilde{\mu}_c := \frac{\mu_2}{\mu_1},\quad \tilde{\mu}_e := \frac{\mu_W}{\mu_1},\\
&\tilde{\chi}_W:= \chi_a\eta_0, \quad \tilde{\gamma}_g: = \frac{\gamma_g}{\mu_1 K_g}, \quad \tilde{\zeta}_g: = \frac{\zeta_g}{\mu_1}, \quad \tilde{\gamma}_h := \frac{\gamma_h}{\mu_1 K_h}, \quad \tilde{\zeta}_h := \frac{\zeta _h}{\mu_1},\\ 
& \tilde{\phi}(\tilde{ h}, \tilde{ M}): = \frac{\gamma_1 \nabla_{\tilde{x}} \tilde{h}}{\sqrt{1 + |\nabla_{\tilde{x}} \tilde{h}|^2}} + \frac{\gamma_2 \nabla_{\tilde{x}} \tilde{M}}{\sqrt{1 + |\nabla_{\tilde{x}} \tilde{M}|^2}},\quad X=\sqrt{\frac{D_h}{\mu_1}}. 
\end{align*}
After dropping the tildes for simplicity of notation we therewith get the following nondimensionalized system:
\begin{subequations}\label{eq:macro-nondim}
\begin{align}
M_t &= \nabla \nabla : (\mathbb{D}_T M) + \alpha \nabla \cdot \left(\left( D_s \mathbb{I}_N - \mathbb{D}_T \right) \phi(h,M)M \right) + M \left(1-M \right)\left(1-h\right) + \mu_c \frac{MW}{1+W}\label{eq:macro-nondim-a}\\
W_t &= \nabla \cdot \left(\mathbb{D}_{EC} \nabla W \right) - \nabla \cdot \left( \chi_W \frac{K_R}{(K_R+g)^2}\mathbb{D}_{EC}  W \nabla g \right) + \mu_e g W\left(1-W\right) \\
h_t & = \nabla h + \gamma_h \frac{M}{1+M} - \zeta_h hW\\
g_t & = D_g \nabla g + \gamma_g \frac{M h}{1+ M} -\zeta_g gW,
\end{align}
\end{subequations}
whose parameters are given by the above scalings and the values provided in Table \ref{table}. 
\subsection{Description of tissue}\label{subsec:tissue}

Typically the brain structure of each patient is assessed via diffusion tensor imaging (DTI) and the therewith provided water diffusion tensor $\mathbb D_{water}$ already occurring in Subsection \ref{subsec:glioma} above in the peanut distribution. The space scale on which glioma patterns are observed is, however, smaller than the best resolution of DTI, which does not go below the size of a voxel (ca. 1 mm$^3$): pseudopalisades are structures with a medium width of $200-400 \mu m$; for more details see the introduction of \cite{kumar2020multiscale} and references therein. Therefore, as in that reference, we use  
an artificially created tissue with a corresponding water diffusion tensor as introduced in \cite{Painter2013}, in order to analyze the effect of tissue anisotropy. We consider the water diffusion tensor as
\begin{equation}
\mathbb D_{water}(x,y) = \begin{pmatrix}
0.5-d(x,y) & 0 \\
0 & 0.5+d(x,y),
\end{pmatrix}
\end{equation}
where $d(x,y) = 0.25e^{-0.005(x-450)^2} - 0.25e^{-0.005(y-450)^2}$.
A combination of uniform and von Mises-Fisher distributions is considered for the orientation distribution of tissue fibers:
\begin{equation}\label{eq_q}
q(\mathbf{x},\thetabf) = \frac{\delta}{2\pi} + (1-\delta)\left(\frac{1}{2\pi I_0(k(\mathbf{x}))}\right) \frac{e^{k(\mathbf{x})\varphi_1(\mathbf{x})\cdot \thetabf} + e^{-k(\mathbf{x})\varphi_1(\mathbf{x})\cdot \thetabf}}{2}
\end{equation}
with $\delta \in \left[0,1\right]$ being a weighting coefficient, $\varphi_1$ denotes the eigenvector corresponding to the leading eigenvalue of $\mathbb D_{water}(\mathbf{ x})$ and $I_0$ is the modified Bessel function of first kind of order $0$. Furthermore, $\thetabf = \left(\cos \xi , \sin \xi \right)$ for $\xi \in \left[0,2\pi\right]$,  and $k(\mathbf{ x}$) = $\kappa$FA$(\mathbf{ x})$, where $\kappa \geq 0$ describes the sensitivity of cells towards orientation of tissue fibers and 
\begin{equation*}
FA(\mathbf{ x}) = \frac{|\lambda_1 -\lambda_2|}{\sqrt{\lambda_1^2 + \lambda_2^2}}
\end{equation*}
denotes the  fractional anisotropy in 2D \cite{Painter2013} with $\lambda_i$ ($i=1,2$) denoting the eigenvalues of $\mathbb D_{water}(\xbf )$.\\

\begin{table}[!ht]	
	\centering
	\begin{tabular}{p{1.3cm} p{6.0cm} p{4.2cm} p{2.8cm}}
		\toprule[1.5pt]
		Parameter &\ Meaning & Value & Reference\\[0.75ex]
		\hline \\[-1.5ex]
		$K_{M}$   & glioma carrying capacity    &$0.3-0.8 \text{ cells/$\mu$m}^2$&this work, \cite{TCdim}\\
		$K_{h}$ &  acidity threshold for cancer cell death   & $10^{-6.4} \text{ mol/L}$  & \cite{Webb2011} \\
		$s$  & speed of glioma cells & $15-20\ \text{$\mu$m/h}$&  
		\cite{diao2019,Prag}\\
		$\lambda$&   turning frequency coefficient   & $360$ 1/h&  \cite{engwer2015glioma,sidani2007cofilin}\\
		$\gamma_h$ & proton production rate & $3.6\cdot 10^{-6}$ mol /(L$\cdot $h)  & this work, \cite{Martin2} \\
		$\zeta_h$ & proton removal rate  & $3.6\cdot 10^{-5}-3.6\cdot 10^{-4}$ /h& this work\\
		$D_h$ & acidity diffusion coefficient & \begin{minipage}{6cm}$1.8\cdot 10^{2}-3.6\cdot 10^{4}$ $\mu$m$^2$/h\end{minipage} & this work, \cite{kumar2020multiscale} \\
		$\mu_1$ & glioma growth/decay rate, influ- enced by acidity & $(4.16-8.3)\cdot 10^{-3}$ /h & \cite{stein2007mathematical, eikenberry2009virtual} \\
		$\mu_2$ & glioma growth rate, influenced by ECs & $(3-3.95)\cdot 10^{-2}$ /h & this work  \\
		$\alpha$ &  advection constant   & $10^{3} $  & \cite{corbin2020modeling} \\
		$\gamma_1$   & weight coefficient related to acidic stress &$3\cdot 10^{-4}-3 \cdot 10^{-5}$& this work\\ 
		$\gamma_2$   & weight coefficient related to cell population pressure& $10^{-4}-10^{-3}$&this work\\ 
		$\sigma$   & speed of ECs  &$15-20$ $\mu$m/h& \cite{czirok2013endothelial,conte2020mathematical}\\ 
		$\eta_0$   & turning rate of ECs  &$36$ /h&this work\\ 
		$\mu_w$   & EC proliferation rate  &$(1.25-2.08)\cdot 10^{-3}$ /h &this work, \cite{conte2020mathematical}\\ 
		$\chi_a$   & coefficient of chemotactic sen\-si- ti\-vi\-ty of ECs     &$0.72-21.6$ h&this work\\ 
		$K_W$   & carrying capacity of ECs     &$0.1-0.3$ cells/$\mu $m$^{2}$& this work, \cite{ECdim}\\ 
		$D_g$ & diffusion coefficient of VEGF & \begin{minipage}{6cm}$3.6 \cdot 10^2 - 1.04\cdot10^{5}$ $\mu$m$^2$/h\end{minipage}& \cite{plank2004mathematical},\cite{shamsi2018mathematical}\\
		$K_g$ & VEGF threshold value & $8\cdot 10^{-8}-10^{-6}$mol/L & this work, \cite{gevertz2006modeling} \\
		$\gamma_g$ & VEGF production rate & $ 3.6 \cdot 10^{-7}$ mol/(L$\cdot$h)& \cite{plank2004mathematical}\\
		$\zeta_g$ & VEGF uptake rate & $3.6 \cdot 10^{-7}- 3.6 \cdot 10^{-6}$/h& this work \\
		\bottomrule[1.5pt]
	\end{tabular}
	\caption{Parameters (dimensional quantities)}\label{table} 
\end{table}
\subsection{Initial conditions}\label{subsec:ICs}
We consider the following initial densities and concentrations for the quantities involved:
\begin{subequations}\label{eq:ICs}
	\begin{align}
	M(0,\mathbf{ x}) &= 0.95 \cdot K_M\cdot \left( e^{\frac{-(x-500)^2- (y-500)^2}{2(25)^2}}    + e^{\frac{-(x-600)^2- (y-500)^2}{2 (20)^2}} + e^{\frac{-(x-300)^2- (y-400)^2}{2(10)^2}}\right)\label{eq:IC-M}\\
	h(0,\mathbf{ x}) & = 10^{-7}e^{\frac{-(x-500)^2- (y-500)^2}{2 (15)^2}} + 10^{-7}e^{\frac{-(x-600)^2- (y-500)^2}{2(10)^2}} + 10^{-6.4} e^{\frac{-(x-300)^2- (y-400)^2}{2(7.5)^2}}\label{eq:IC-h}\\
	g(0,\mathbf{ x}) & = 10^{-14}e^{\frac{-(x-500)^2- (y-500)^2}{2 (15)^2}} + 10^{-12}e^{\frac{-(x-600)^2- (y-500)^2}{2(10)^2}} + 10^{-11} e^{\frac{-(x-300)^2- (y-400)^2}{2(7.5)^2}}\label{eq:IC-G}\\	
	W(0,\mathbf{ x}) &= 0.02 \cdot K_W \cdot  sin^{10}\left(0.002\pi y\right) \left( e^{ \frac{-(x-50)^2}{0.002} } + e^{ \frac{-(x-950)^2}{0.002} } \right). \label{eq:IC-W}
	\end{align}
\end{subequations}

\begin{figure}[!htbp]
	\centering
	\begin{subfigure}[b]{0.25\textwidth}
		{\includegraphics[width=1\linewidth]{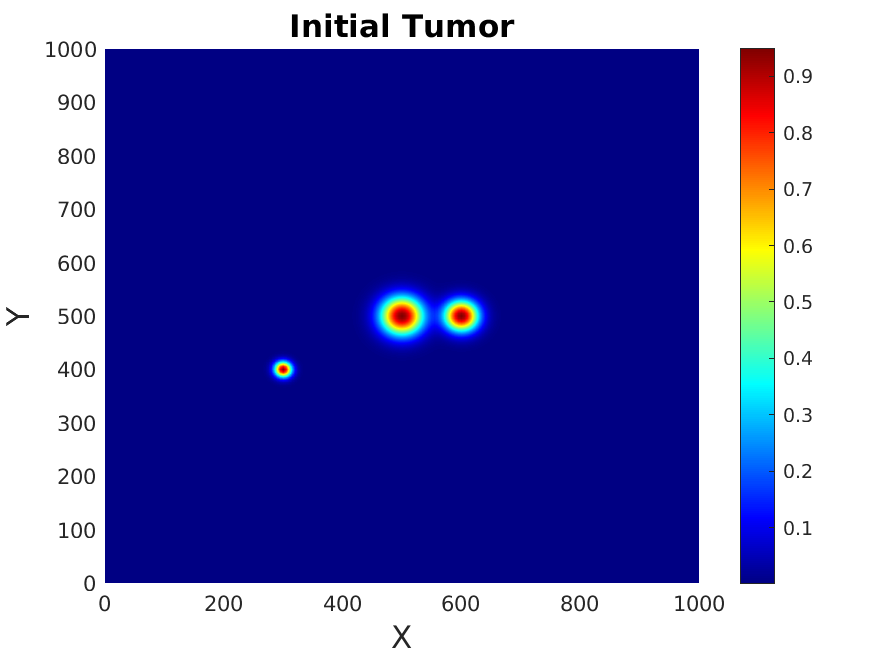}\subcaption{Tumor}}
	\end{subfigure}\hfill
	\begin{subfigure}[b]{0.25\textwidth}
		{\includegraphics[width=1\linewidth]{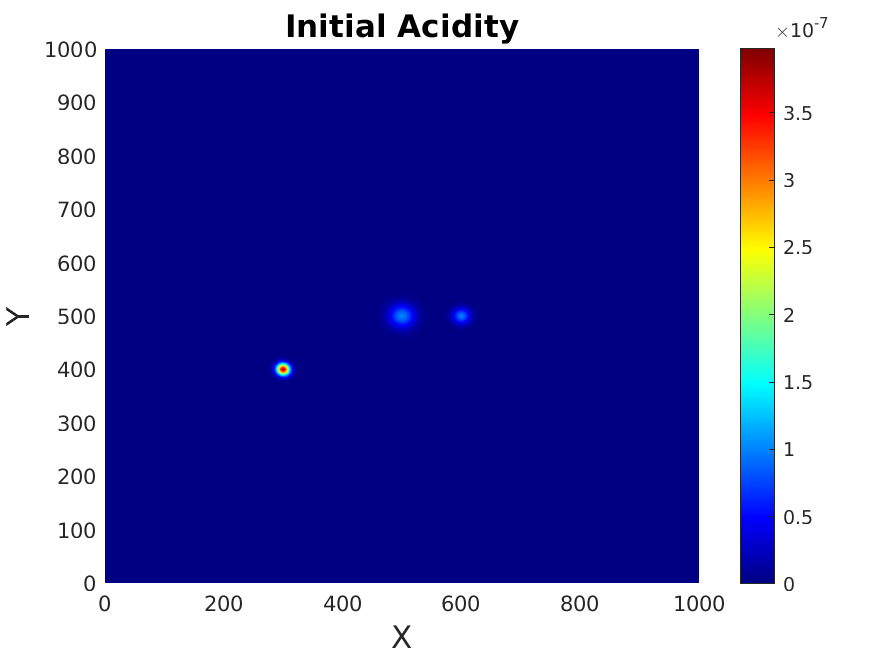}\subcaption{Acidity}}
	\end{subfigure}\hfill
\begin{subfigure}[b]{0.25\textwidth}
		{\includegraphics[width=1\linewidth]{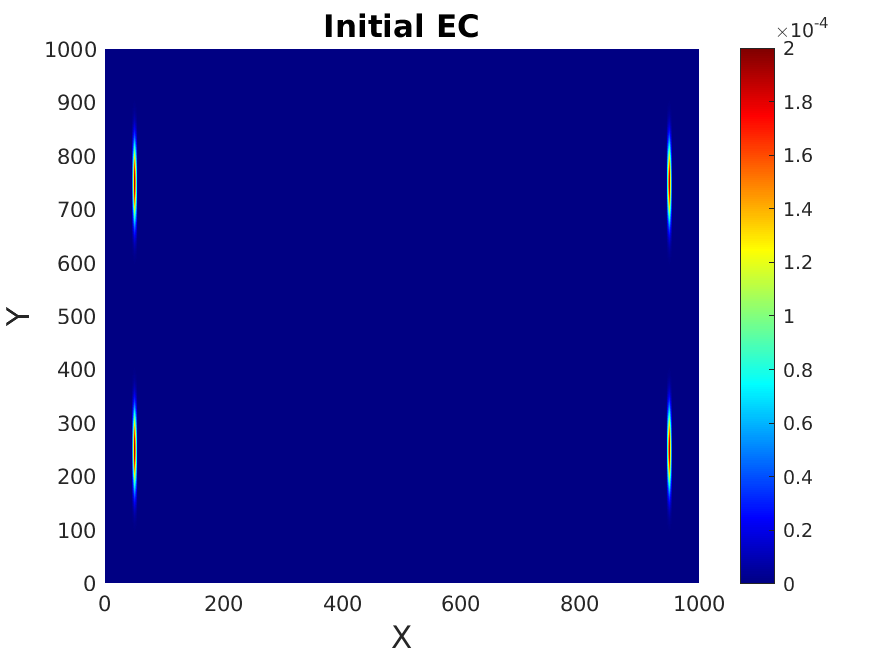}\subcaption{ECs}}
	\end{subfigure}\hfill
\begin{subfigure}[b]{0.25\textwidth}
		{\includegraphics[width=1\linewidth]{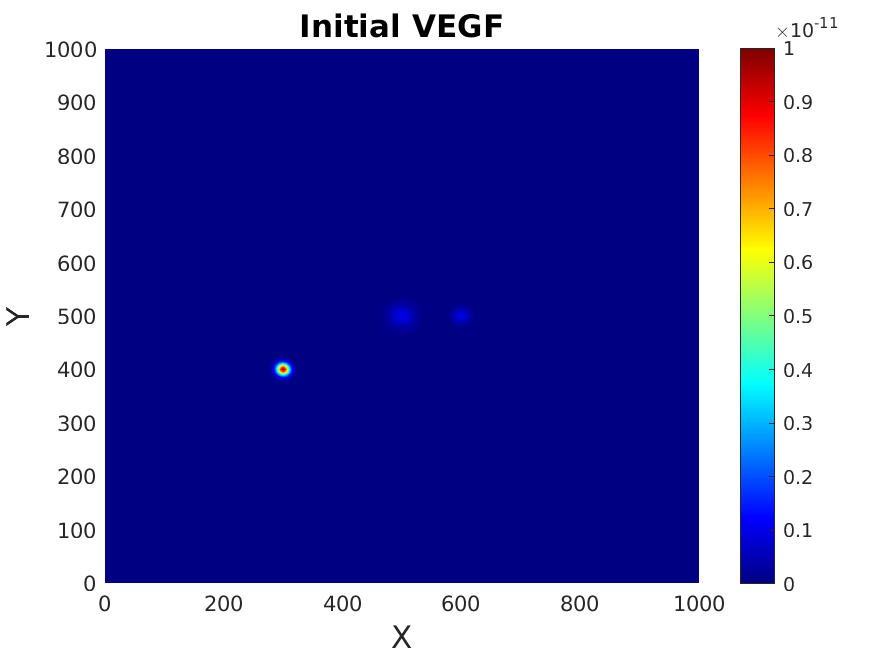}\subcaption{VEGF}}%
	\end{subfigure}
\caption{Initial conditions}
\end{figure}
\subsection{Numerical method}
We solve system \eqref{eq:macro-nondim} with no-flux boundary conditions and with the initial data given above on a square domain $[0,1000] \times [0,1000]$ (in $\mu m$) by using a finite difference method. The diffusion parts of acidity\footnote{here and subsequently we will mean by 'acidity' the concentration of protons, keeping in mind its relationship to the corresponding pH value; the latter is usually refer to 'acidity' in medical practice}, ECs and VEGF are discretized using a standard central difference scheme,  while a nonnegativity-preserving discretization scheme, as proposed in \cite{weickert1998anisotropic}, is used for the anisotropic diffusion of glioma cells. We apply a first order upwind scheme for the taxis terms in glioma and EC equations. For time discretization, an IMEX scheme is used for the acidity, EC, and VEFG equations, thereby treating the diffusion parts implicitly and discretizing the taxis and source terms with an explicit Euler method. We use an explicit Euler scheme for each term in the glioma equation.

\subsection{Numerical experiments}

We addressed in \cite{kumar2020multiscale} the case with isotropic tissue ($\delta =1$) as well as that with anisotropy ($\delta =0.2$, $k=3$); here similar observations can be made in this respect, thus we only consider the latter case. The parameters we use are given in Table \ref{table}, with changes correspondingly specified for each scenario.

\begin{figure}[!htbp]
	\begin{minipage}[hstb]{.24\linewidth}
		\raisebox{1.2cm}{\rotatebox[origin=t]{90}{30 days}}{\includegraphics[width=1\linewidth, height = 3cm]{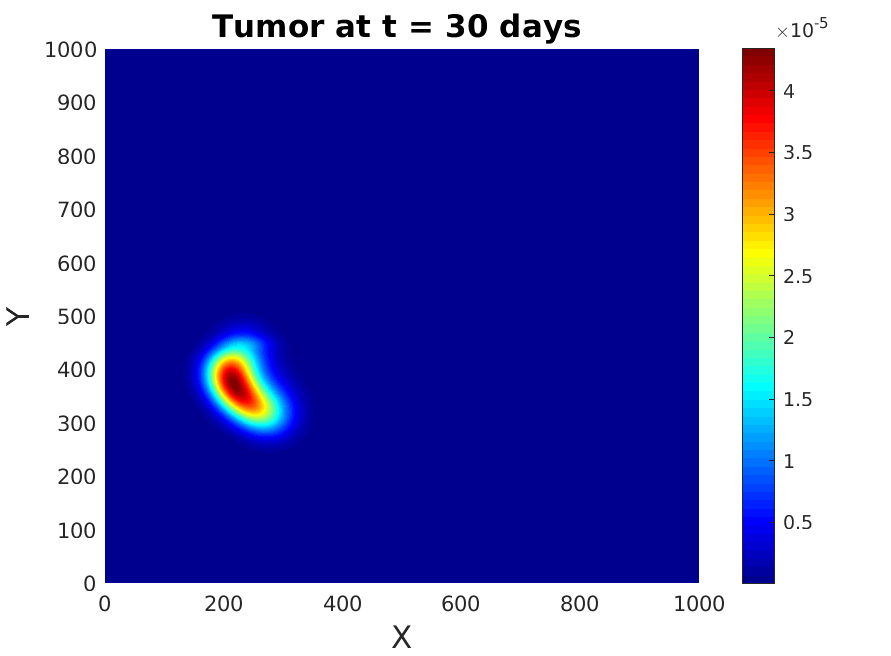}}\\
		\raisebox{1.2cm}{\rotatebox[origin=t]{90}{60 days}}{\includegraphics[width=1\linewidth, height = 3cm]{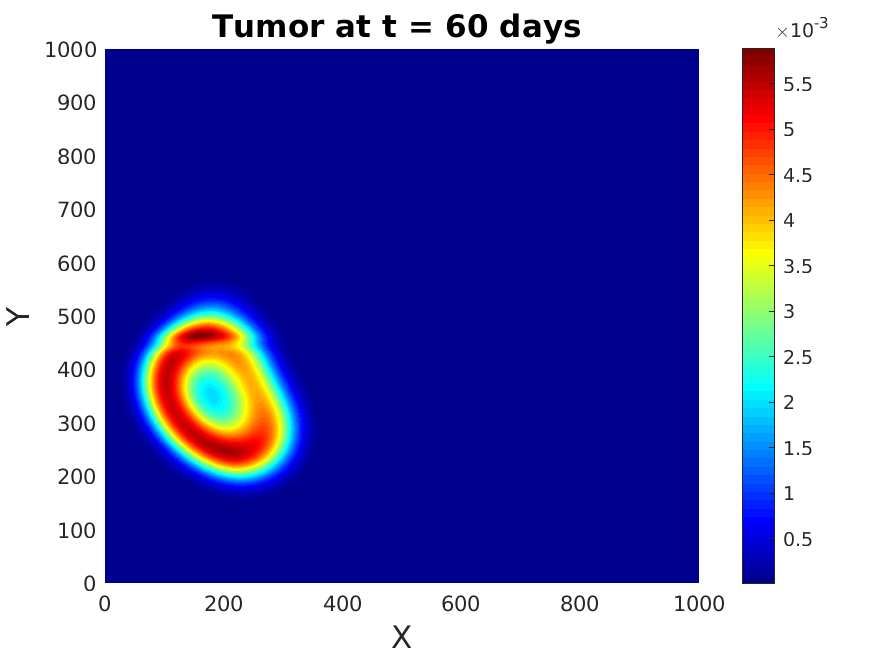}}\\
		\raisebox{1.2cm}{\rotatebox[origin=t]{90}{90 days}}{\includegraphics[width=1\linewidth, height = 3cm]{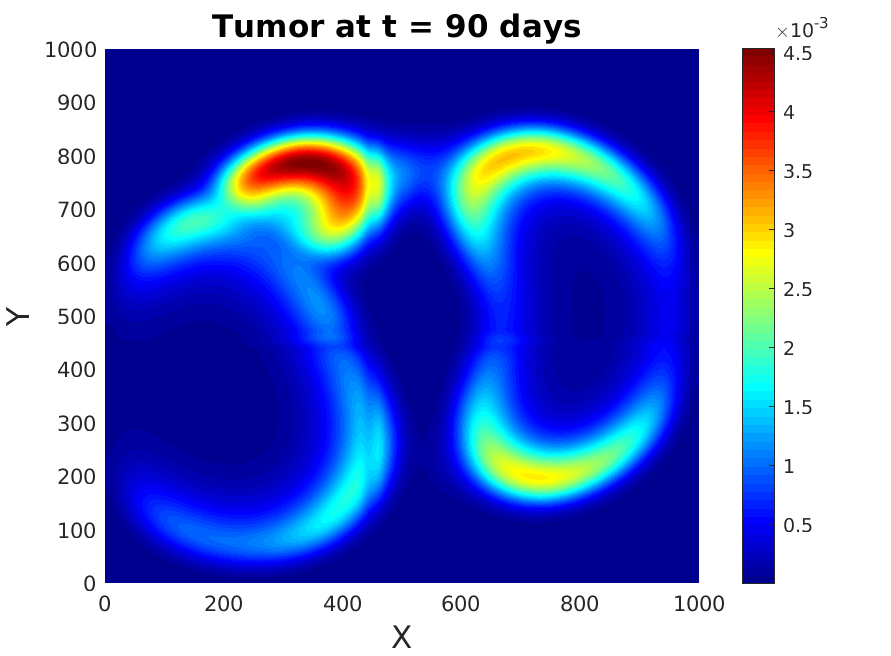}}\\
		\raisebox{1.2cm}{\rotatebox[origin=t]{90}{300 days}}{\includegraphics[width=1\linewidth, height = 3cm]{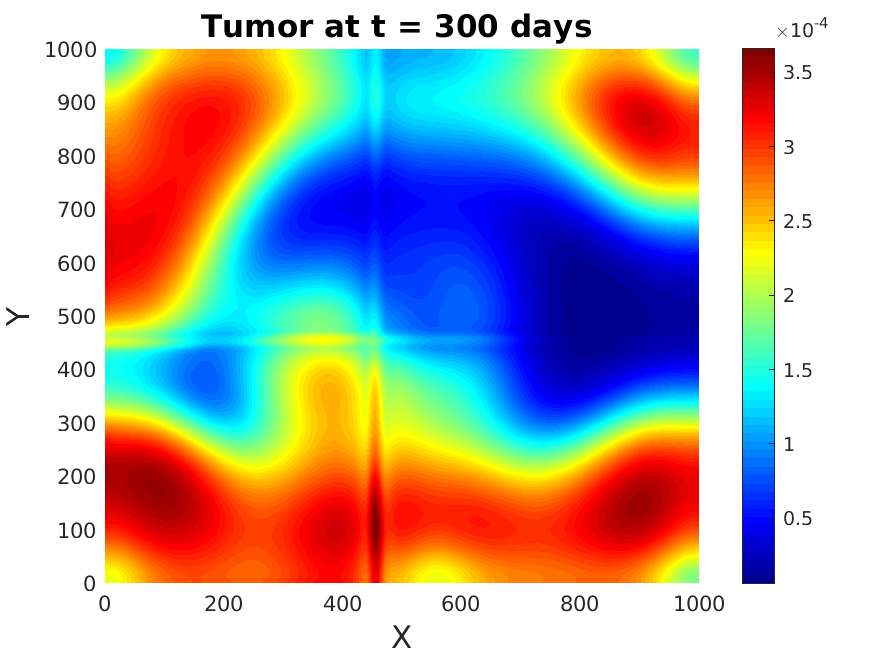}}\\
		\raisebox{1.2cm}{\rotatebox[origin=t]{90}{330 days}}{\includegraphics[width=1\linewidth, height = 3cm]{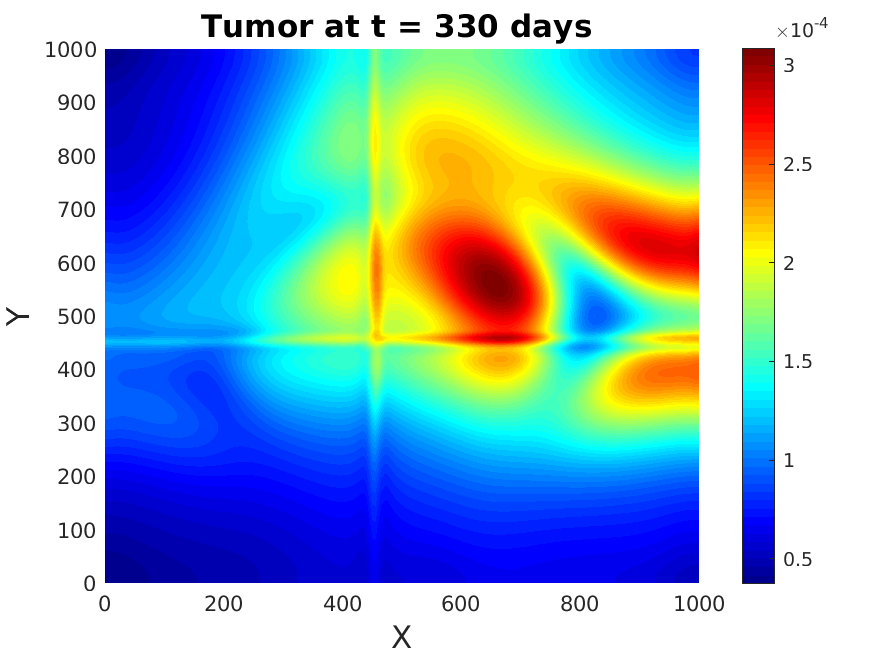}}\\
		\raisebox{1.2cm}{\rotatebox[origin=t]{90}{390 days}}{\includegraphics[width=1\linewidth, height = 3cm]{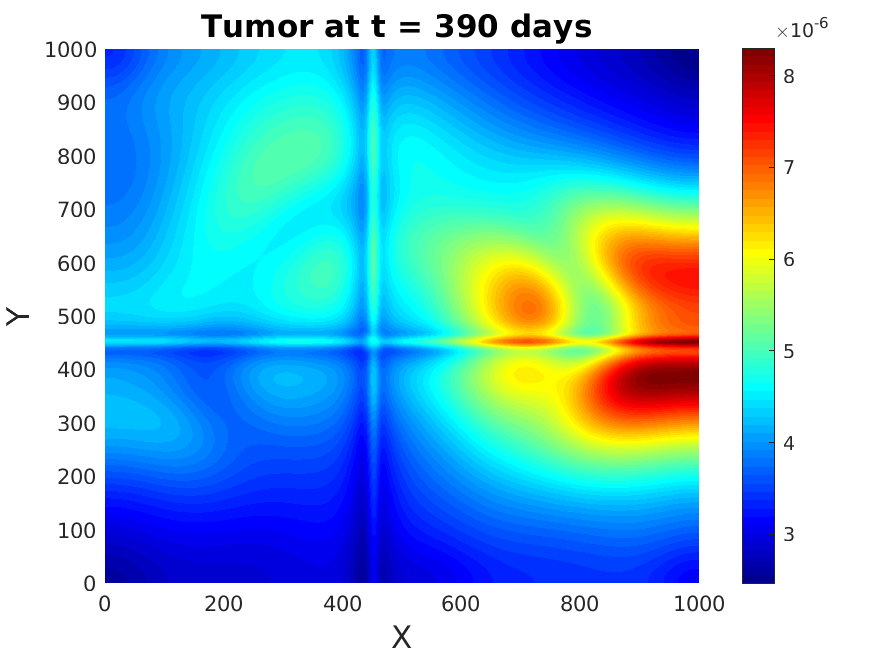}}\\
		\raisebox{1.2cm}{\rotatebox[origin=t]{90}{420 days}}{\includegraphics[width=1\linewidth, height = 3cm]{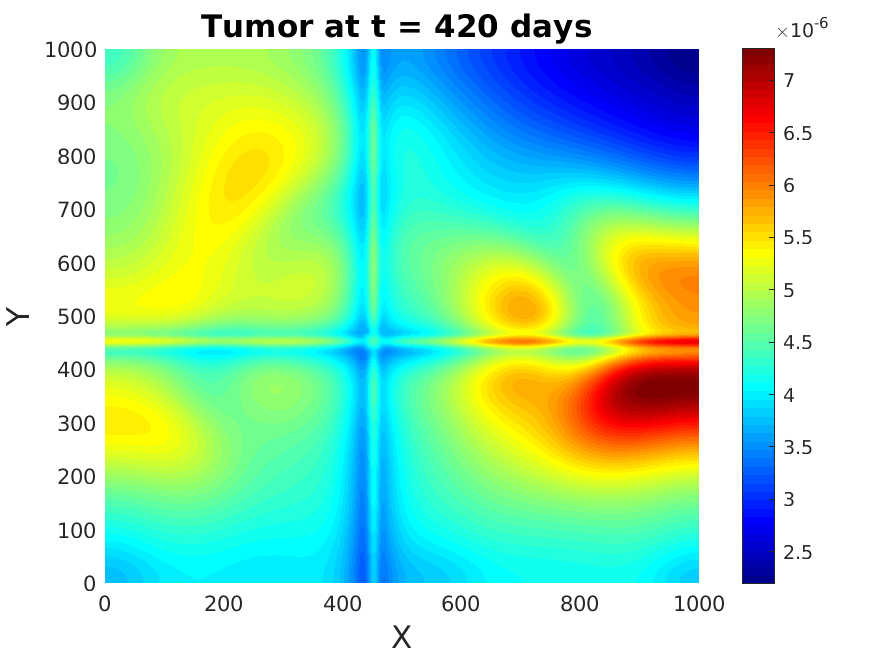}}%
		\subcaption{Tumor}
	\end{minipage}%
	\hspace{0.2cm}
	\begin{minipage}[hstb]{.24\linewidth}
		{\includegraphics[width=1\linewidth, height = 3cm]{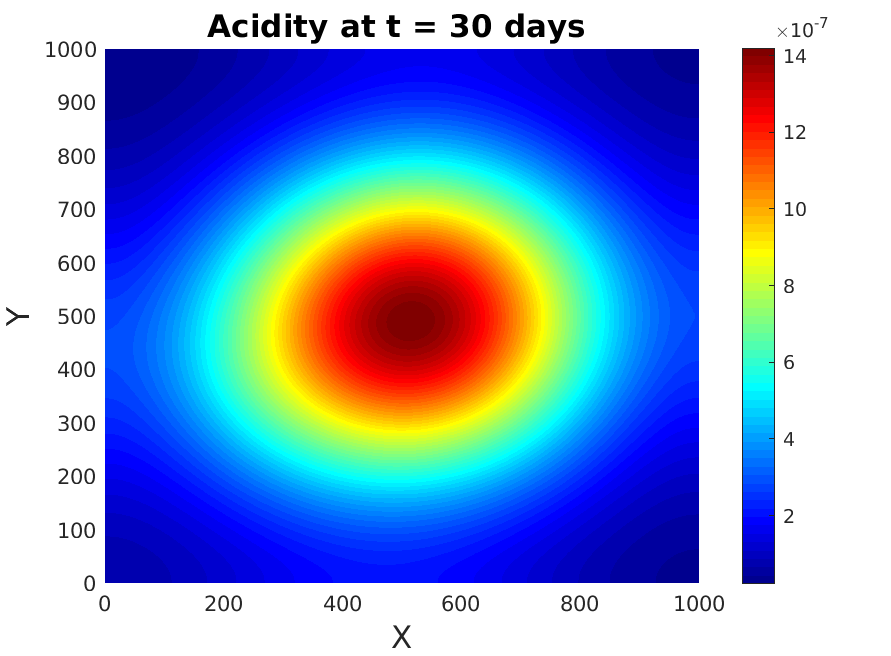}}\\
		{\includegraphics[width=1\linewidth, height = 3cm]{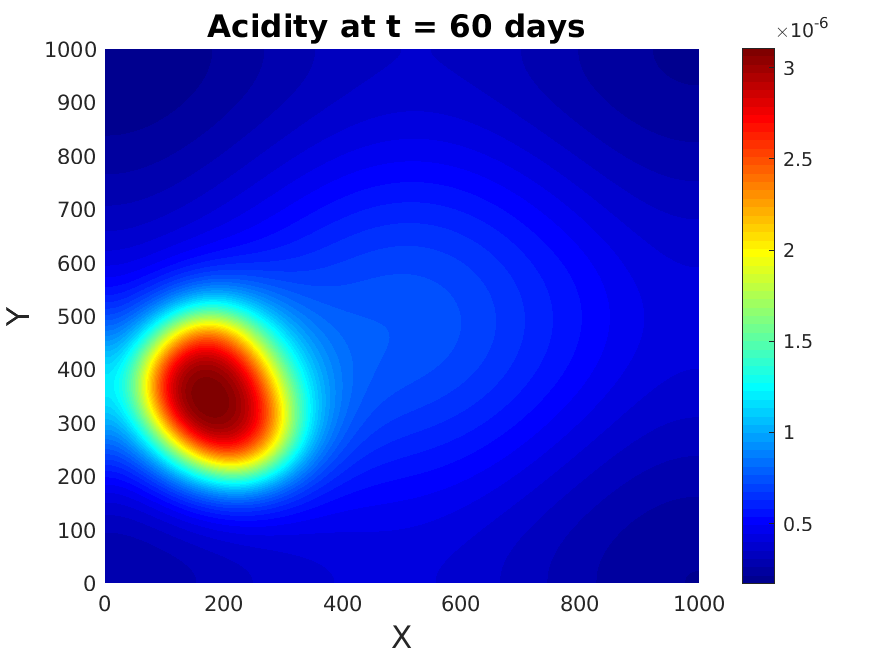}}\\
		{\includegraphics[width=1\linewidth, height = 3cm]{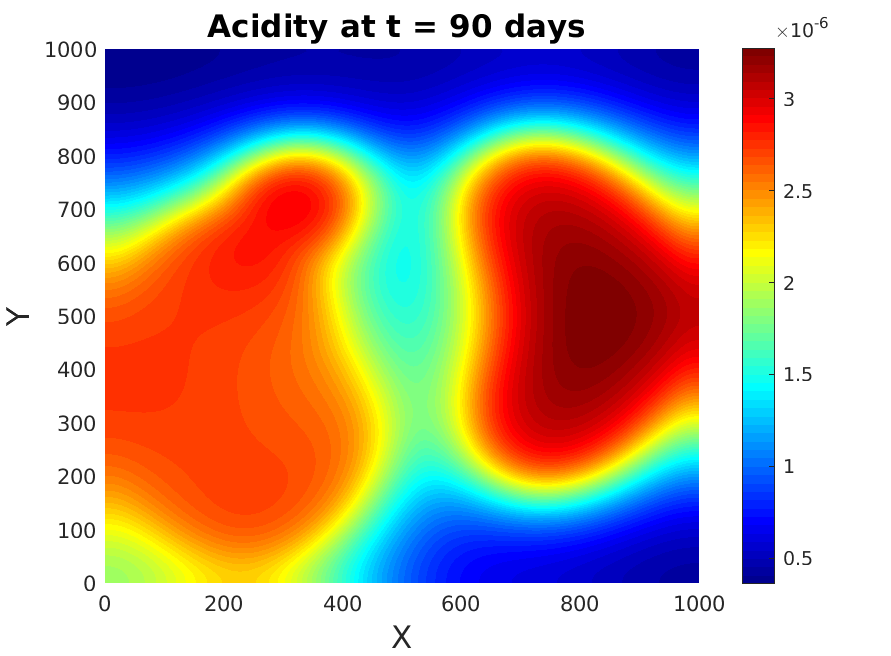}}\\
		{\includegraphics[width=1\linewidth, height = 3cm]{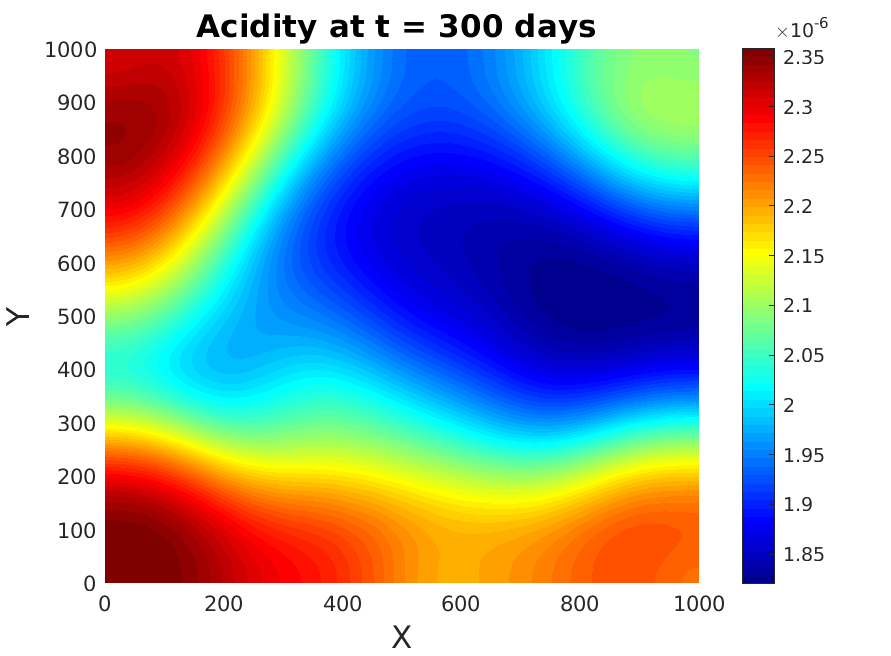}}\\
		{\includegraphics[width=1\linewidth, height = 3cm]{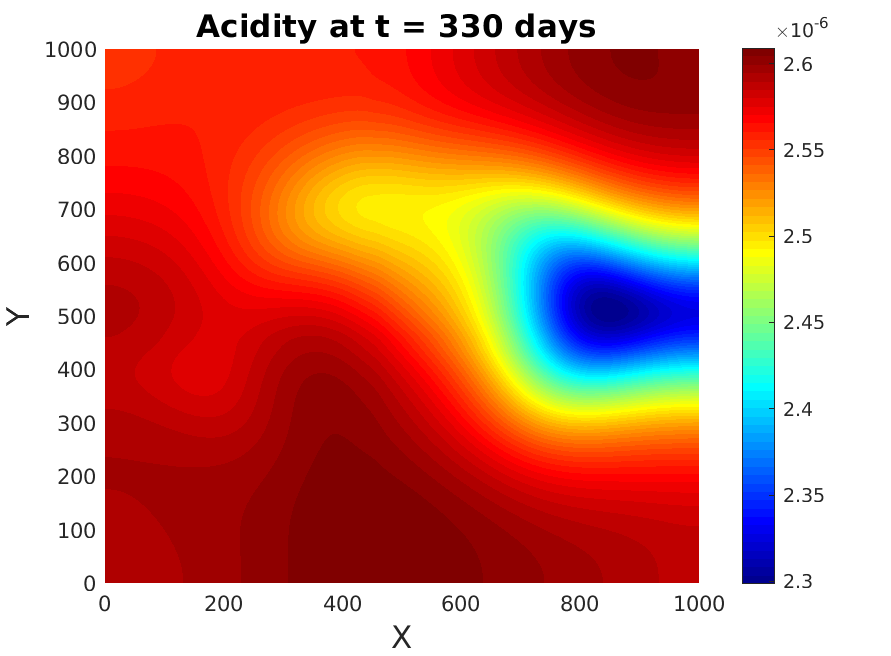}}\\
		{\includegraphics[width=1\linewidth, height = 3cm]{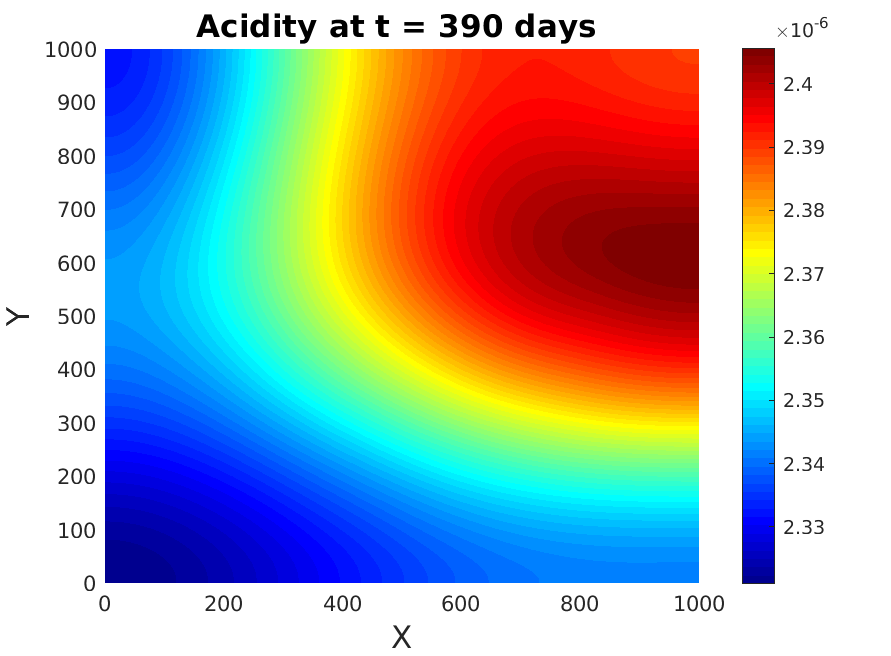}}\\
		{\includegraphics[width=1\linewidth, height = 3cm]{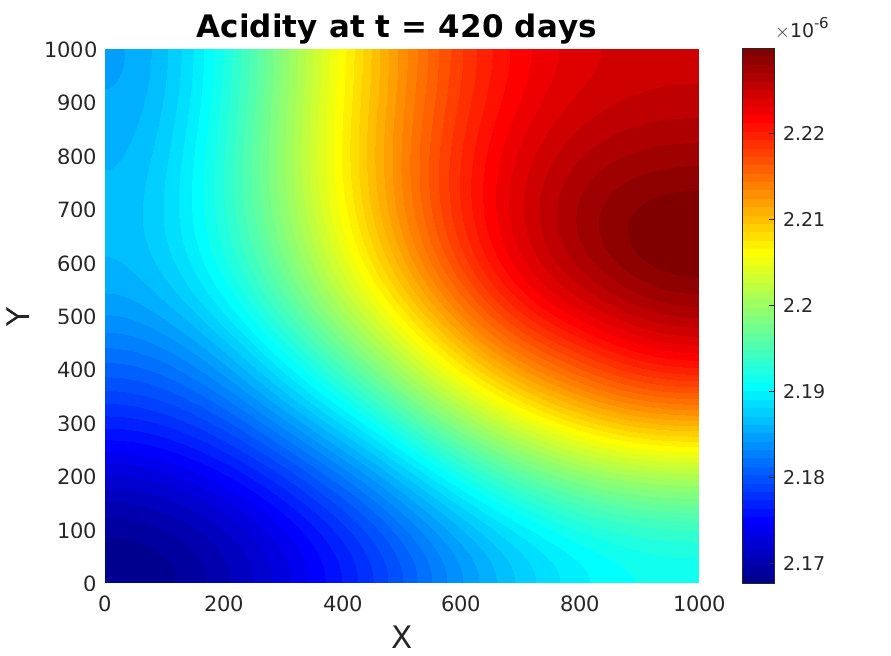}}%
		\subcaption{Acidity}
	\end{minipage}%
	\hspace{0.01cm}
	\begin{minipage}[hstb]{.24\linewidth}
		{\includegraphics[width=1\linewidth, height = 3cm]{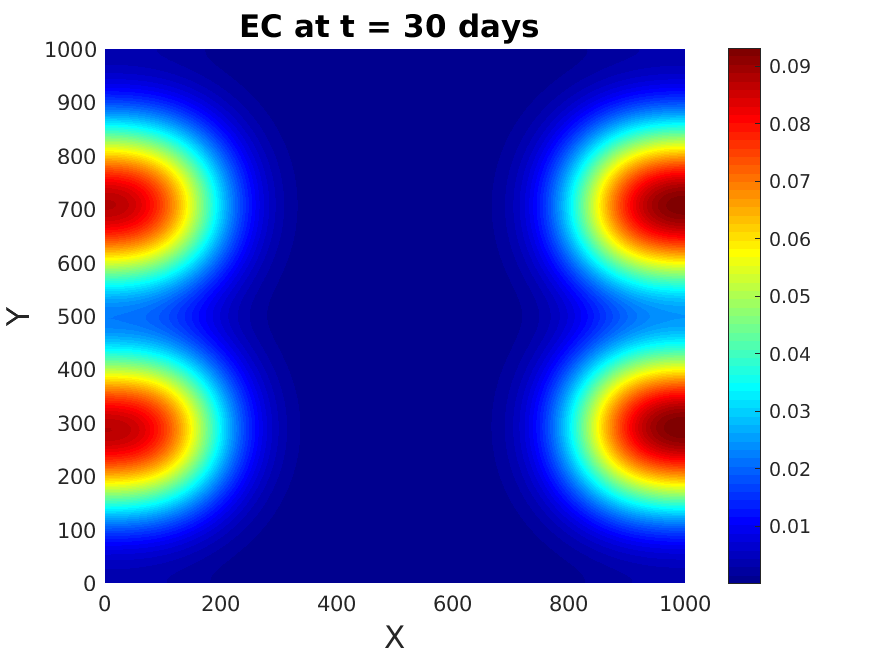}}\\
		{\includegraphics[width=1\linewidth, height = 3cm]{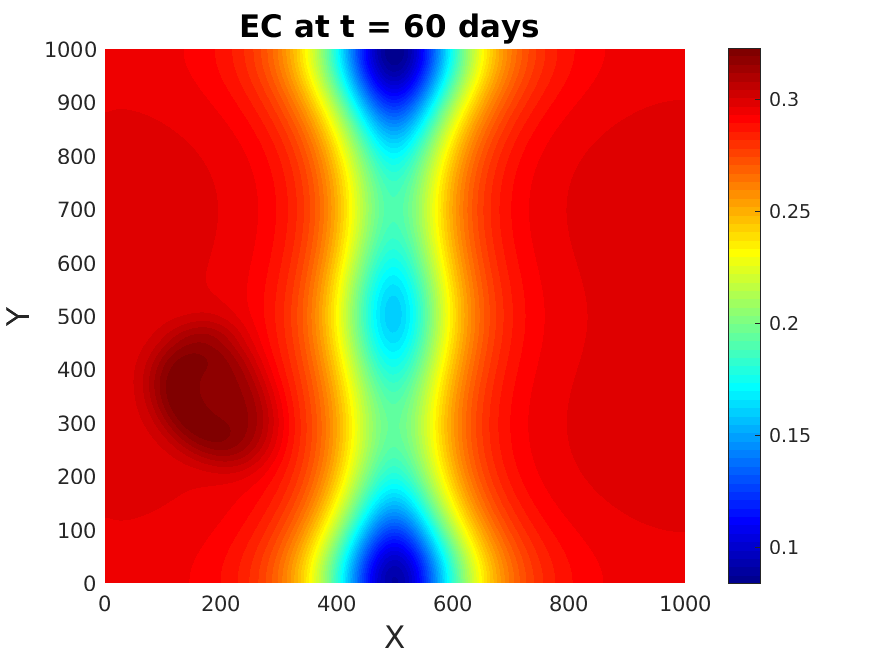}}\\
		{\includegraphics[width=1\linewidth, height = 3cm]{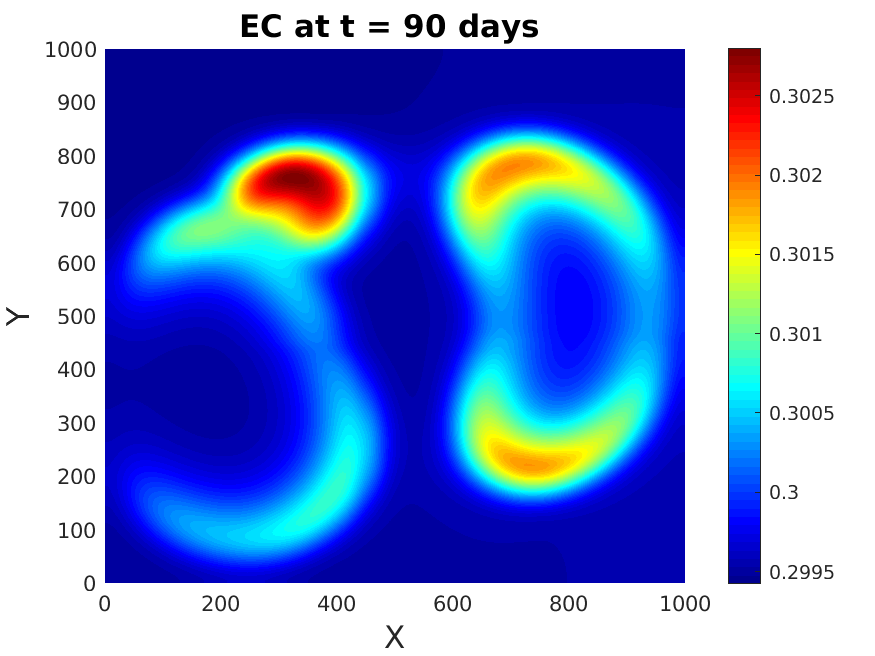}}\\
		{\includegraphics[width=1\linewidth, height = 3cm]{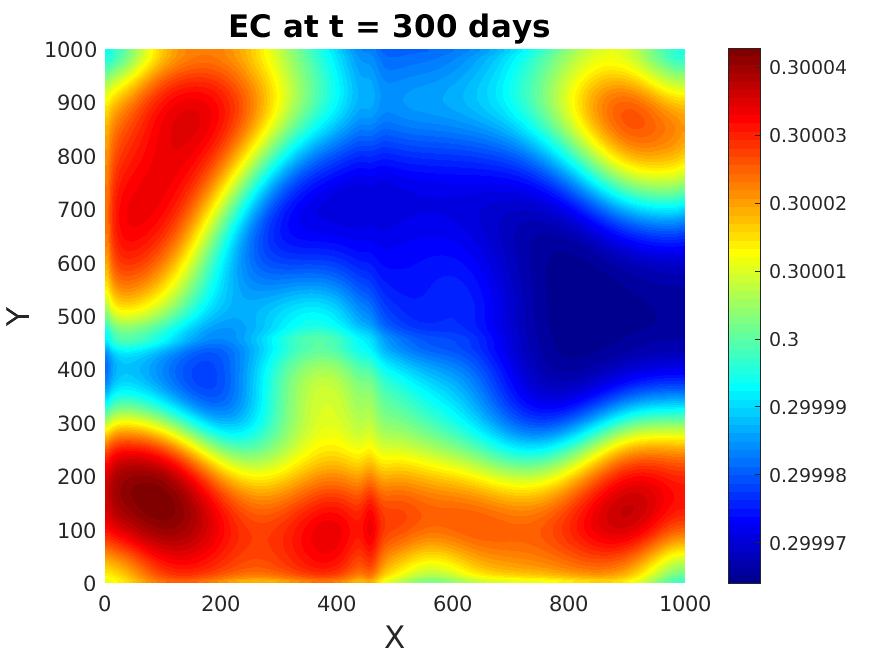}}\\
		{\includegraphics[width=1\linewidth, height = 3cm]{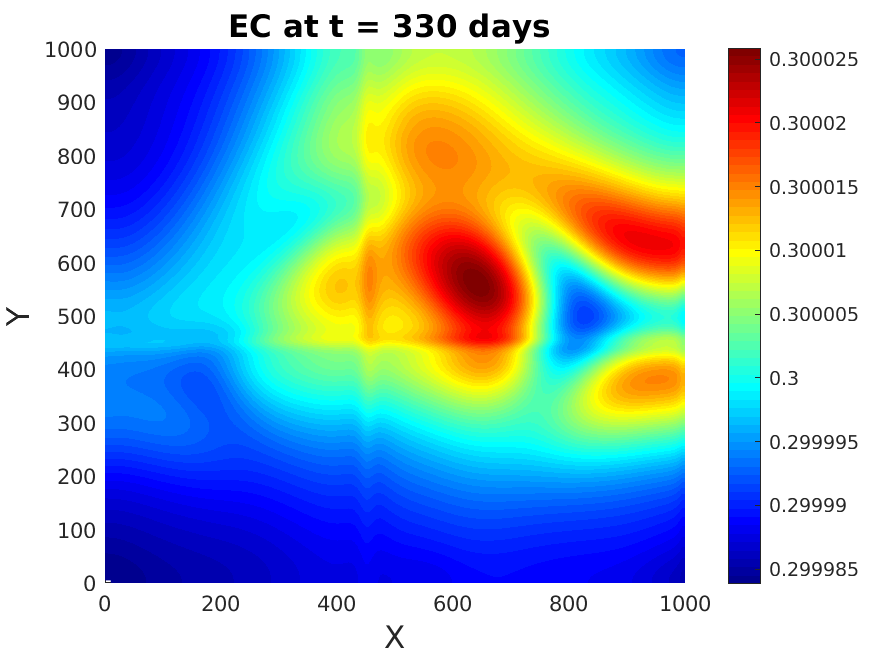}}\\
		{\includegraphics[width=1\linewidth, height = 3cm]{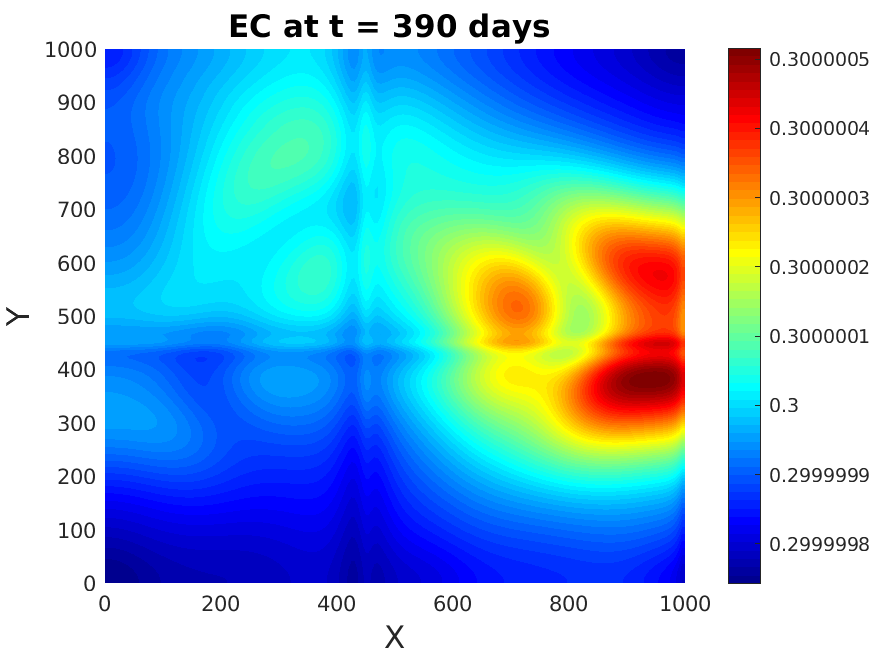}}\\
		{\includegraphics[width=1\linewidth, height = 3cm]{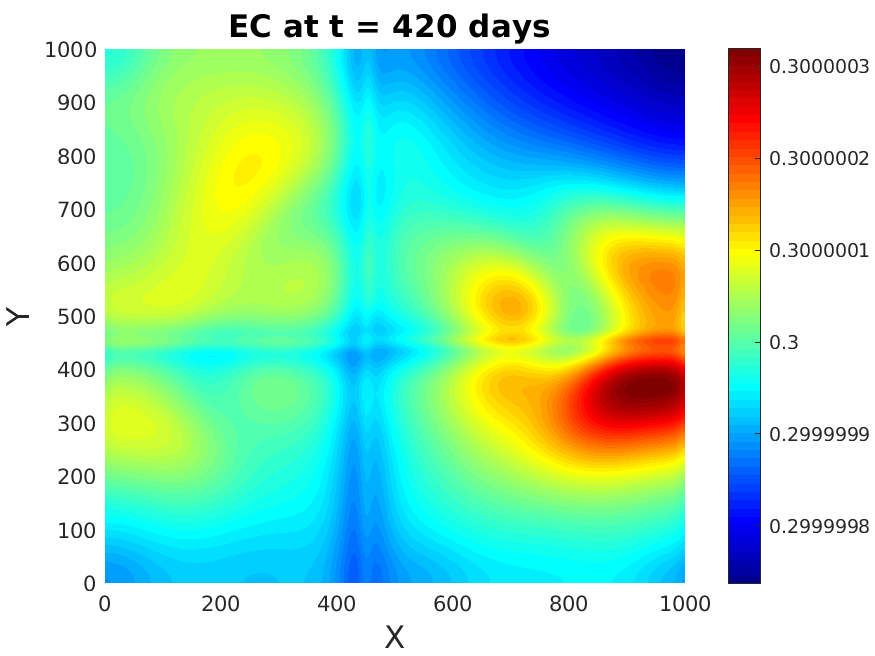}}%
		\subcaption{ECs}
	\end{minipage}%
	\hspace{0.01cm}
	\begin{minipage}[hstb]{.24\linewidth}
		{\includegraphics[width=1\linewidth, height = 3cm]{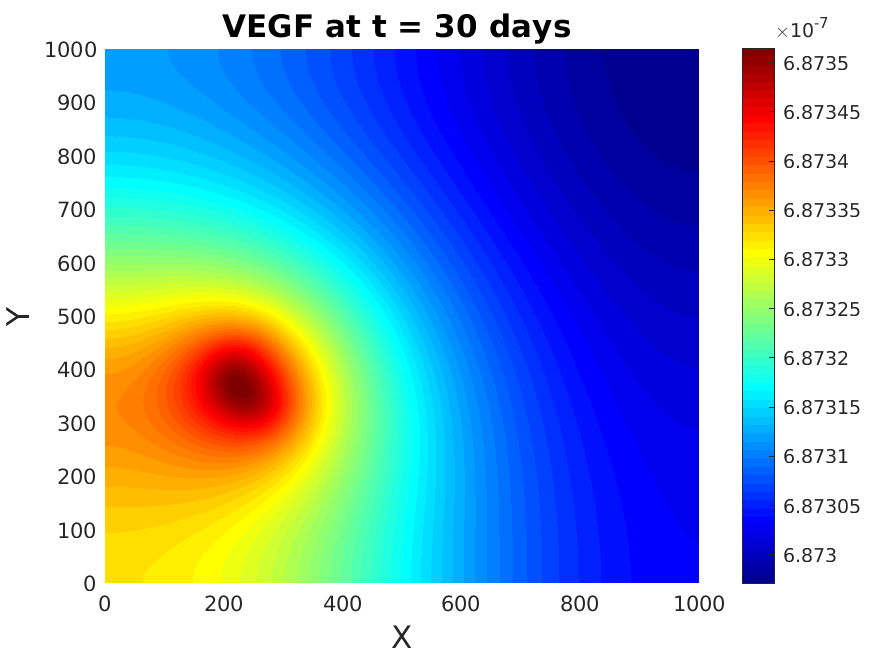}}\\
		{\includegraphics[width=1\linewidth, height = 3cm]{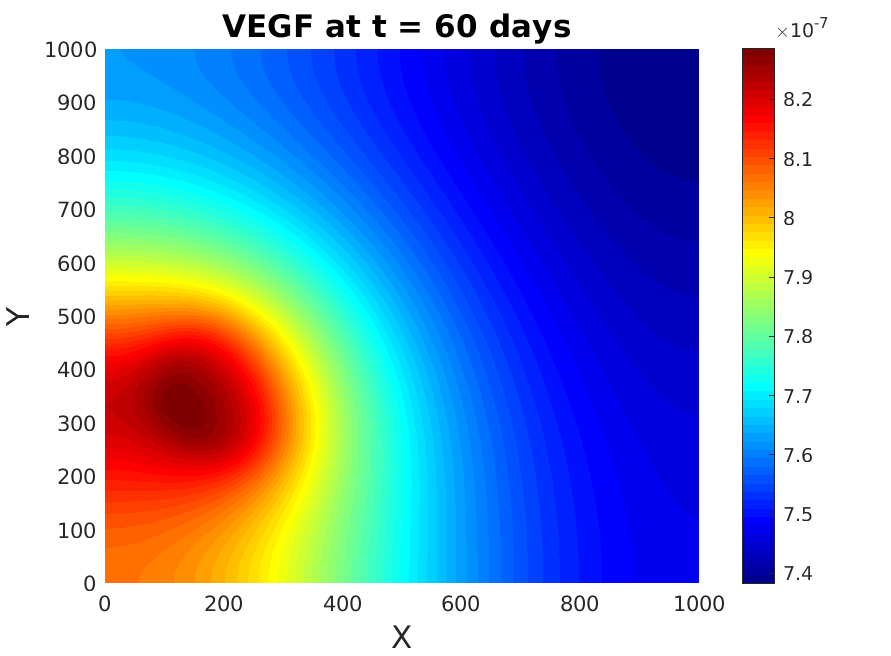}}\\
		{\includegraphics[width=1\linewidth, height = 3cm]{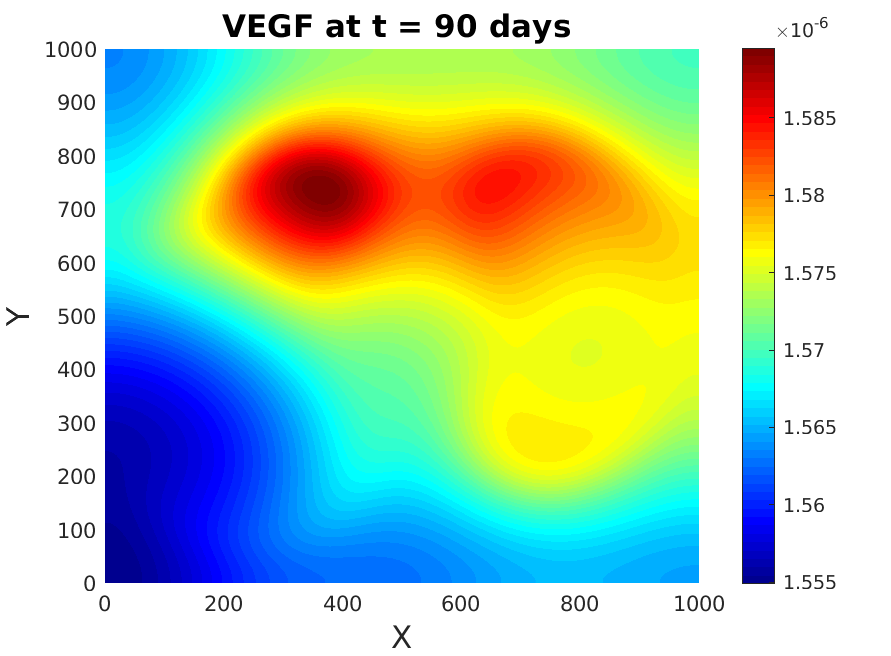}}\\
		{\includegraphics[width=1\linewidth, height = 3cm]{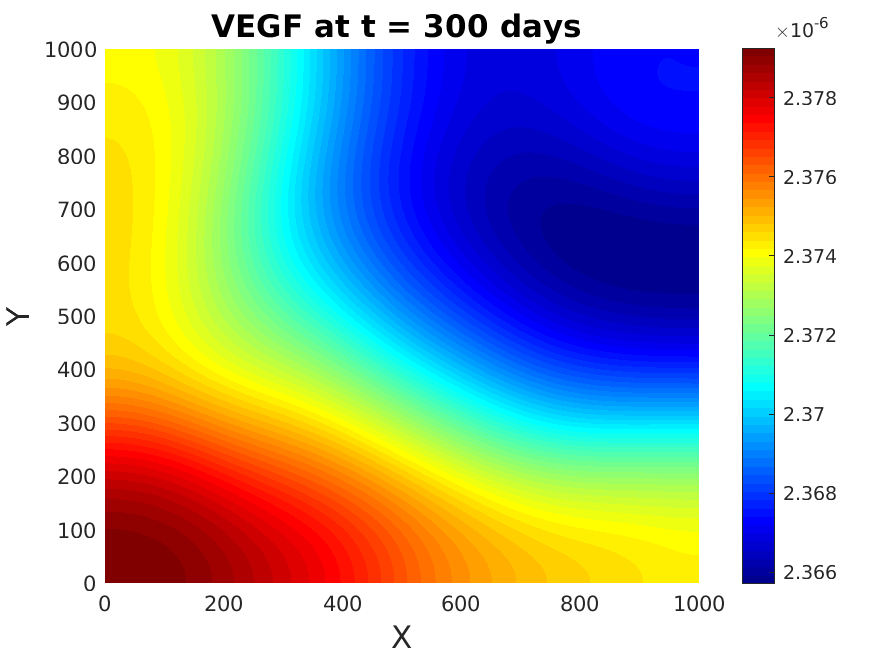}}\\
		{\includegraphics[width=1\linewidth, height = 3cm]{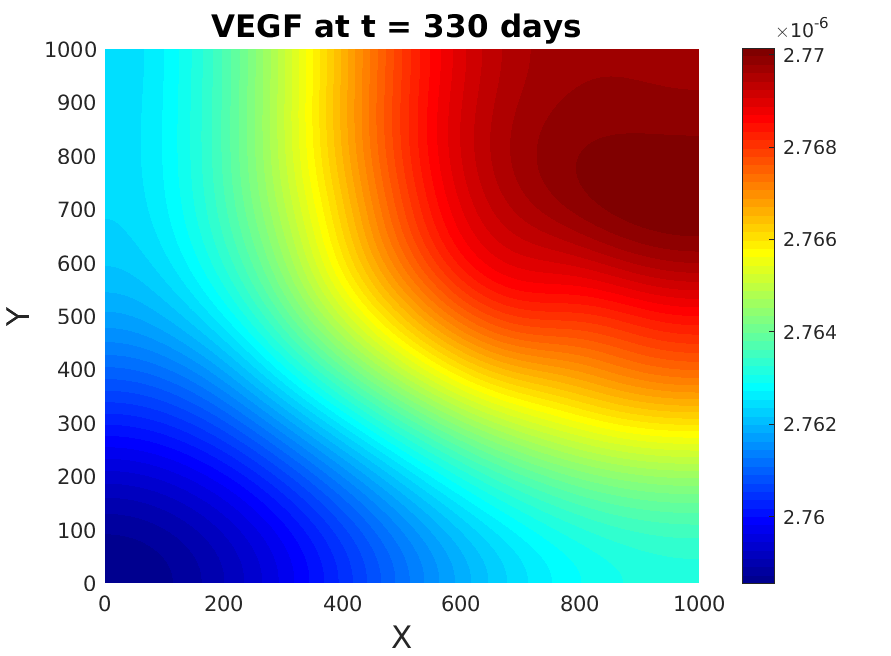}}\\
		{\includegraphics[width=1\linewidth, height = 3cm]{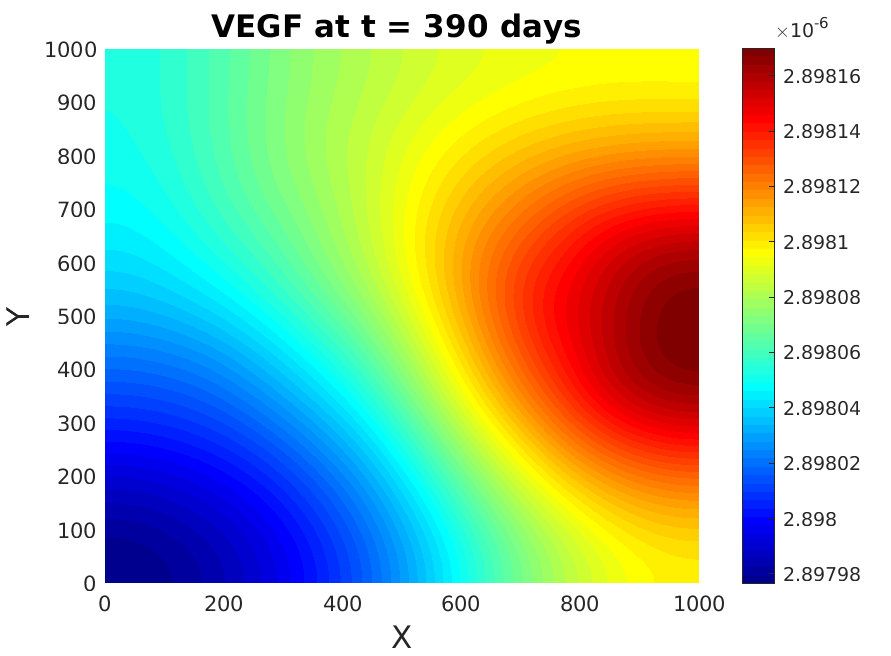}}\\
		{\includegraphics[width=1\linewidth, height = 3cm]{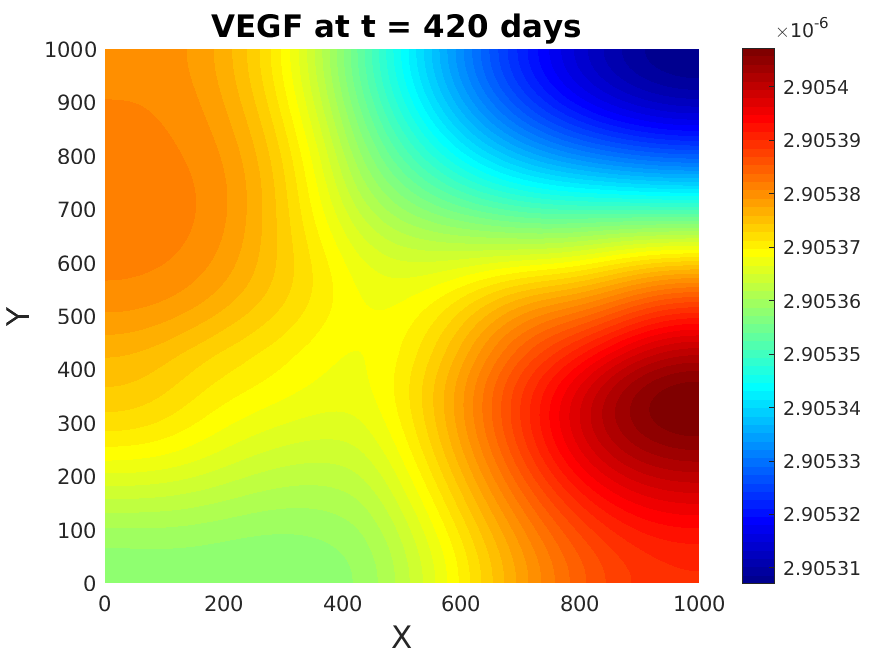}}%
		\subcaption{VEGF}
	\end{minipage}%
	\caption{Evolution of the solution components at several times in the framework of \textbf{Experiment 1} (full model \eqref{eq:macro-nondim}).}\label{fig:scenario1}
\end{figure}

\begin{figure}[!htbp]
	\begin{minipage}[stb]{.24\linewidth}
		\raisebox{1.2cm}{\rotatebox[origin=t]{90}{30 days}}{\includegraphics[width=1\linewidth, height = 3cm]{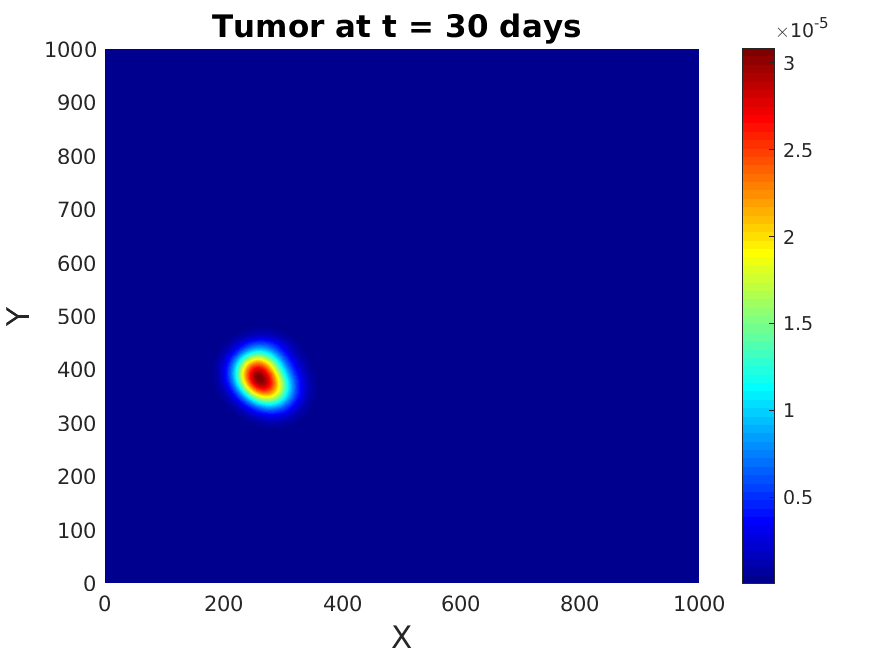}}\\
		\raisebox{1.2cm}{\rotatebox[origin=t]{90}{60 days}}{\includegraphics[width=1\linewidth, height = 3cm]{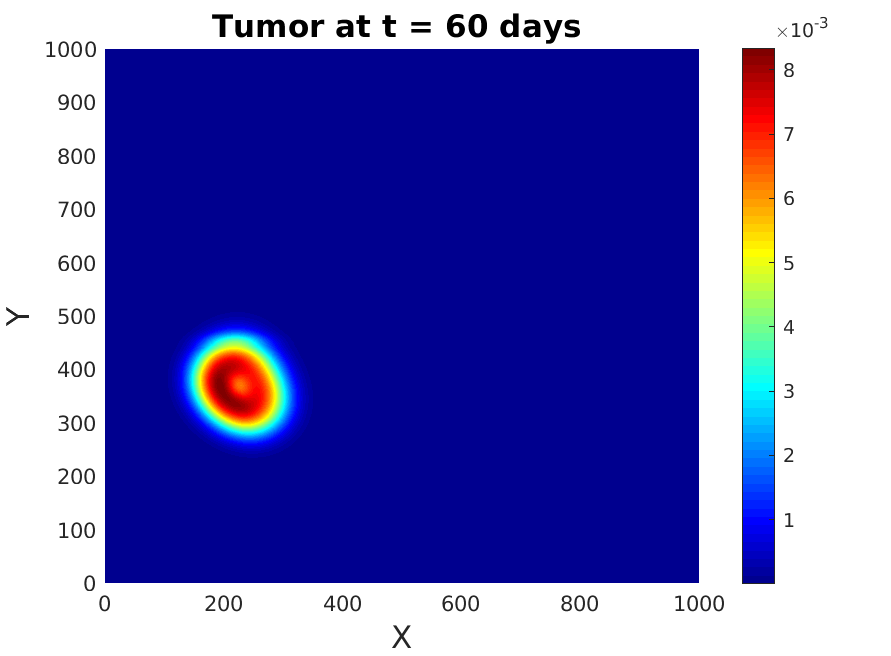}}\\
		\raisebox{1.2cm}{\rotatebox[origin=t]{90}{90 days}}{\includegraphics[width=1\linewidth, height = 3cm]{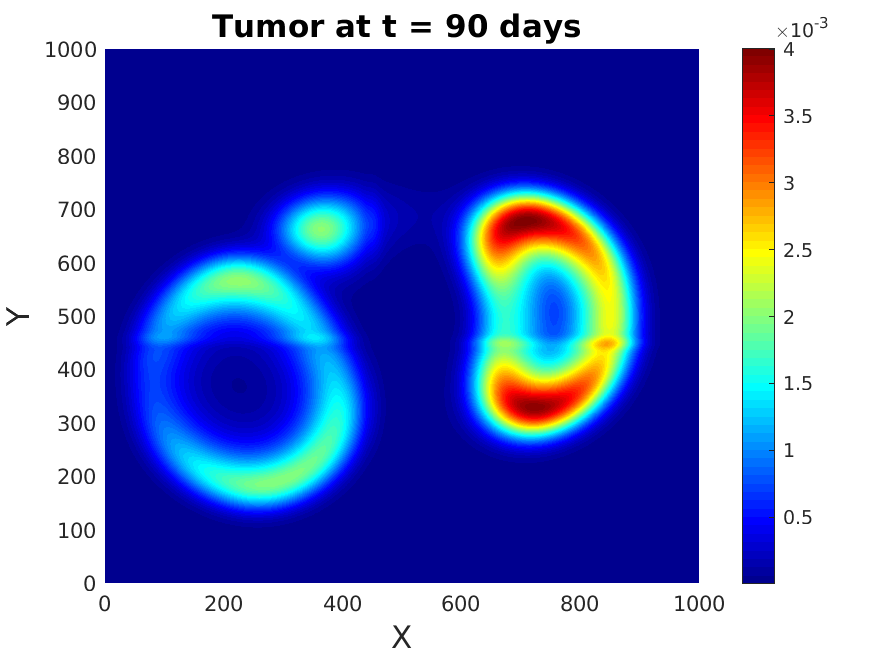}}\\
		\raisebox{1.2cm}{\rotatebox[origin=t]{90}{300 days}}{\includegraphics[width=1\linewidth, height = 3cm]{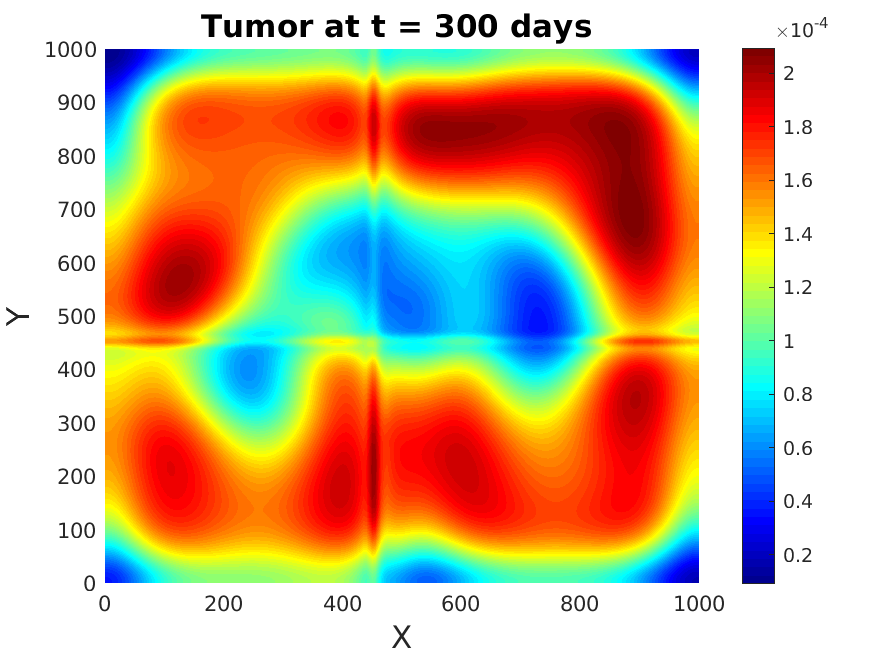}}\\
		\raisebox{1.2cm}{\rotatebox[origin=t]{90}{330 days}}{\includegraphics[width=1\linewidth, height = 3cm]{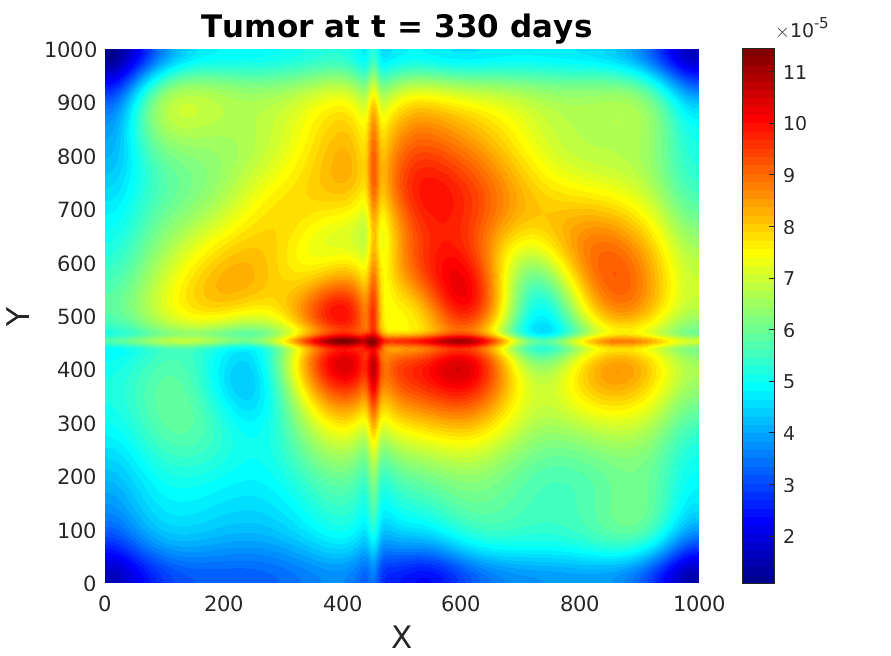}}\\
		\raisebox{1.2cm}{\rotatebox[origin=t]{90}{390 days}}{\includegraphics[width=1\linewidth, height = 3cm]{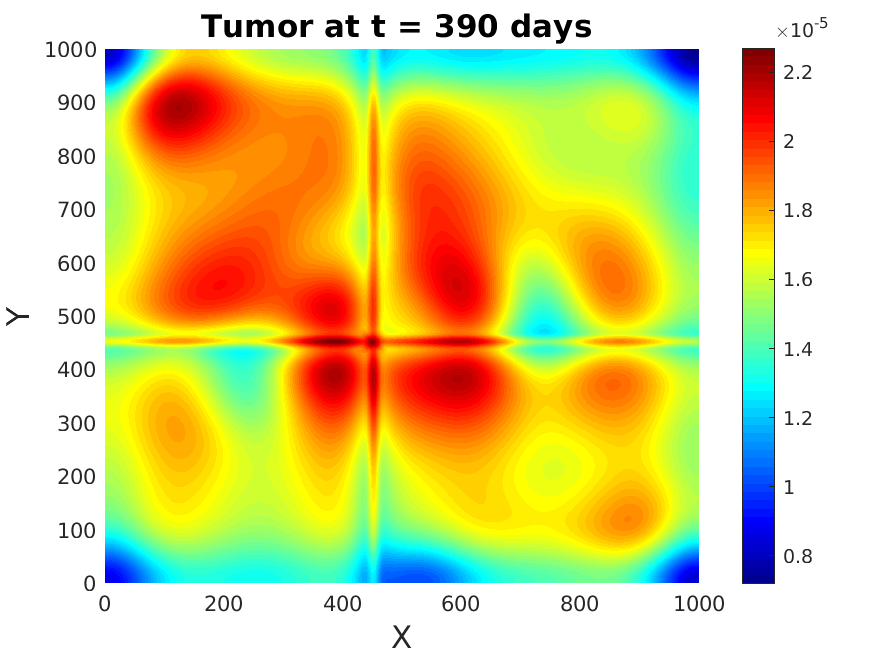}}\\
		\raisebox{1.2cm}{\rotatebox[origin=t]{90}{420 days}}{\includegraphics[width=1\linewidth, height = 3cm]{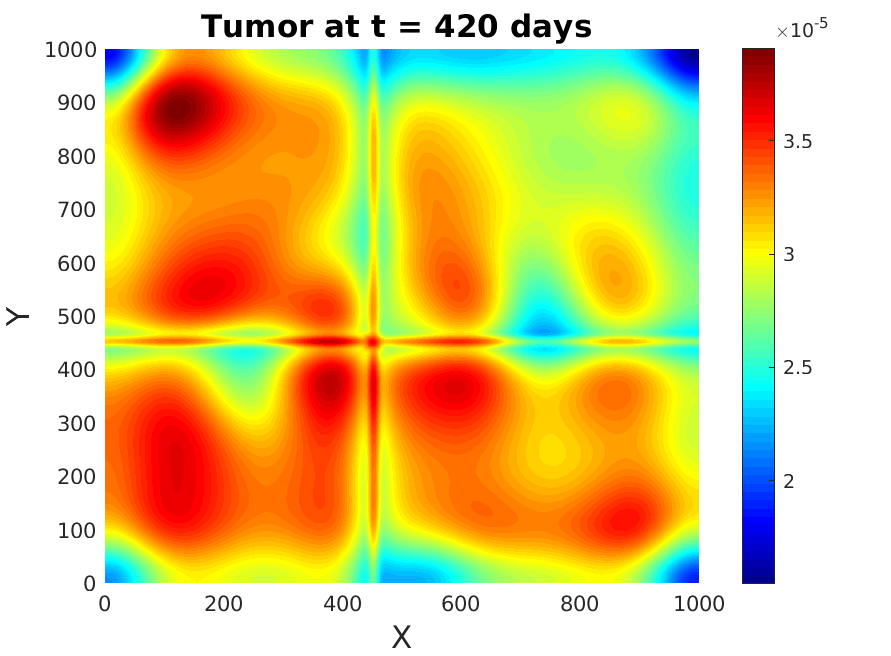}}%
		\subcaption{Tumor}
	\end{minipage}%
	\hspace{0.2cm}
	\begin{minipage}[stb]{.24\linewidth}
		{\includegraphics[width=1\linewidth, height = 3cm]{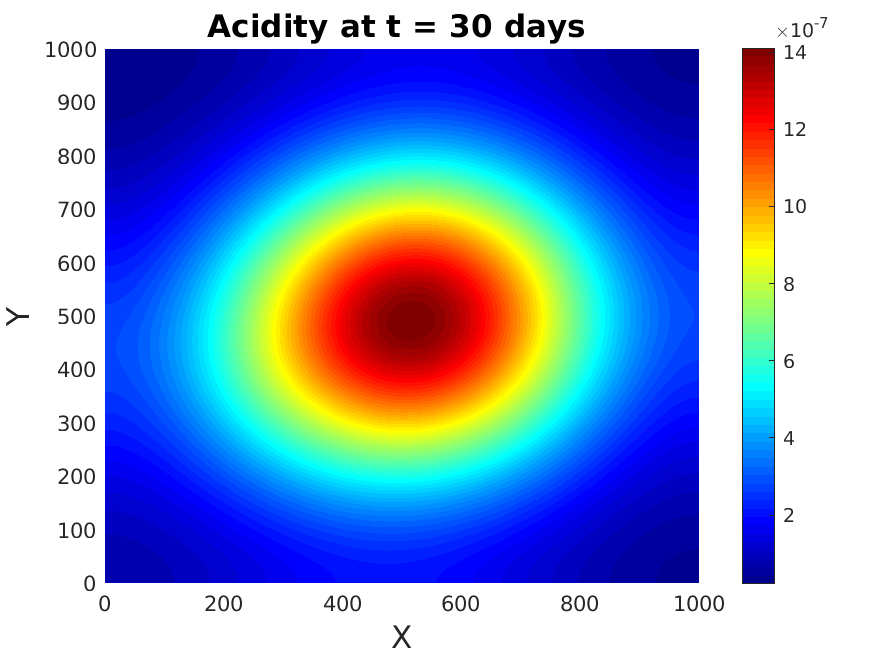}}\\
		{\includegraphics[width=1\linewidth, height = 3cm]{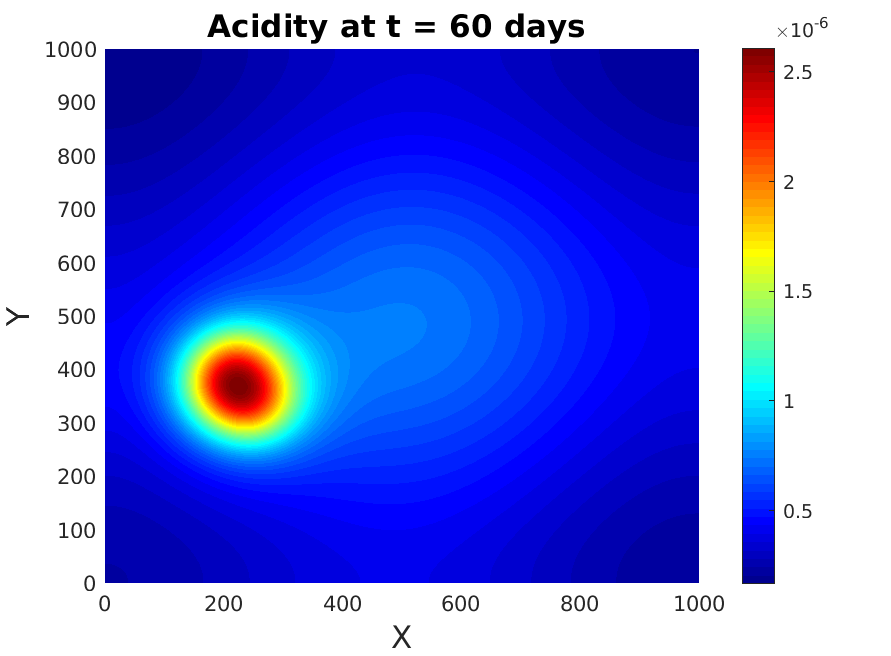}}\\
		{\includegraphics[width=1\linewidth, height = 3cm]{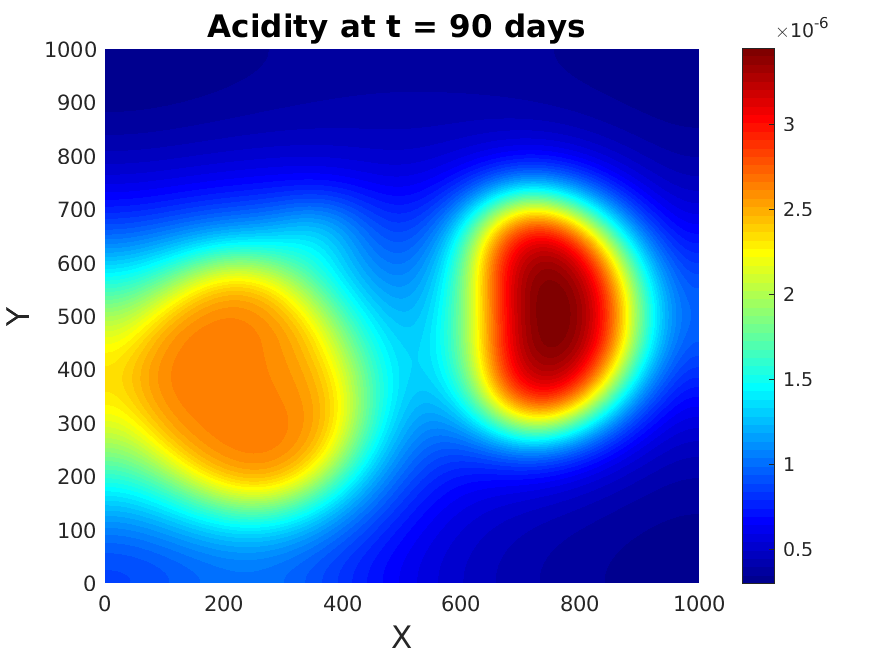}}\\
		{\includegraphics[width=1\linewidth, height = 3cm]{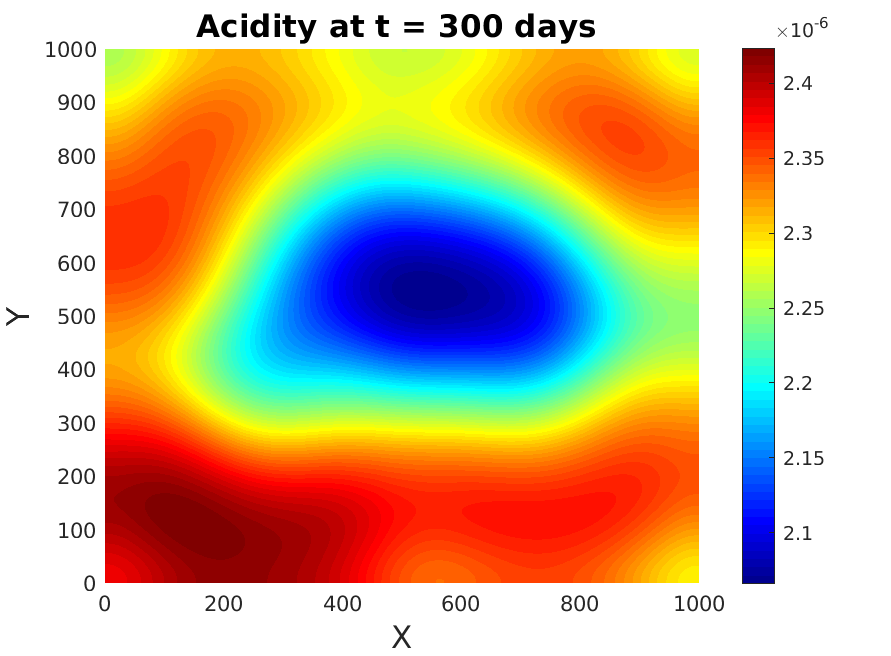}}\\
		{\includegraphics[width=1\linewidth, height = 3cm]{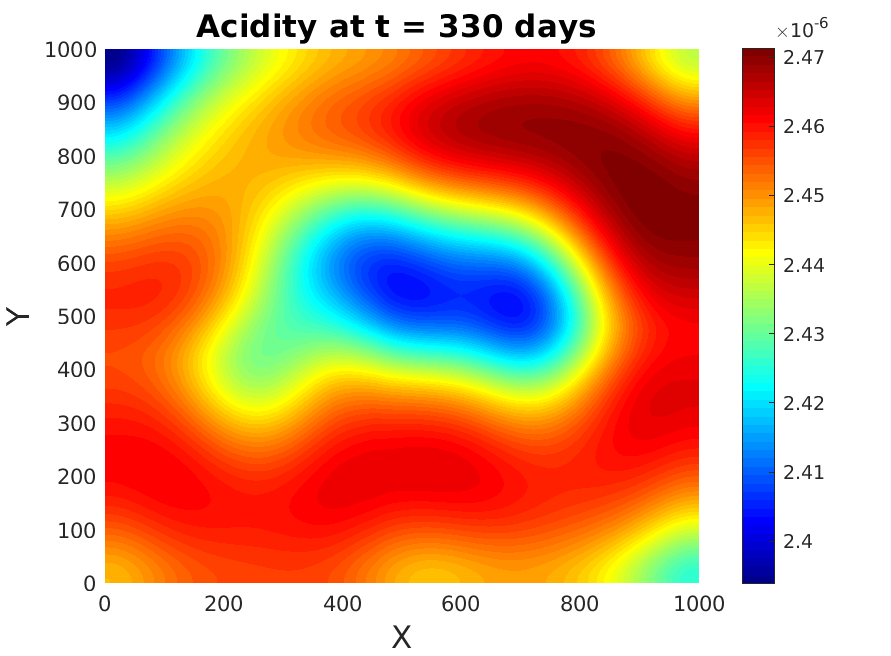}}\\
		{\includegraphics[width=1\linewidth, height = 3cm]{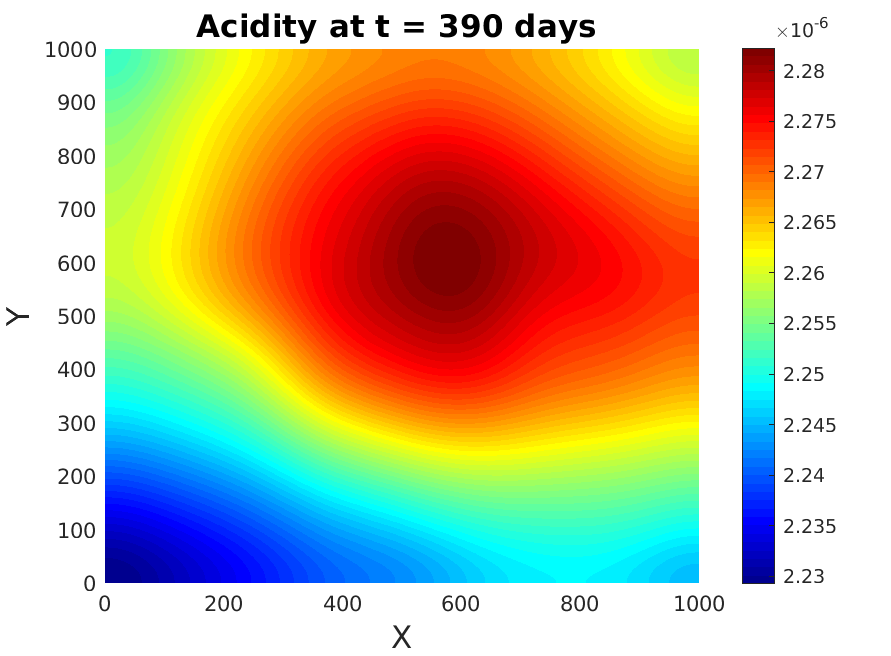}}\\
		{\includegraphics[width=1\linewidth, height = 3cm]{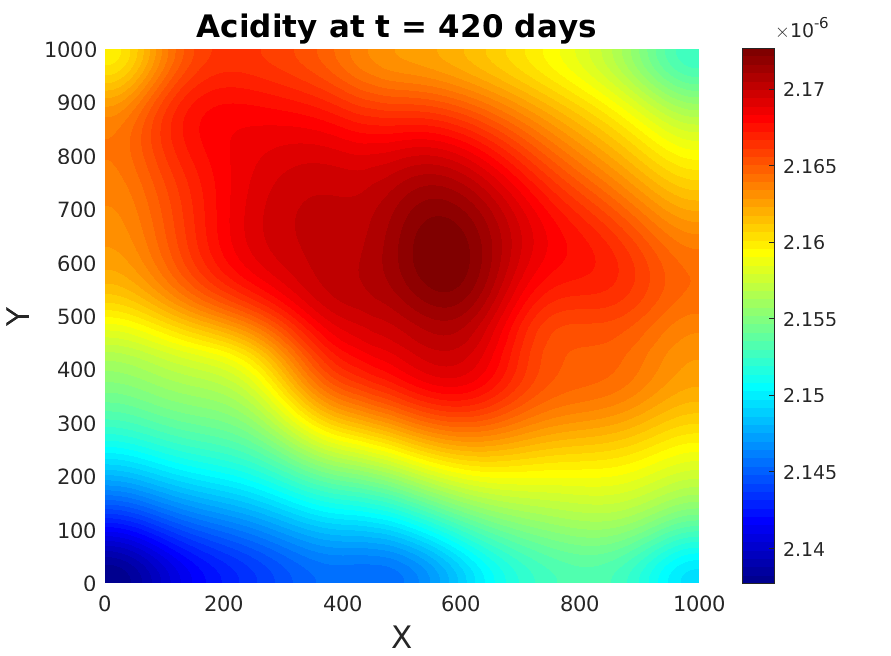}}%
		\subcaption{Acidity}
	\end{minipage}%
	\hspace{0.01cm}
	\begin{minipage}[stb]{.24\linewidth}
		{\includegraphics[width=1\linewidth, height = 3cm]{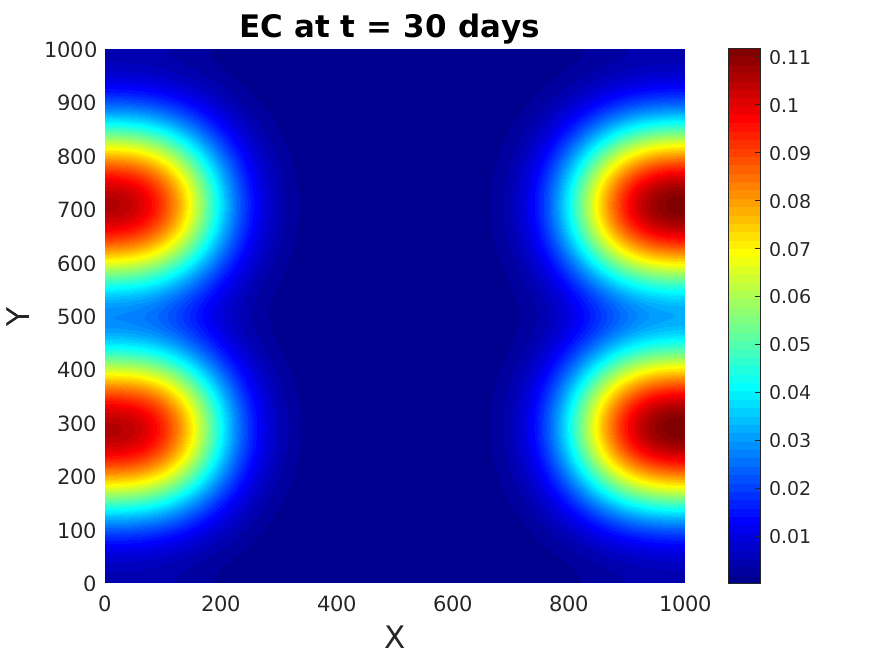}}\\
		{\includegraphics[width=1\linewidth, height = 3cm]{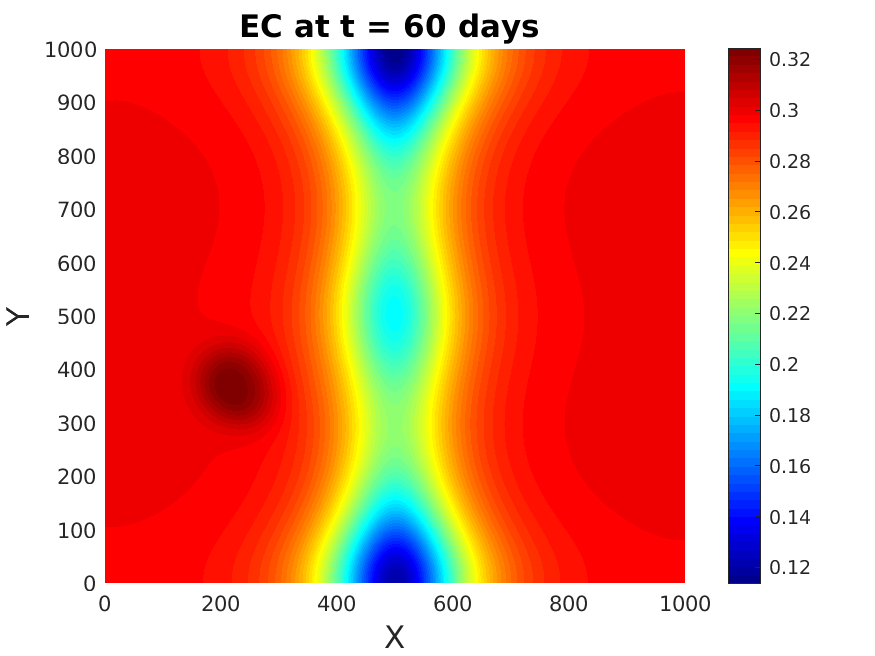}}\\
		{\includegraphics[width=1\linewidth, height = 3cm]{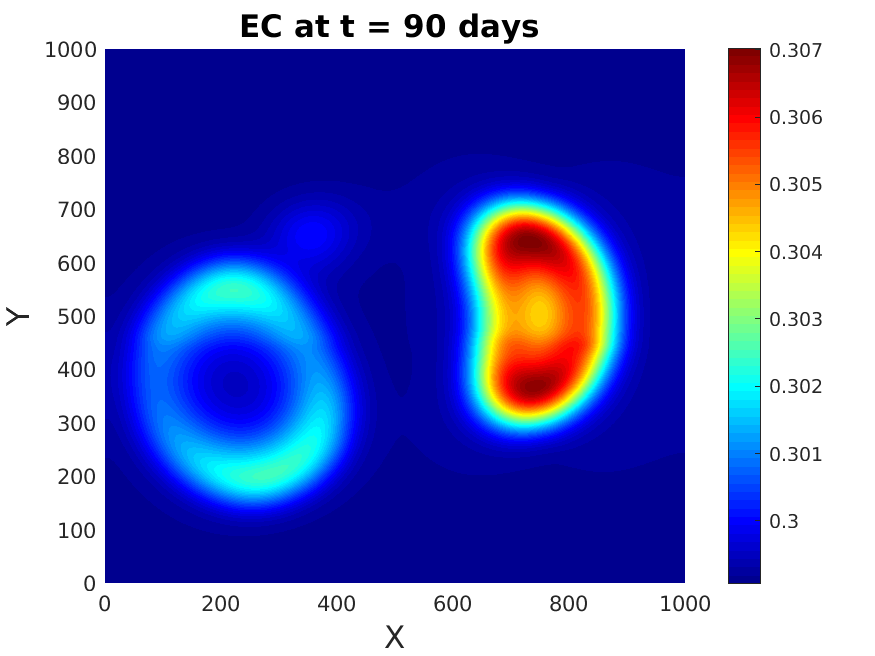}}\\
		{\includegraphics[width=1\linewidth, height = 3cm]{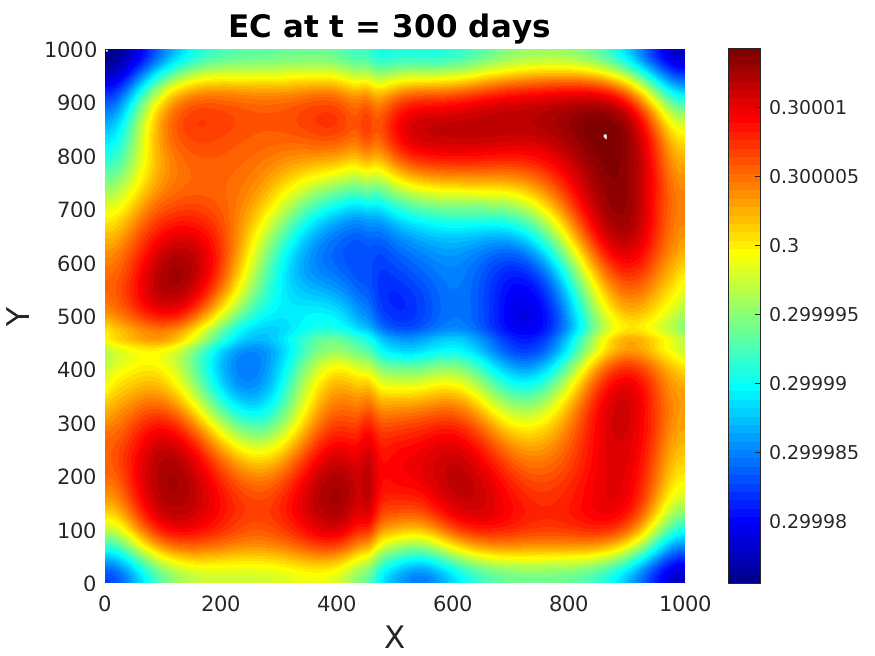}}\\
		{\includegraphics[width=1\linewidth, height = 3cm]{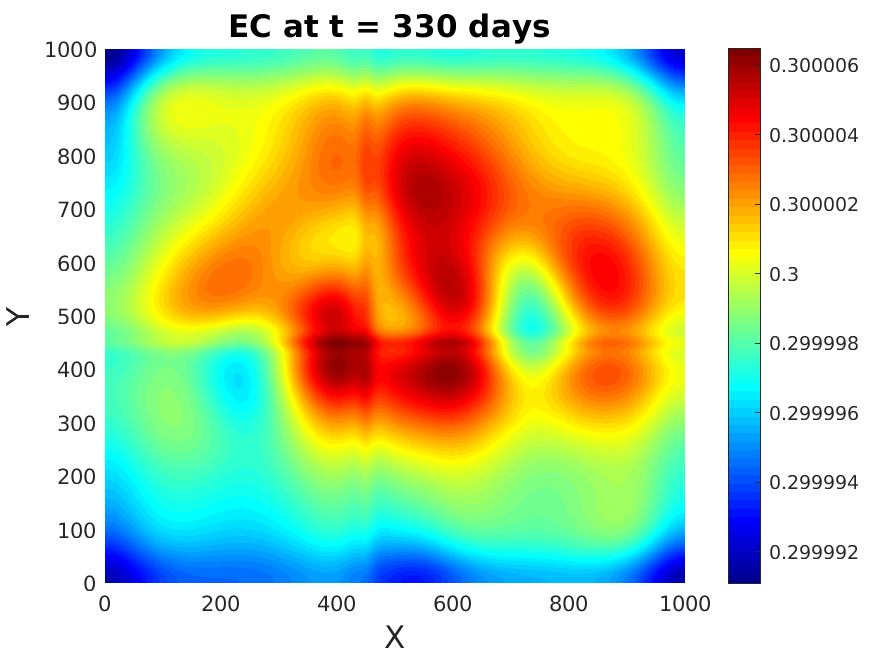}}\\
		{\includegraphics[width=1\linewidth, height = 3cm]{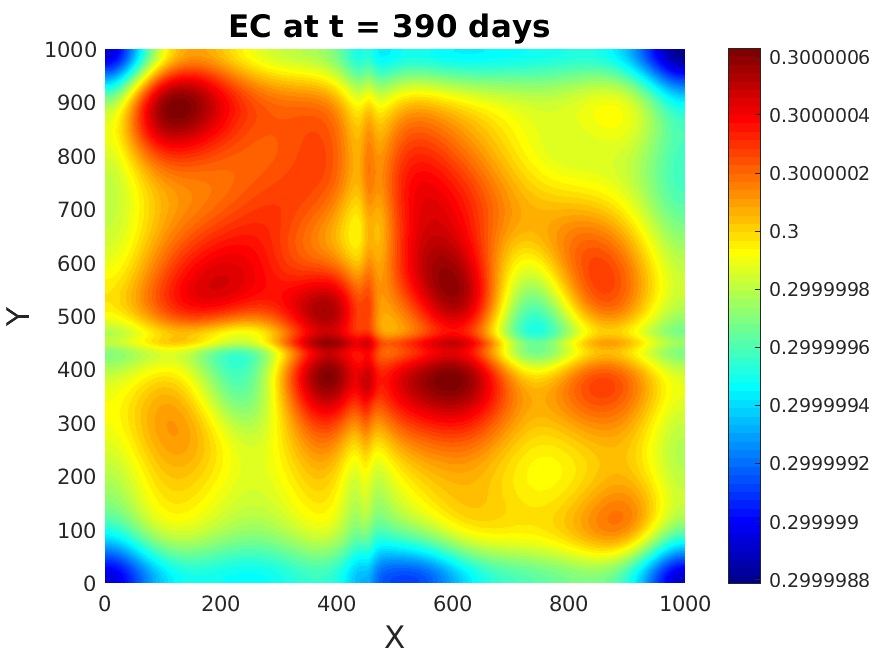}}\\
		{\includegraphics[width=1\linewidth, height = 3cm]{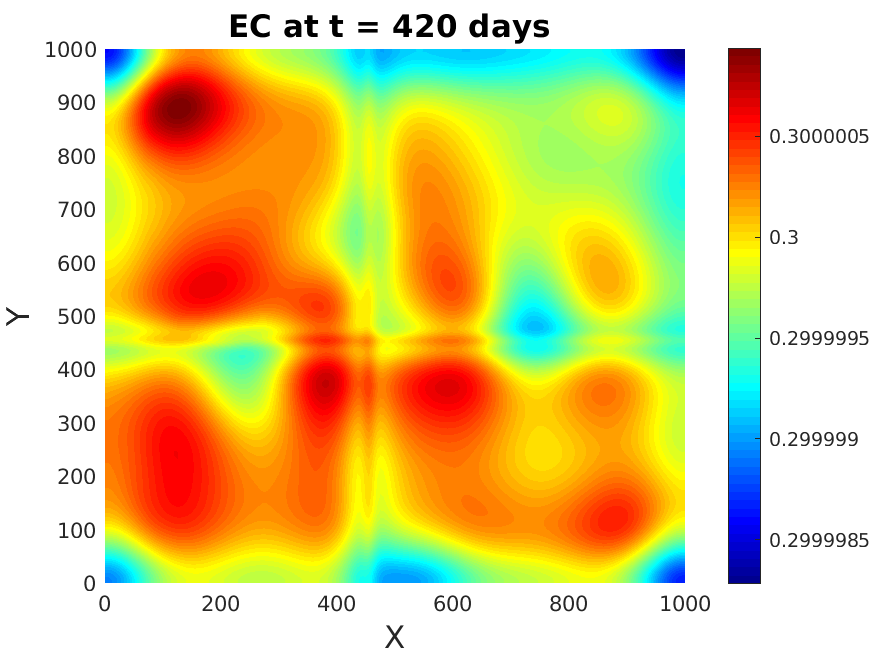}}%
		\subcaption{ECs}
	\end{minipage}%
	\hspace{0.01cm}
	\begin{minipage}[stb]{.24\linewidth}
		{\includegraphics[width=1\linewidth, height = 3cm]{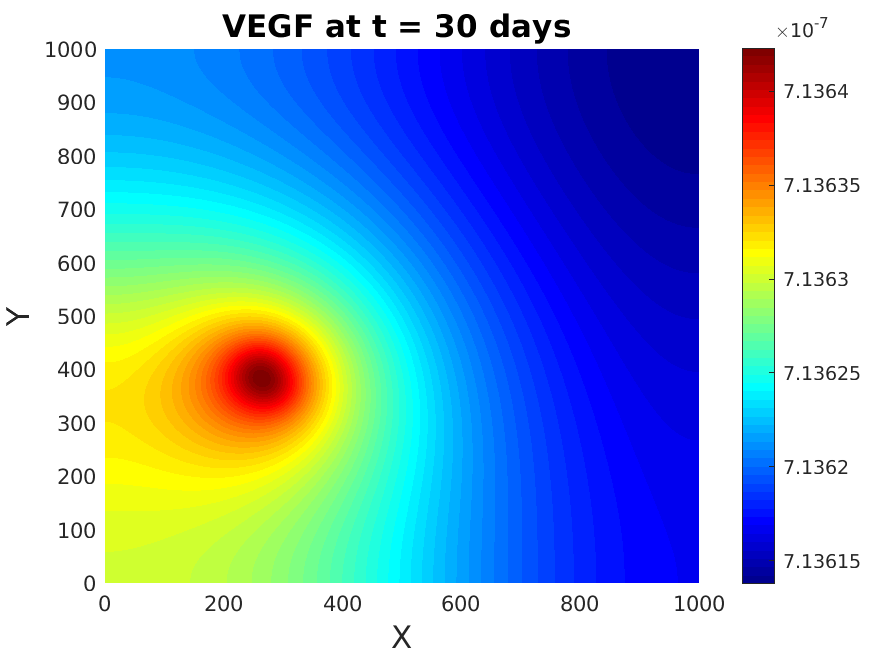}}\\
		{\includegraphics[width=1\linewidth, height = 3cm]{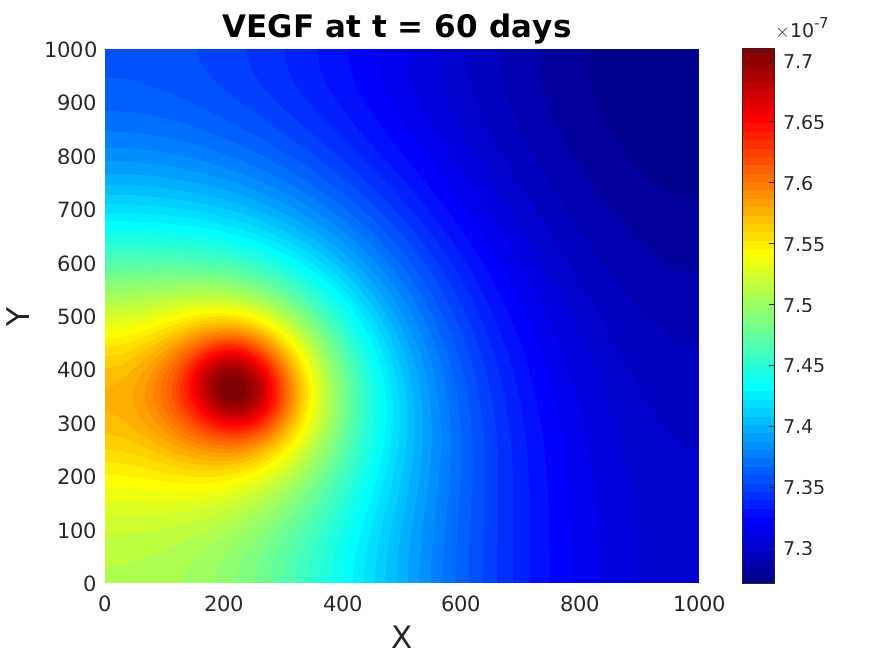}}\\
		{\includegraphics[width=1\linewidth, height = 3cm]{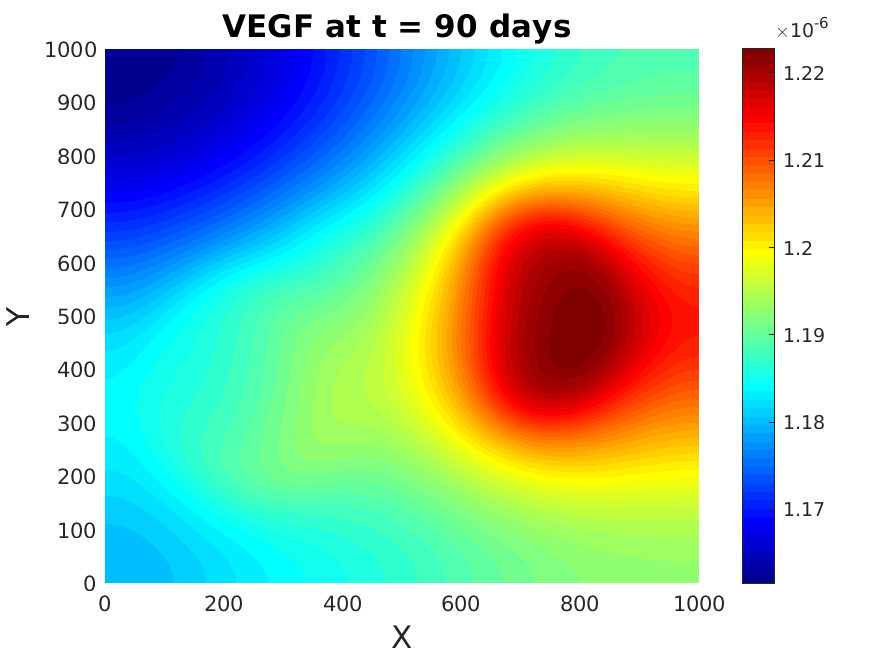}}\\
		{\includegraphics[width=1\linewidth, height = 3cm]{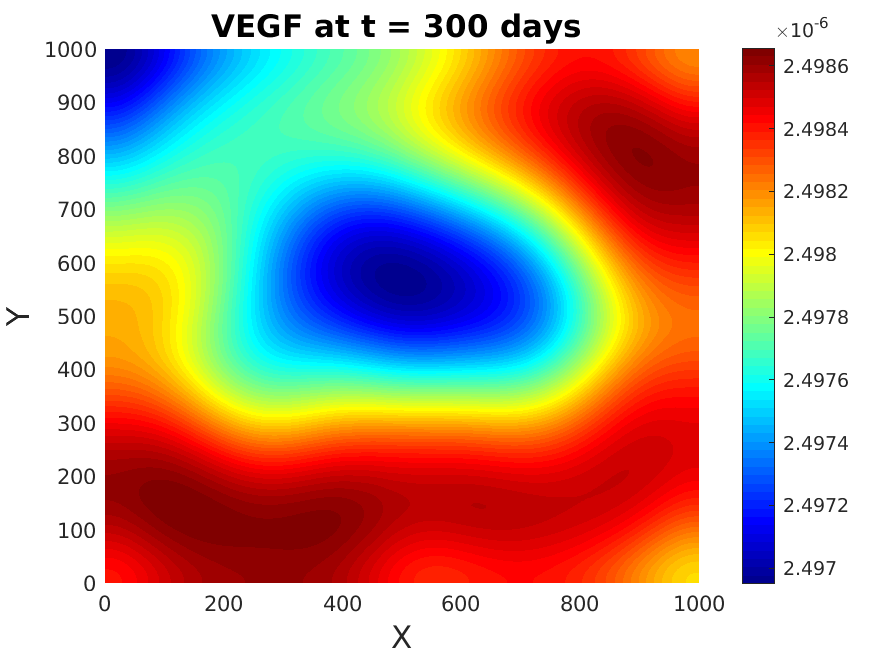}}\\
		{\includegraphics[width=1\linewidth, height = 3cm]{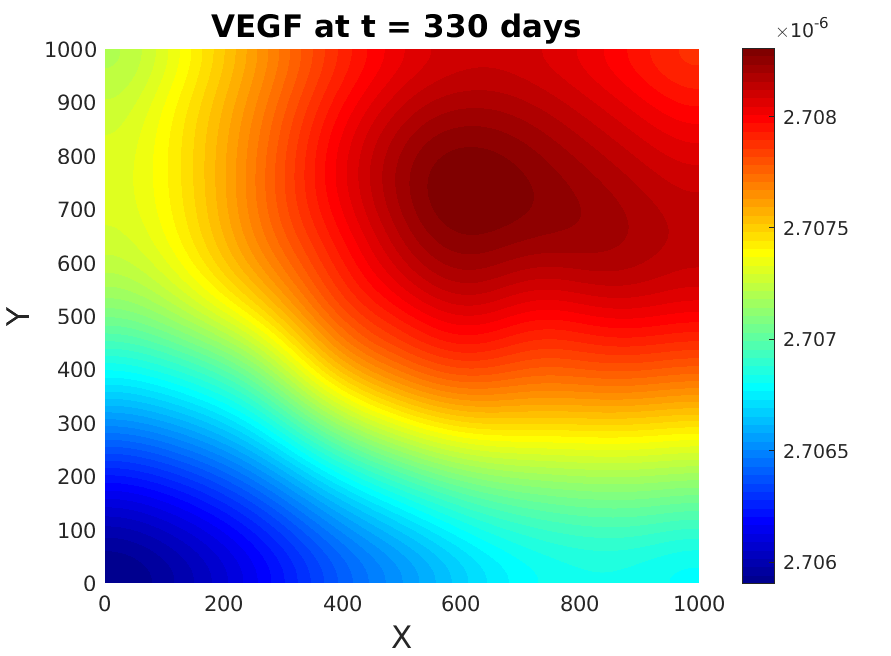}}\\
		{\includegraphics[width=1\linewidth, height = 3cm]{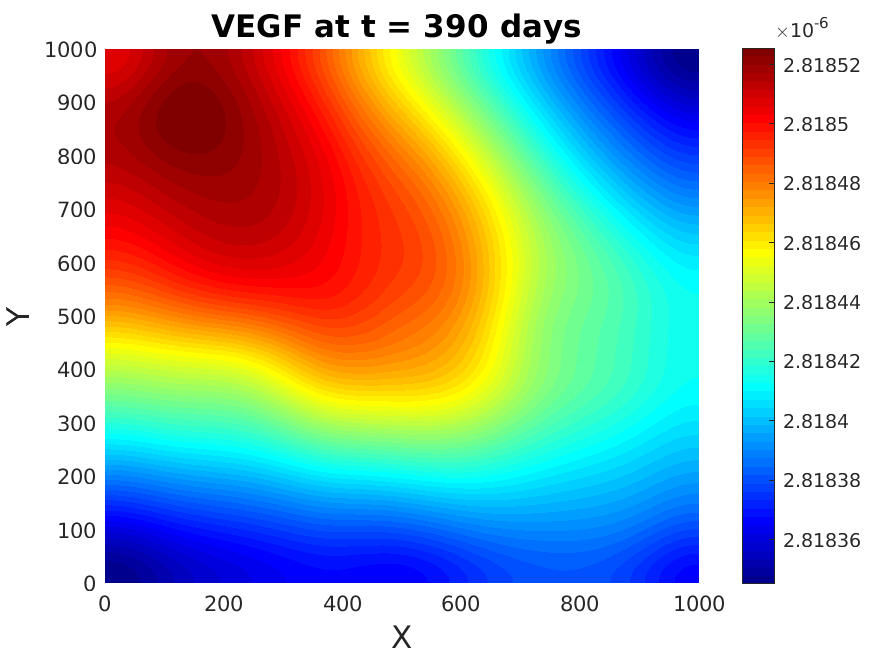}}\\
		{\includegraphics[width=1\linewidth, height = 3cm]{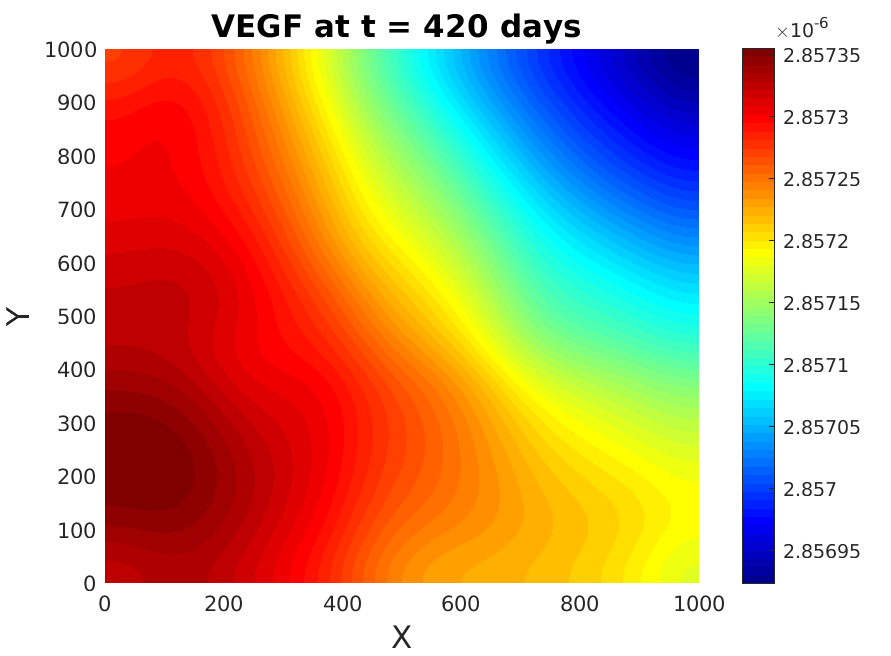}}%
		\subcaption{VEGF}
	\end{minipage}%
	\caption{Evolution of the solution components at several times in the framework of \textbf{Experiment 2} (equally weighted influence of flux-limited repellent pH-taxis and self-diffusion).}\label{fig:scenario2}
\end{figure}

\begin{figure}[!htbp]
\begin{minipage}[stb]{.24\linewidth}
	\raisebox{1.2cm}{\rotatebox[origin=t]{90}{30 days}}{\includegraphics[width=1\linewidth, height = 3cm]{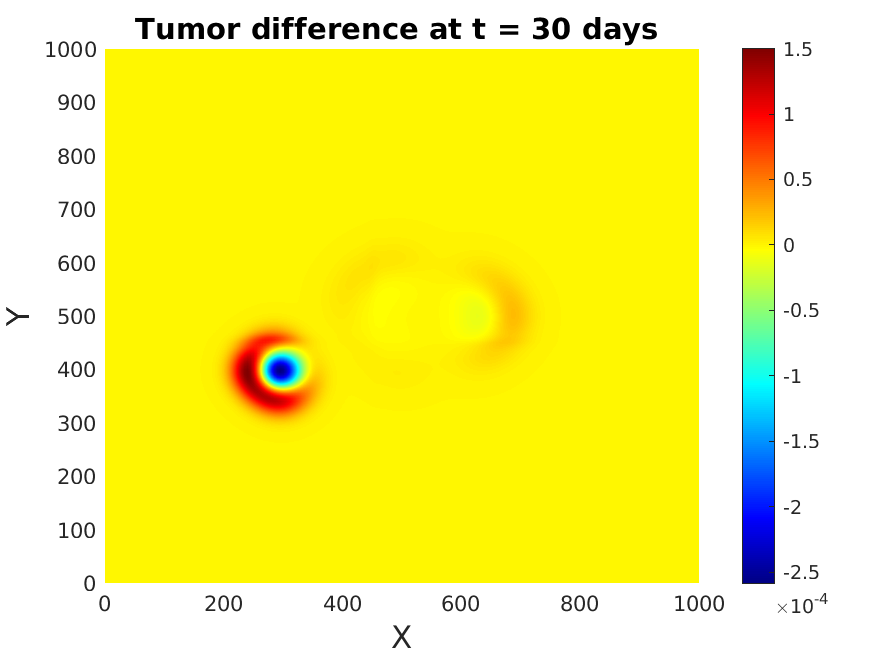}}\\
	\raisebox{1.2cm}{\rotatebox[origin=t]{90}{60 days}}{\includegraphics[width=1\linewidth, height = 3cm]{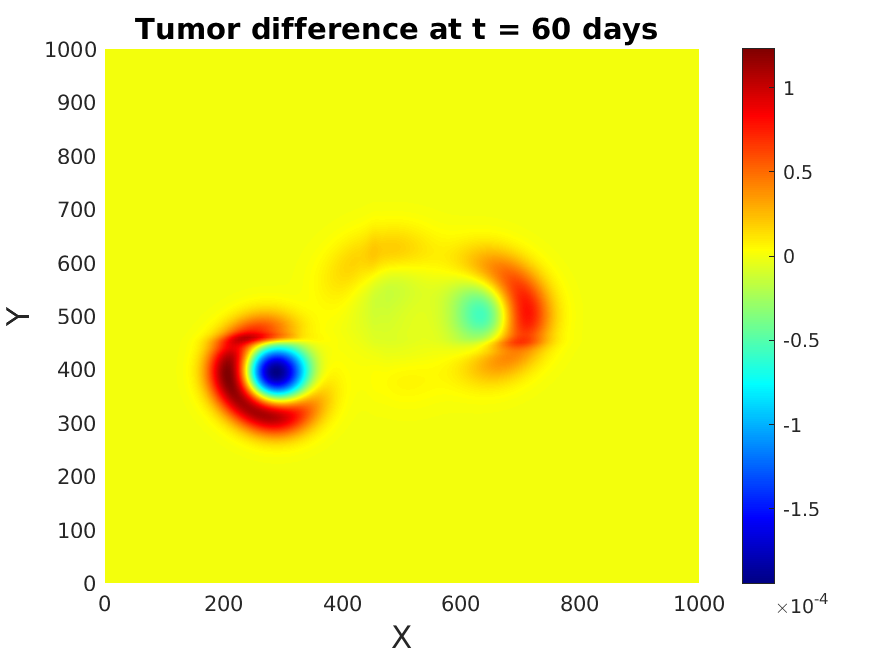}}\\
	\raisebox{1.2cm}{\rotatebox[origin=t]{90}{90 days}}{\includegraphics[width=1\linewidth, height = 3cm]{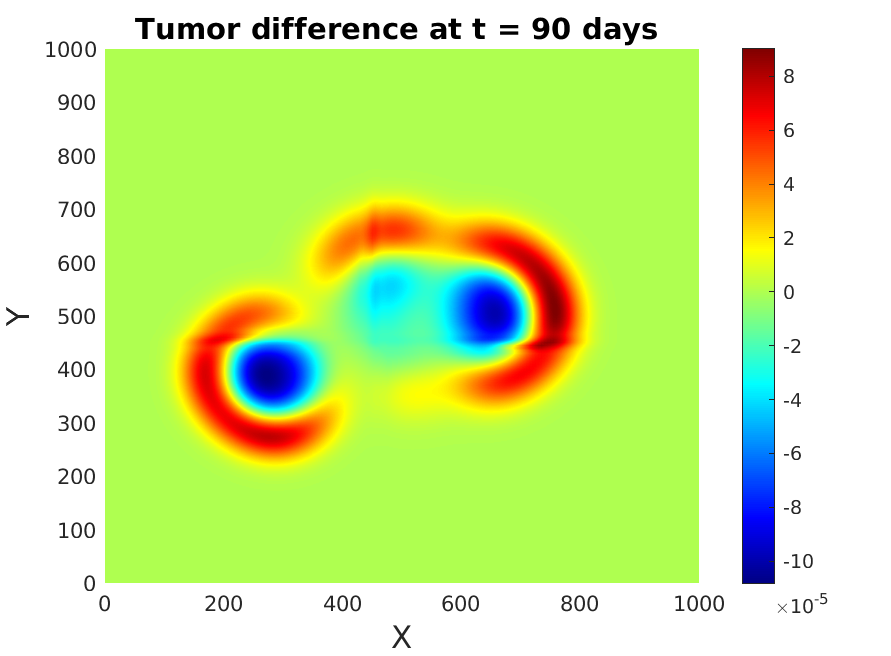}}\\
	\raisebox{1.2cm}{\rotatebox[origin=t]{90}{300 days}}{\includegraphics[width=1\linewidth, height = 3cm]{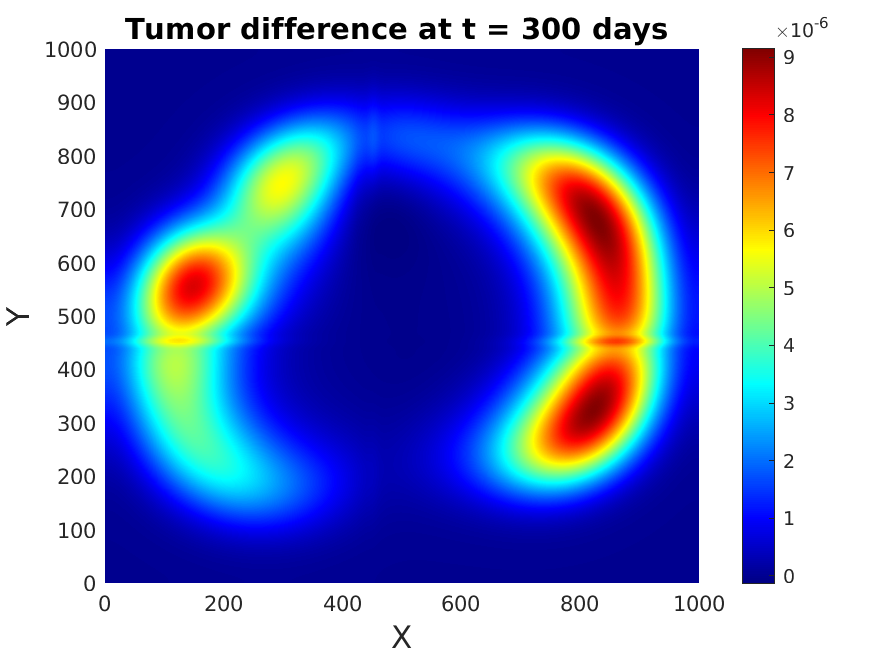}}\\
	\raisebox{1.2cm}{\rotatebox[origin=t]{90}{330 days}}{\includegraphics[width=1\linewidth, height = 3cm]{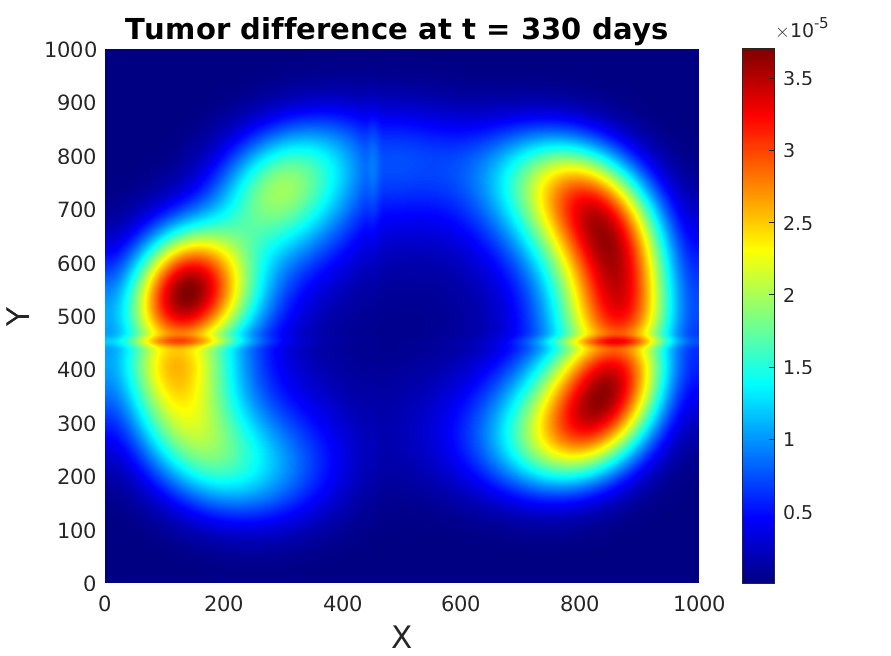}}\\
	\raisebox{1.2cm}{\rotatebox[origin=t]{90}{390 days}}{\includegraphics[width=1\linewidth, height = 3cm]{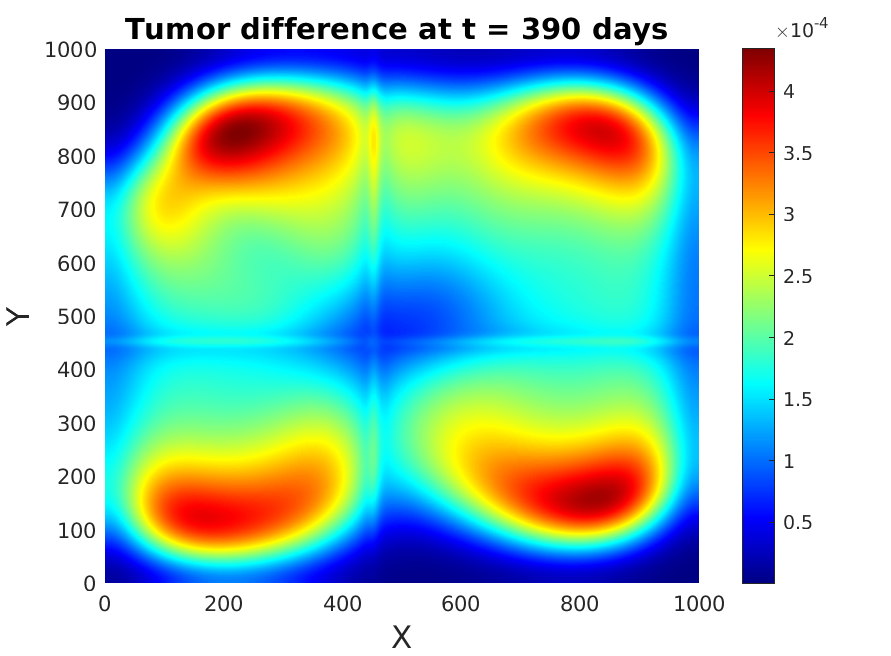}}\\
	\raisebox{1.2cm}{\rotatebox[origin=t]{90}{420 days}}{\includegraphics[width=1\linewidth, height = 3cm]{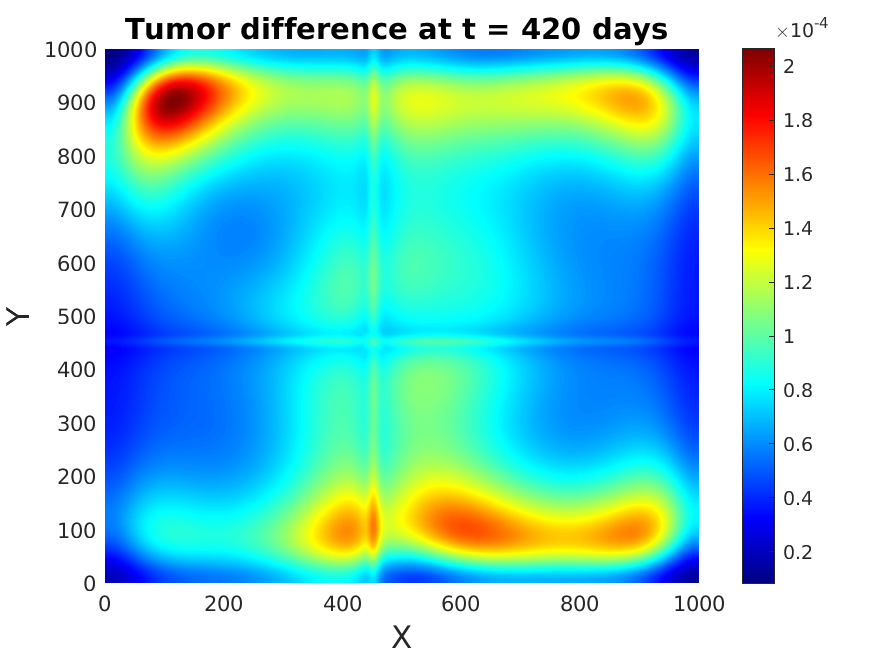}}%
	\subcaption{Tumor difference}
\end{minipage}%
\hspace{0.2cm}
\begin{minipage}[stb]{.24\linewidth}
	{\includegraphics[width=1\linewidth, height = 3cm]{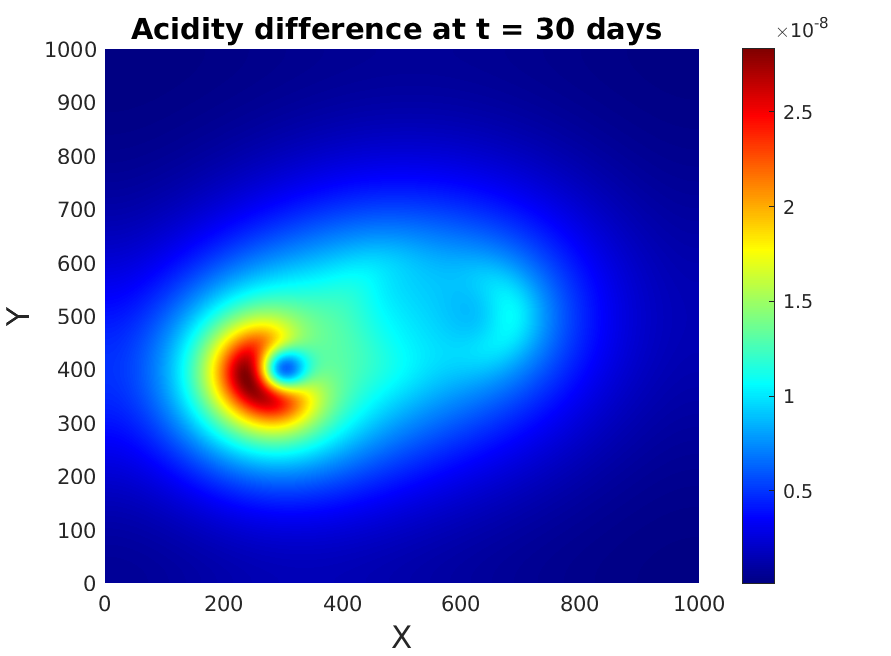}}\\
	{\includegraphics[width=1\linewidth, height = 3cm]{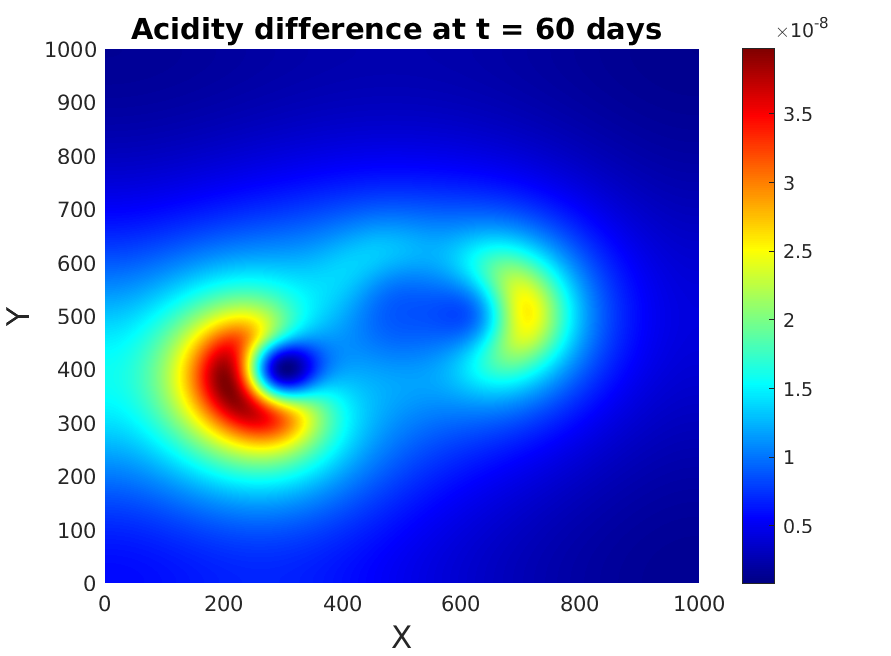}}\\
	{\includegraphics[width=1\linewidth, height = 3cm]{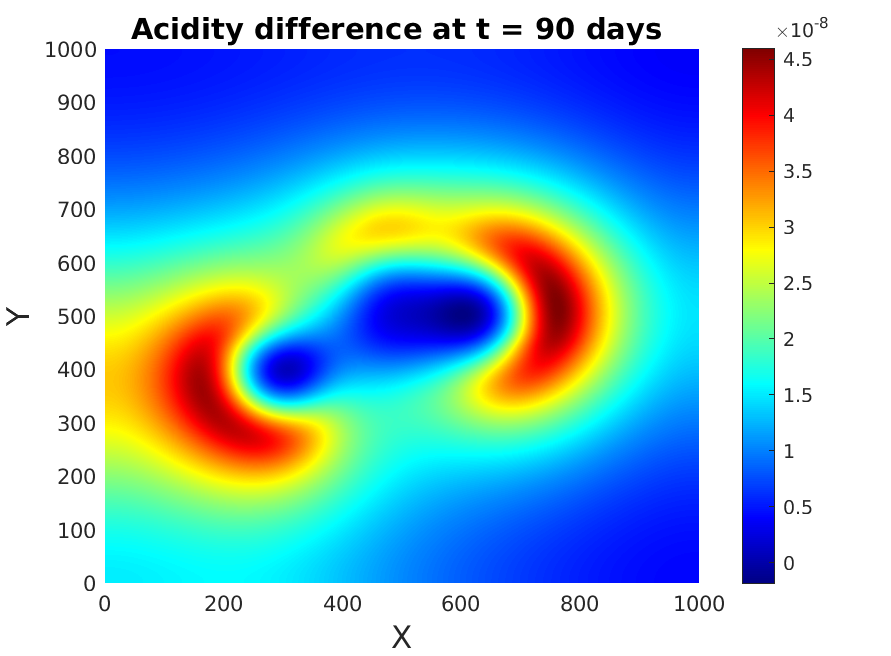}}\\
	{\includegraphics[width=1\linewidth, height = 3cm]{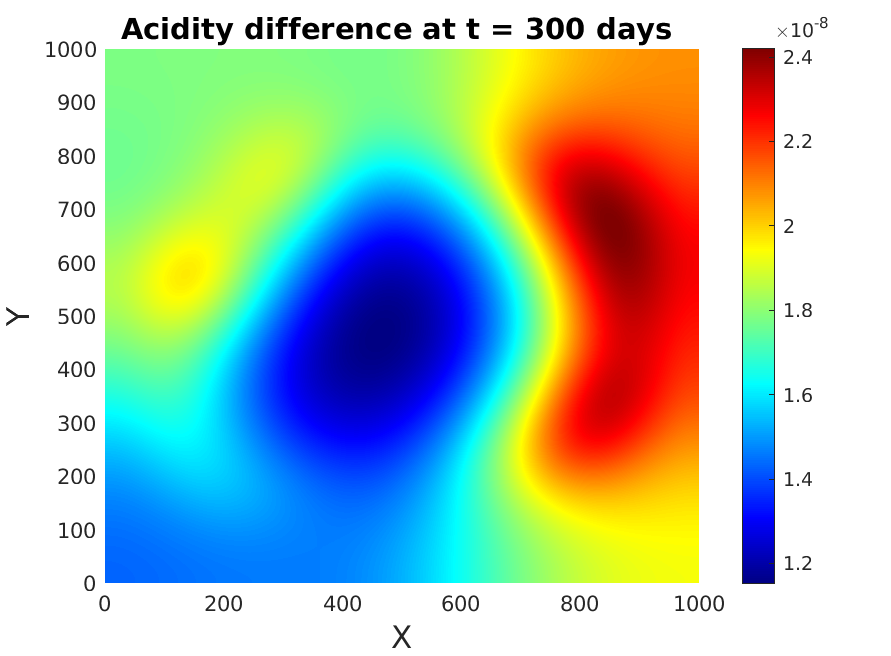}}\\
	{\includegraphics[width=1\linewidth, height = 3cm]{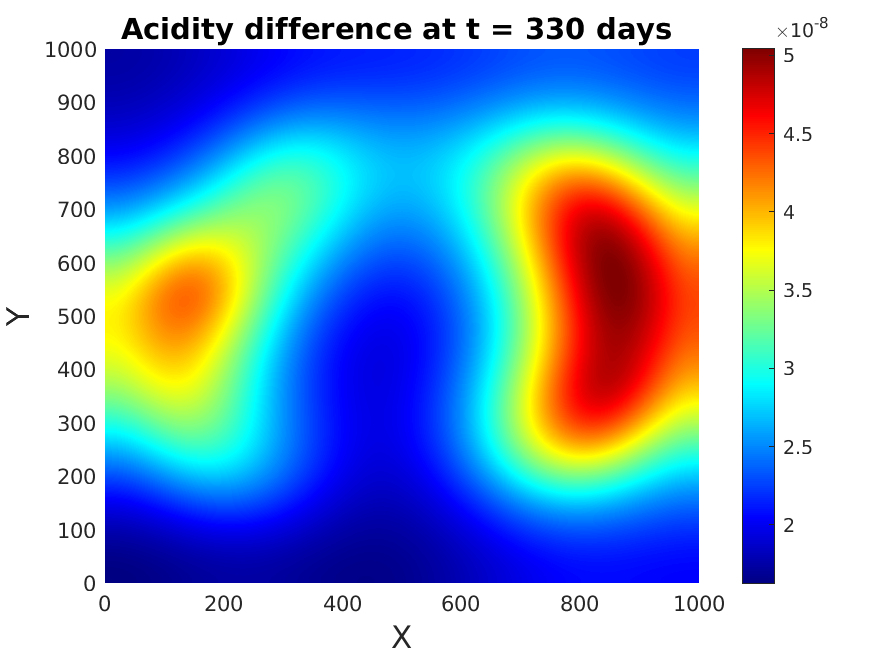}}\\
	{\includegraphics[width=1\linewidth, height = 3cm]{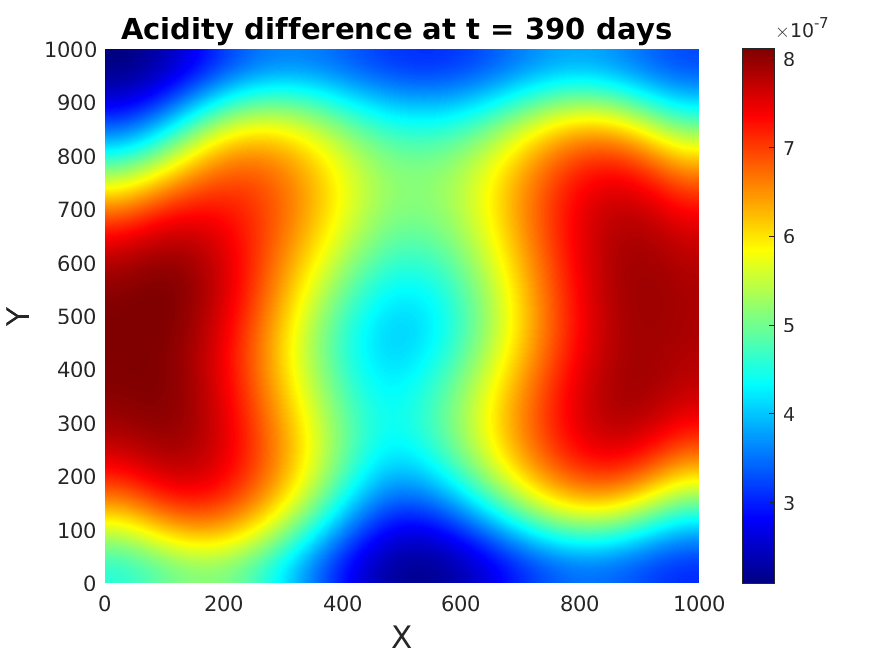}}\\
	{\includegraphics[width=1\linewidth, height = 3cm]{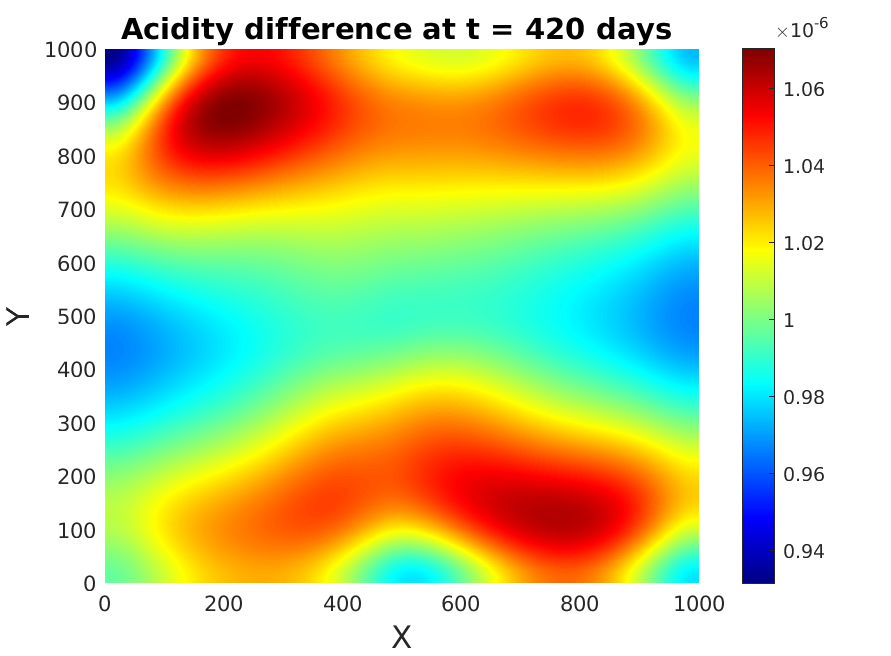}}%
	\subcaption{Acidity difference}
\end{minipage}%
\hspace{0.1cm}
\begin{minipage}[stb]{.24\linewidth}
	{\includegraphics[width=1\linewidth, height = 3cm]{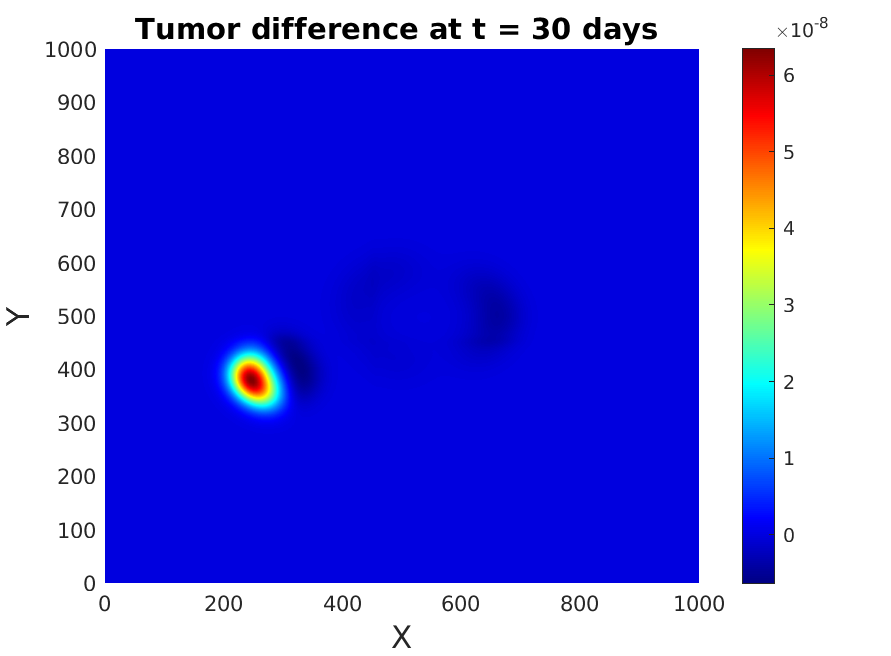}}\\
	{\includegraphics[width=1\linewidth, height = 3cm]{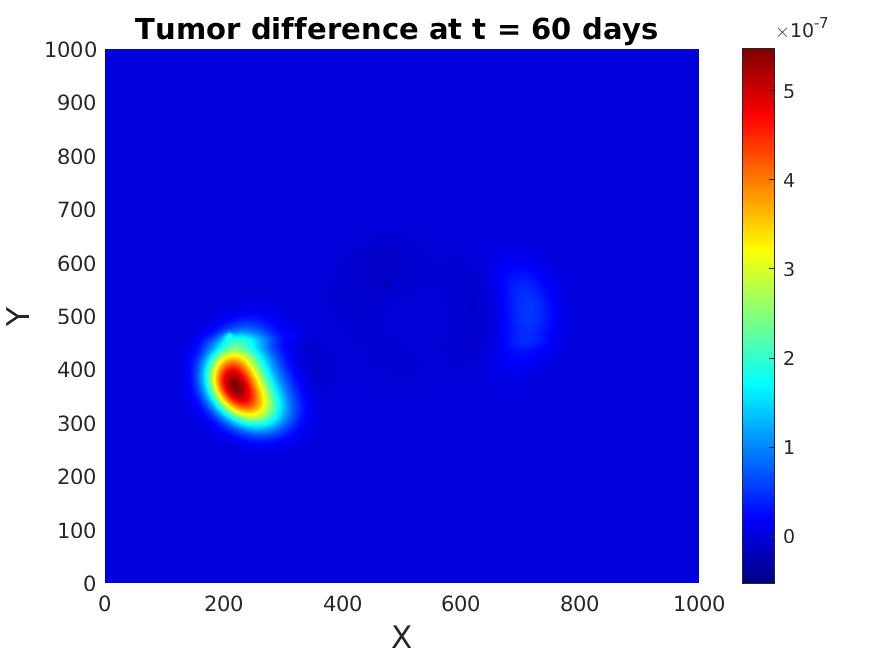}}\\
	{\includegraphics[width=1\linewidth, height = 3cm]{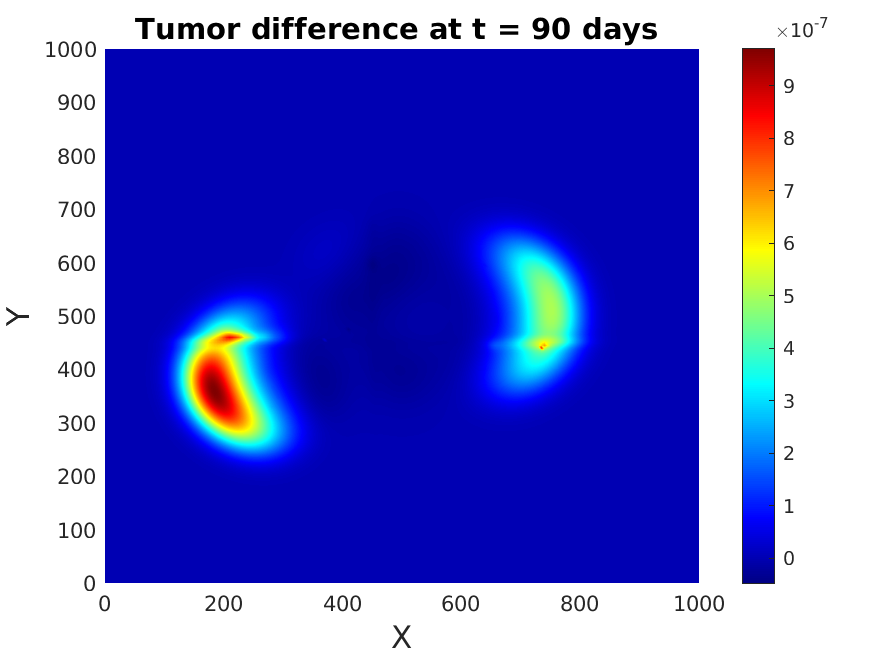}}\\
	{\includegraphics[width=1\linewidth, height = 3cm]{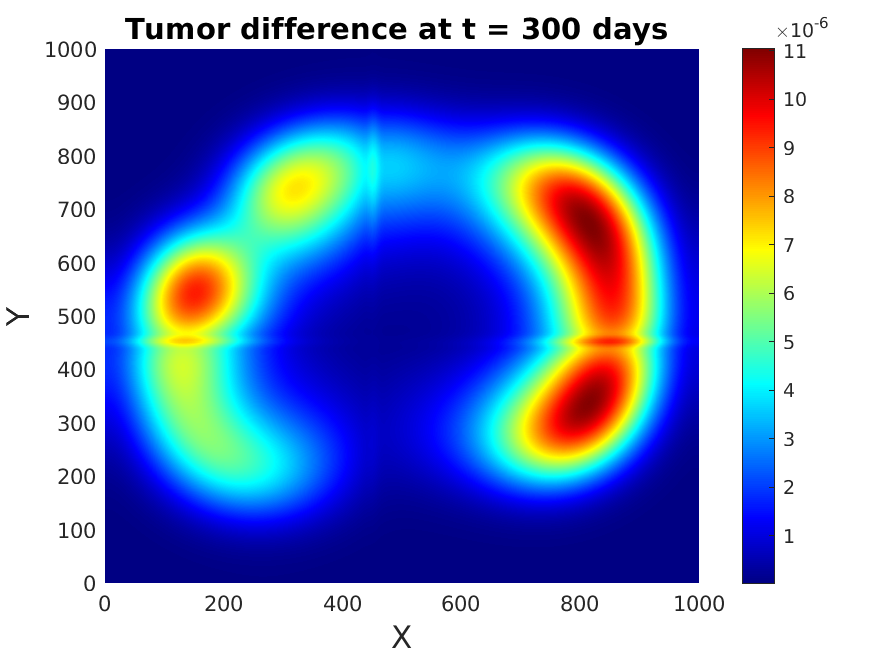}}\\
	{\includegraphics[width=1\linewidth, height = 3cm]{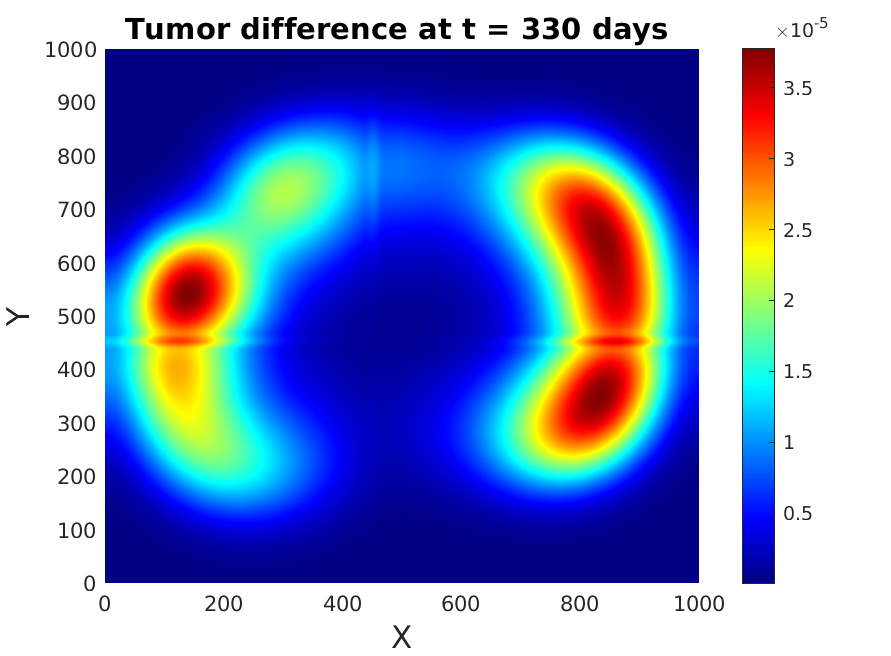}}\\
	{\includegraphics[width=1\linewidth, height = 3cm]{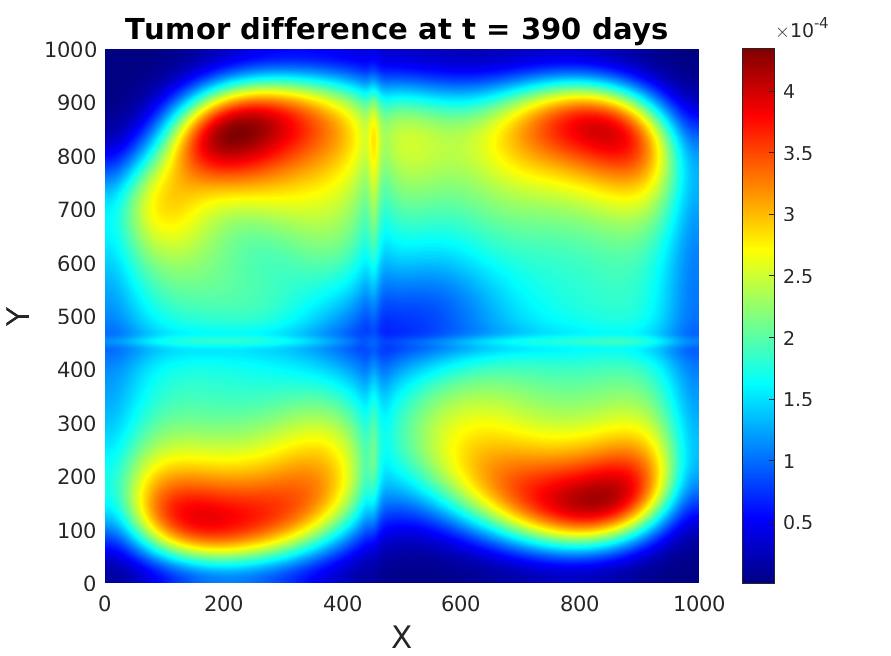}}\\
	{\includegraphics[width=1\linewidth, height = 3cm]{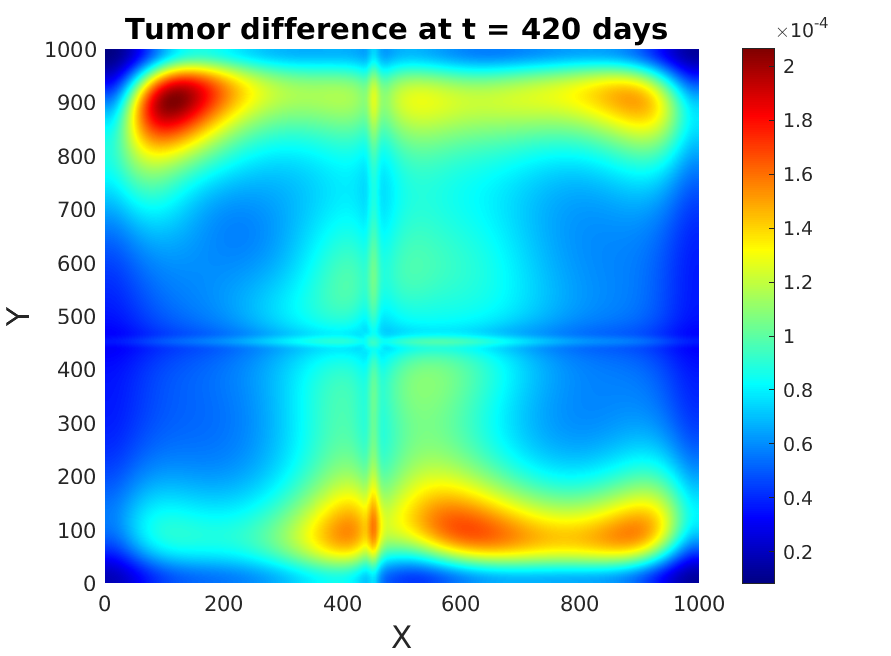}}
	\subcaption{Tumor difference}
\end{minipage}
\begin{minipage}[stb]{.24\linewidth}
	{\includegraphics[width=1\linewidth, height = 3cm]{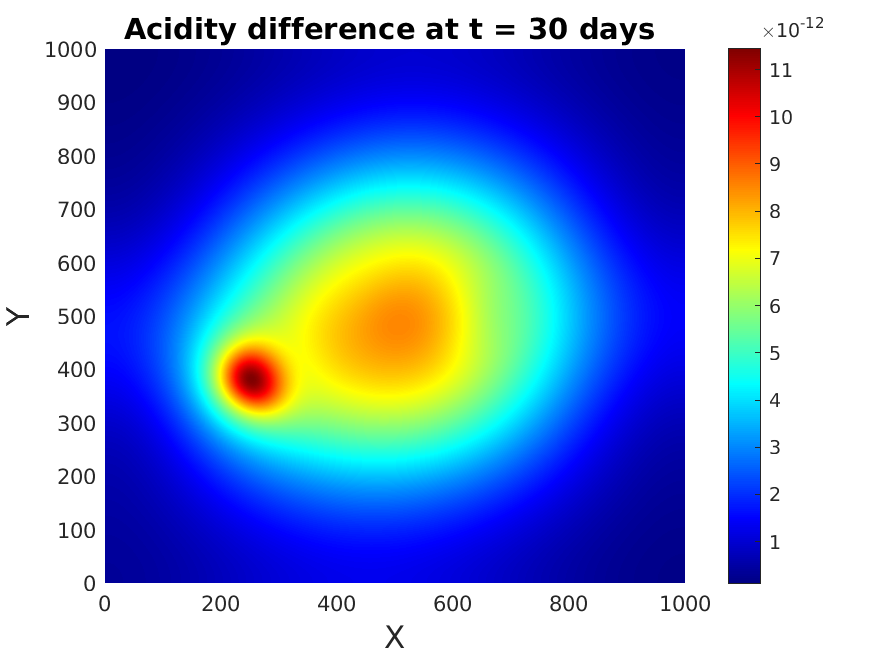}}\\
	{\includegraphics[width=1\linewidth, height = 3cm]{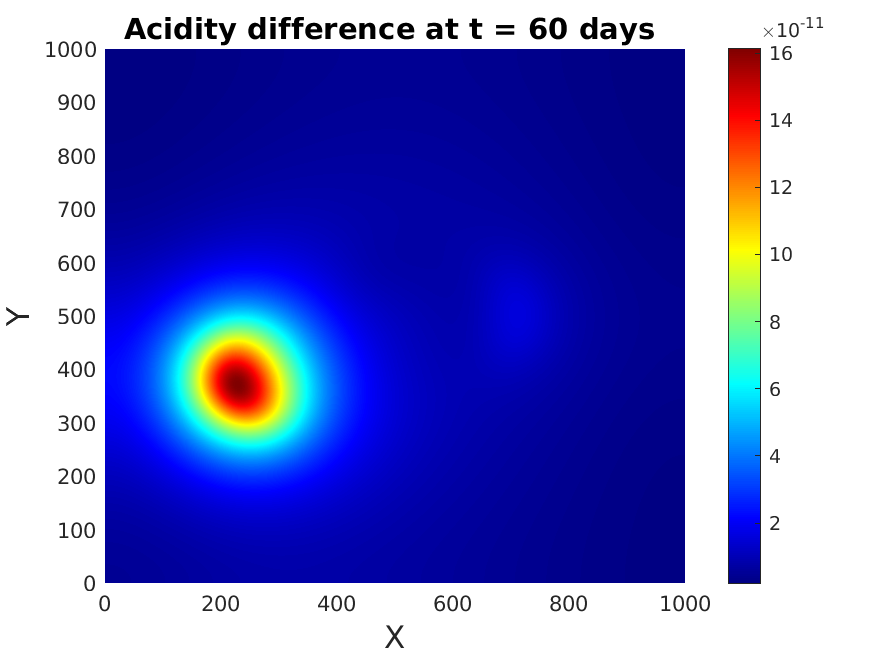}}\\
	{\includegraphics[width=1\linewidth, height = 3cm]{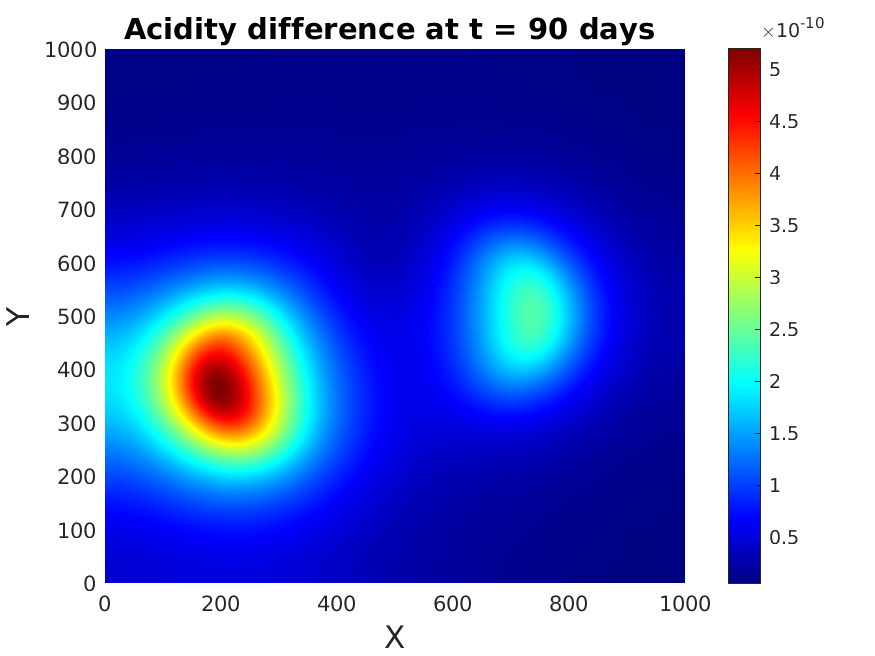}}\\
	{\includegraphics[width=1\linewidth, height = 3cm]{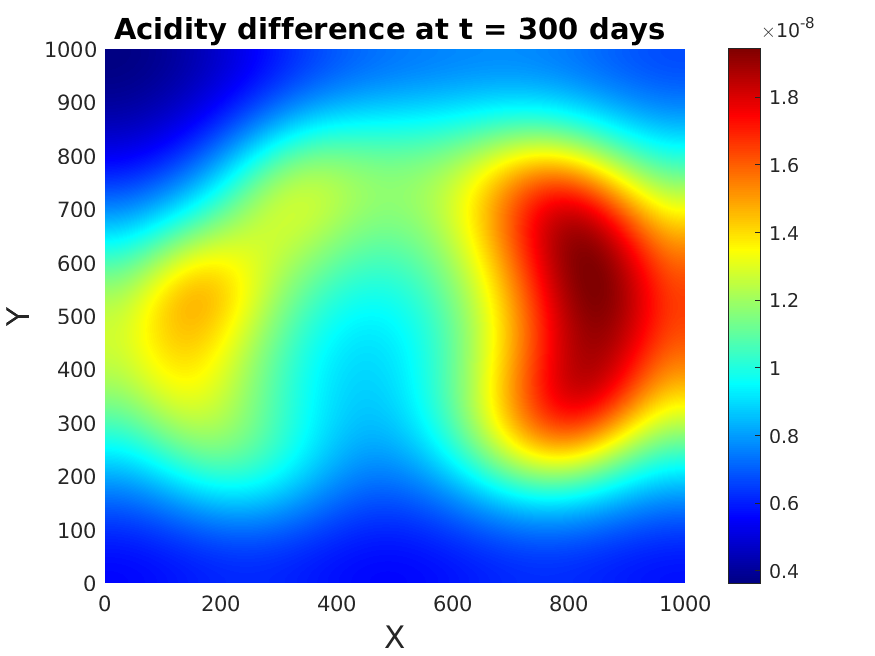}}\\
	{\includegraphics[width=1\linewidth, height = 3cm]{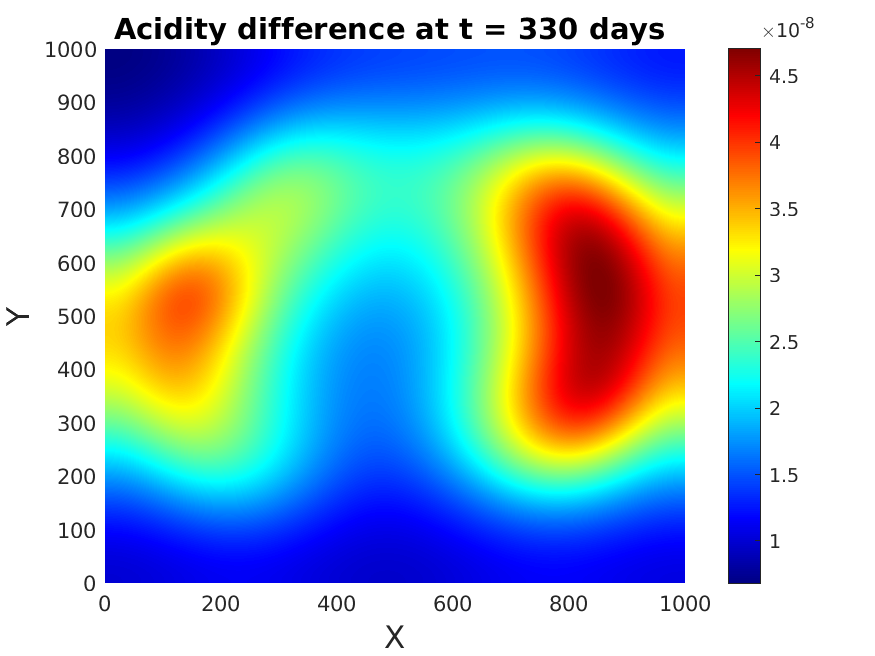}}\\
	{\includegraphics[width=1\linewidth, height = 3cm]{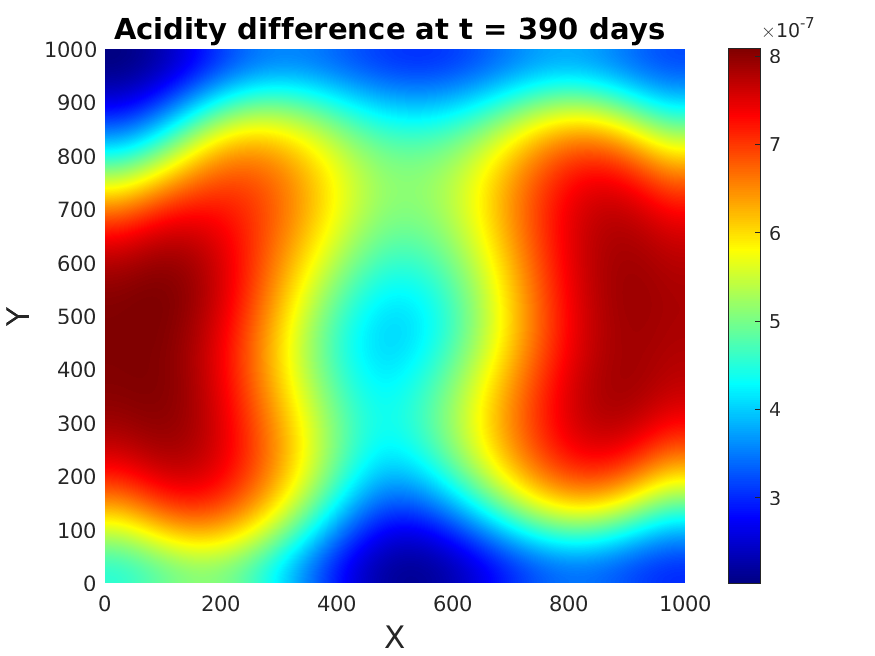}}\\
	{\includegraphics[width=1\linewidth, height = 3cm]{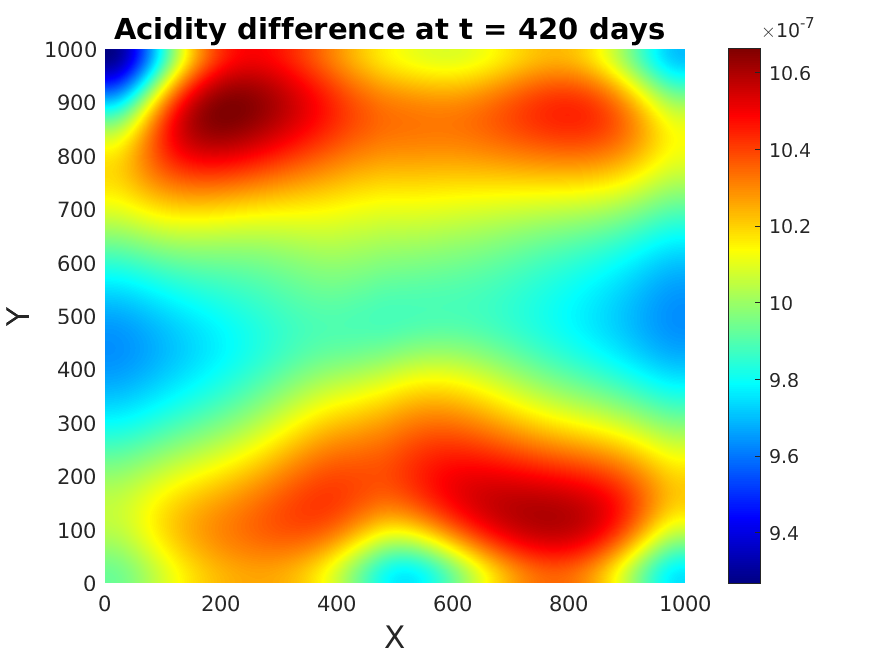}}
	\subcaption{Acidity difference}
\end{minipage}%
\caption{\textbf{Experiment 3}: effect of angiogenesis. Left columns: Difference between tumor density and acidity concentration, as obtained with full system \eqref{eq:macro-nondim}, and those computed from the model variant \eqref{eq:diff_old-model} with proton uptake rate $ \zeta_h = 3.6 \cdot 10^{-8}$/h. Right columns: Same differences, but between computations done with \eqref{eq:macro-nondim} and with \eqref{eq:diff_comp_neu}.  }\label{fig:diff_nonangio}
\end{figure}

\paragraph{Experiment 1} 
To start with, we solve the full system \eqref{eq:macro-nondim} with no-flux boundary conditions and with initial data as in Subsection \ref{subsec:ICs}. The results are shown in Figure \ref{fig:scenario1}. The garland-shaped glioma pattern forms (and is well visible after 60 simulated days) due to the high proton concentration in its central region, in spite of ECs being attracted by the glioma producing VEGF. In fact, ECs spread over almost the entire domain of simulation, coming from the left and right margins; the EC plot at $t=60$ days exhibits large densities (except in the narrow central region, where the cells did not yet arrived), a phenomenon known as hyperplasia. Subsequently, ECs have accumulated enough near the tumor to uptake acidity and release nutrients, which leads among others to the tumor cells being able to move away from the acidic area and toward more favorable regions (due to $\gamma_1\gg \gamma_2$ the influence of acidity dominates the correction of self-diffusion), where they start producing acidity and VEGF.
The pseudopalisades develop larger diameters and involve glioma garlands with lower densities. The cancer cells are followed by ECs, which again reduce the proton concentration, thus changing the glioma behavior. The pseudopalisades are disrupted and at later times both ECs and glioma cells can repopulate the previously too acidic regions. 
\paragraph{Experiment 2} Same as \textbf{Experiment 1}, but with $\gamma_1=\gamma_2$, i.e. the flux-limited self-diffusion and pH-taxis being equally weighted. The results are illustrated in Figure \ref{fig:scenario2}. The behavior of solution components is qualitatively similar to the previous scenario, except the pseudopalisade-like patterns persisting for a longer time, being succeeded by a more uniform tumor- and EC spread occupying the whole domain (also migrating into the formerly most hypoxic areas) and exhibiting higher cell densities. 
\paragraph{Experiment 3} Here we want to assess the effects of extending the model in \cite{kumar2020multiscale}:
\begin{subequations}\label{eq:diff_old-model}
	\begin{align}
	M_t &= \nabla \nabla : (\mathbb{D}_T M) + \nabla \cdot \left(G(h)M\mathbb{D}_T \nabla h\right) + M \left(1-M \right)\left(1-h\right)\label{eq:diff_old-model-1}\\
	h_t & = \nabla h + \gamma_h \frac{M}{1+M} - \zeta_h h,
	\end{align}
\end{subequations}
where $G(h)=\frac{\lambda_1k_D}{(h+k_D)^2(h+k_D+\lambda_0)}$,  $\lambda_0,\lambda_1,k_D>0$ constants, 
by dynamics of quantities involved in angiogenesis, i.e. ECs and VEGF. Figure \ref{fig:diff_nonangio} shows differences between glioma density and proton concentration computed with model \eqref{eq:macro-nondim} and with model versions without vascularization, namely \eqref{eq:diff_old-model} (for the left part of Figure \ref{fig:diff_nonangio}) and respectively
\begin{subequations}\label{eq:diff_comp_neu}
	\begin{align}
	M_t &= \nabla \nabla : (\mathbb{D}_T M) + \alpha \nabla \cdot \left(\left( D_s \mathbb{I}_N - \mathbb{D}_T \right) \phi(h,M)M \right) + M \left(1-M \right)\left(1-h\right)\label{eq:diff_comp_neu-1}\\
	h_t & = \nabla h + \gamma_h \frac{M}{1+M} - \zeta_h h.
	\end{align}
\end{subequations}
for the right part of Figure \ref{fig:diff_nonangio}. \\[-2ex]

\noindent
The plots on the left column exhibit enhanced tumor growth and spread in the presence of ECs and VEGF. Earlier stages feature larger glioma densities (and therewith associated proton concentrations) distributed in a ring-shaped manner, which suggests more pronounced pseudopalisades. At later times, in the presence of vascularization, the tumor cells are not only able to proliferate, but also to reoccupy the formerly too acidic areas near the center of the simulation domain, while still responding to the spatial distribution of protons. This leads (as observed also in the previous numerical experiments) to patterns no longer having the pseudopalisade-like appearance. Similar effects can be observed in the plots on the right column, although less pronounced, due to the smaller structural difference between the two models compared therein. In fact, at later times there are only minimal differences between the plots on the left and on the right columns, which suggests that on the long run the effects of flux limitation in pH-taxis and additional self-diffusion from \eqref{eq:diff_comp_neu-1} are increasingly diminished in comparison to the simple pH-taxis from \eqref{eq:diff_old-model}, most probably because of the diffusion dominance in all these models. \\[-2ex]

\begin{figure}[!htbp]
\centering
\begin{subfigure}[b]{0.31\textwidth}
	{\includegraphics[width=1\linewidth]{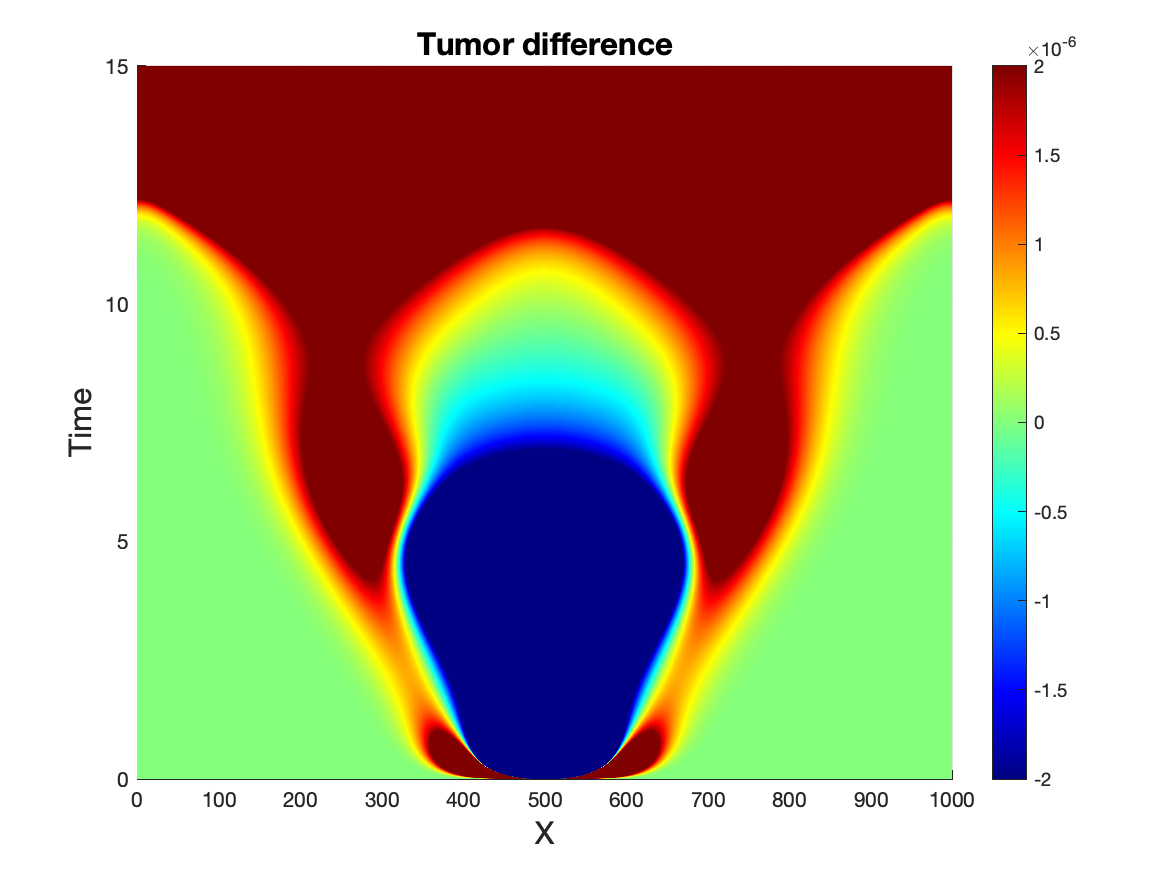}
	}
\end{subfigure}\quad
\begin{subfigure}[b]{0.31\textwidth}
	{\includegraphics[width=1\linewidth]{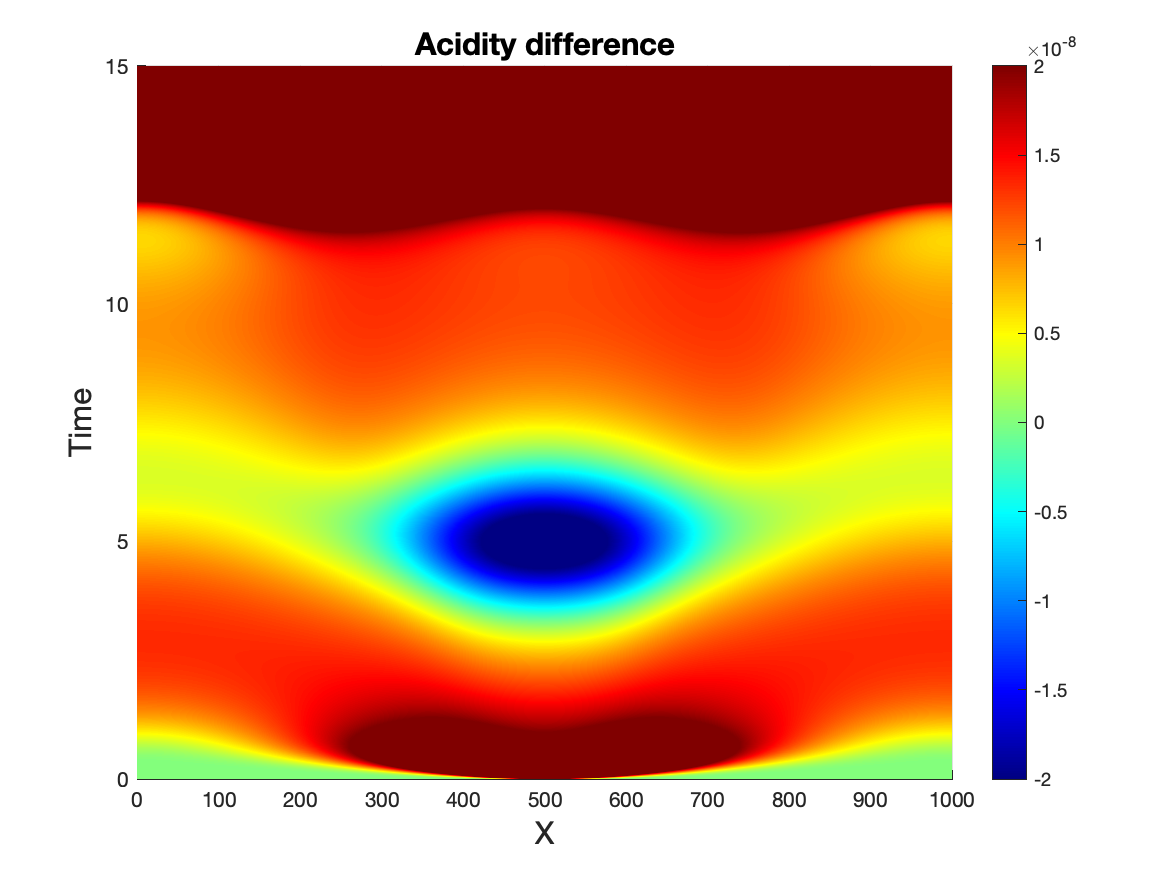}
	}
\end{subfigure}
\quad
\begin{subfigure}[b]{0.31\textwidth}
	{\includegraphics[width=1\linewidth]{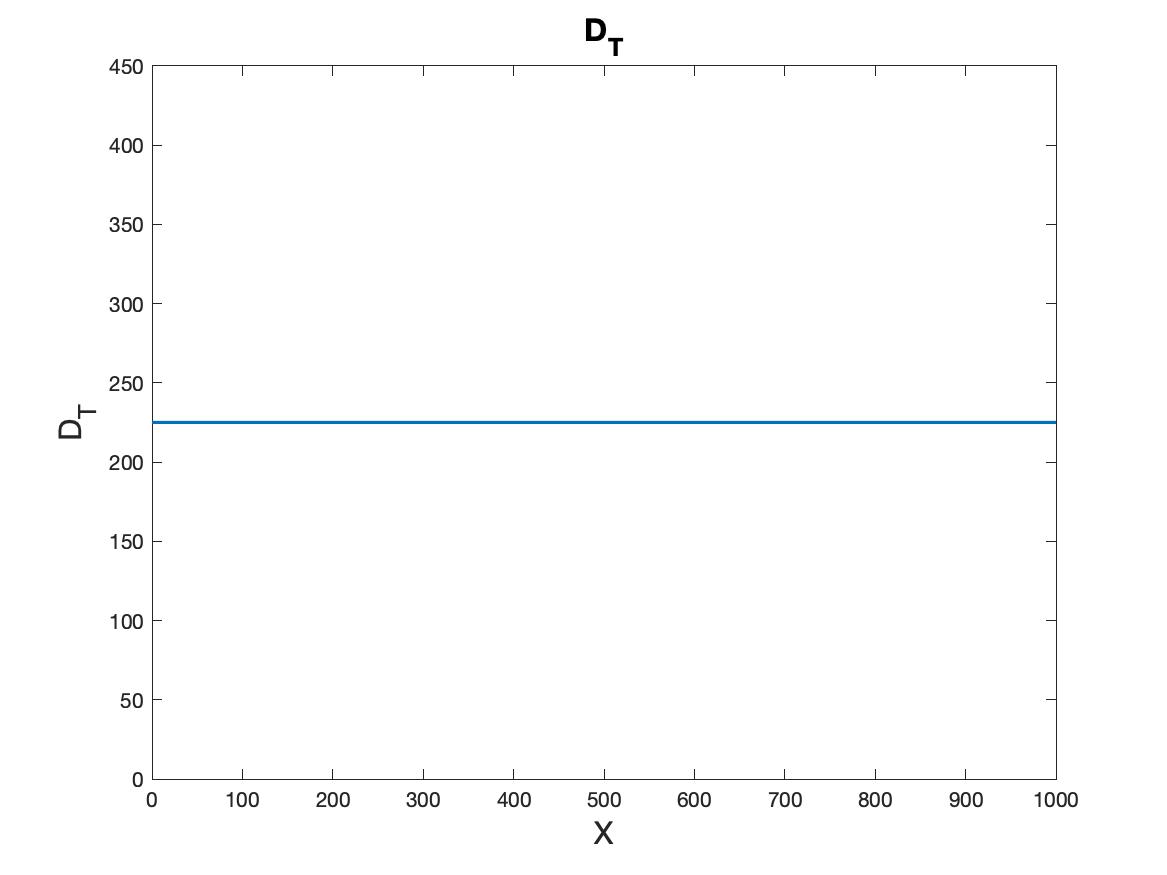}
	}
\end{subfigure}\\
\begin{subfigure}[b]{0.31\textwidth}
	{\includegraphics[width=1\linewidth]{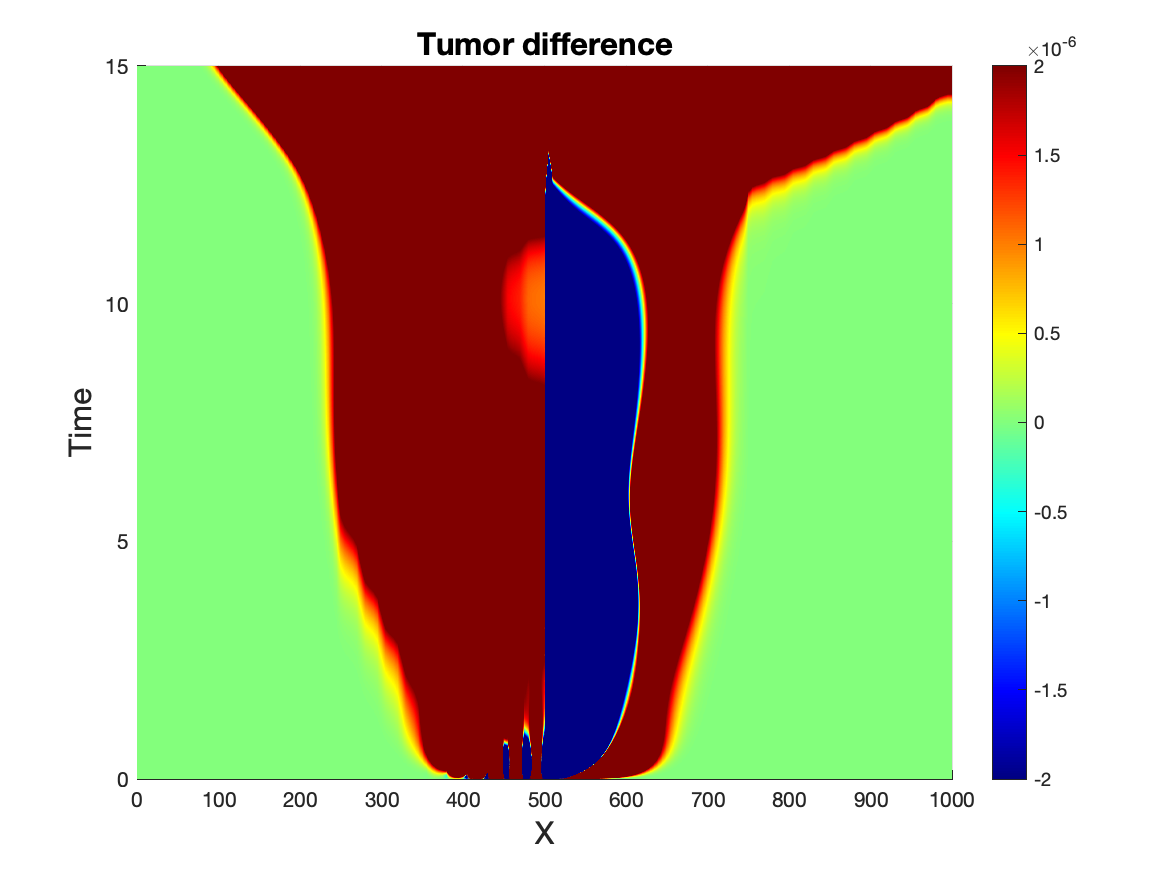}
	}
\end{subfigure}\hfill
\begin{subfigure}[b]{0.31\textwidth}
	{\includegraphics[width=1\linewidth]{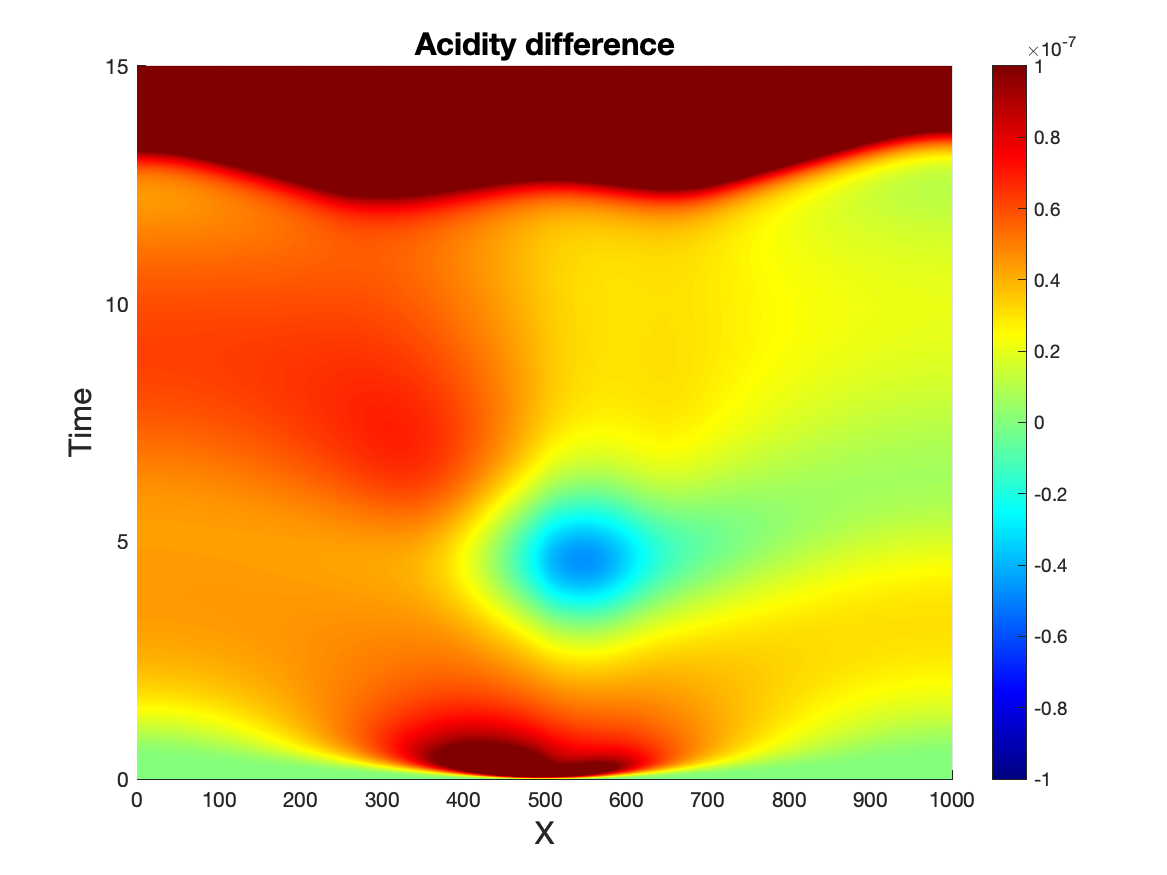}
	}
\end{subfigure}
\hfill
\begin{subfigure}[b]{0.31\textwidth}
	{\includegraphics[width=1\linewidth]{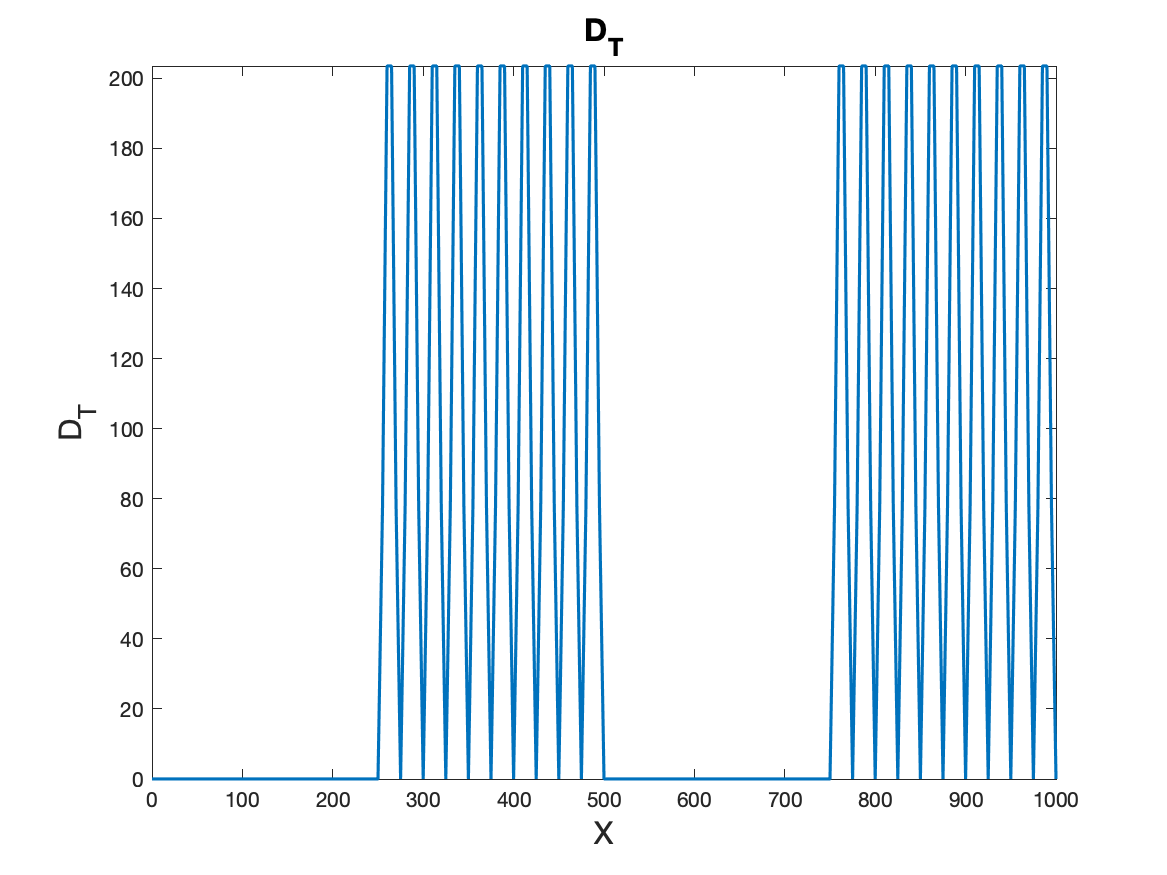}
	}
\end{subfigure}\\
\caption{Differences between tumor and acidity patterns (1D) for the model comparison done in the left column of Figure \ref{fig:diff_nonangio} (framework of \textbf{Experiment 3}). Different choices are made for the glioma diffusion coefficient $D_T(x)$, as shown in the rightmost column.}\label{fig:patterns-1D-Exp3}
\end{figure}

\noindent
Figure \ref{fig:patterns-1D-Exp3} shows differences between 1D tumor and acidity patterns for the model comparison done in the left column of Figure \ref{fig:diff_nonangio} and for two different choices of the diffusion coefficient $D_T$ (which is in this situation a scalar function of position): constant (upper row) and involving strong (over two intervals) or milder (at a finite number of positions) degenerations. The plots suggest on the one hand that the availability of vasculature permits an enhanced tumor spread in regions with higher diffusivity, but also enables invasion in areas where the cells modeled by \eqref{eq:diff_old-model} infer strongly degenerate diffusion and therewith associated annihilation of pH-taxis. Moreover, the model version \eqref{eq:diff_old-model} predicts high cell accumulation near the interface with highest diffusivity difference, whereas model \eqref{eq:macro-nondim} eludes such tendency potentially leading to singularities. This is presumably due to the flux-limitations in the taxis and additional self-diffusion terms (for more comments on this issue we refer to Section \ref{sec:discussions}). 

\begin{figure}[!htbp]
	\begin{minipage}[t]{.24\linewidth}
		\raisebox{1.2cm}{\rotatebox[origin=t]{90}{30 days}}{\includegraphics[width=1\linewidth, height = 3cm]{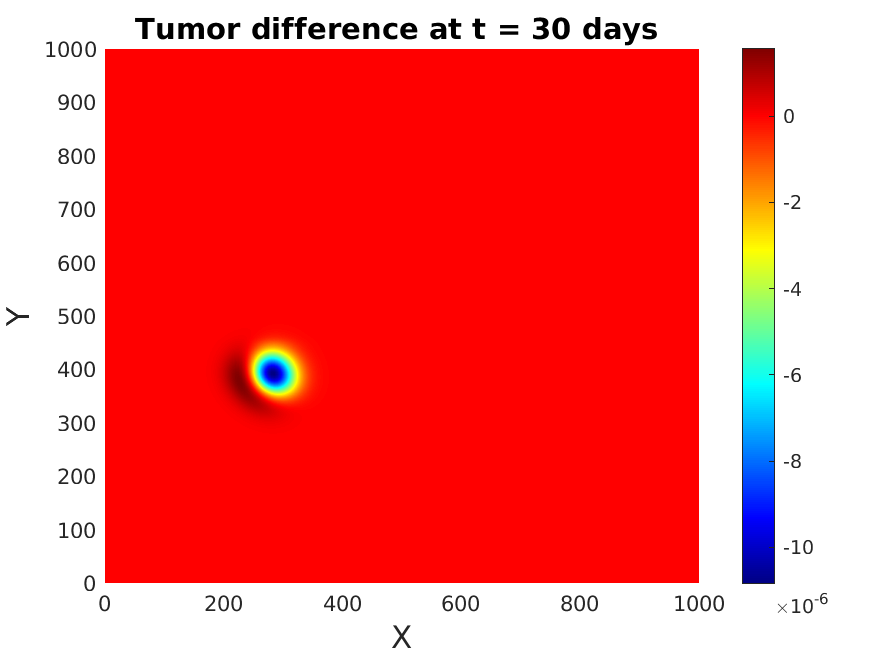}}\\
		\raisebox{1.2cm}{\rotatebox[origin=t]{90}{60 days}}{\includegraphics[width=1\linewidth, height = 3cm]{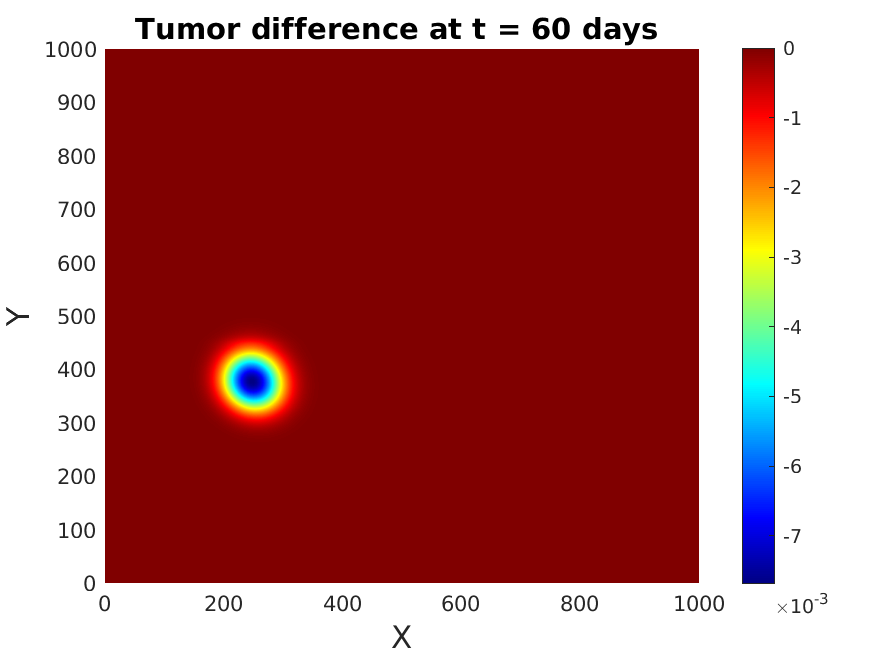}}\\
		\raisebox{1.2cm}{\rotatebox[origin=t]{90}{90 days}}{\includegraphics[width=1\linewidth, height = 3cm]{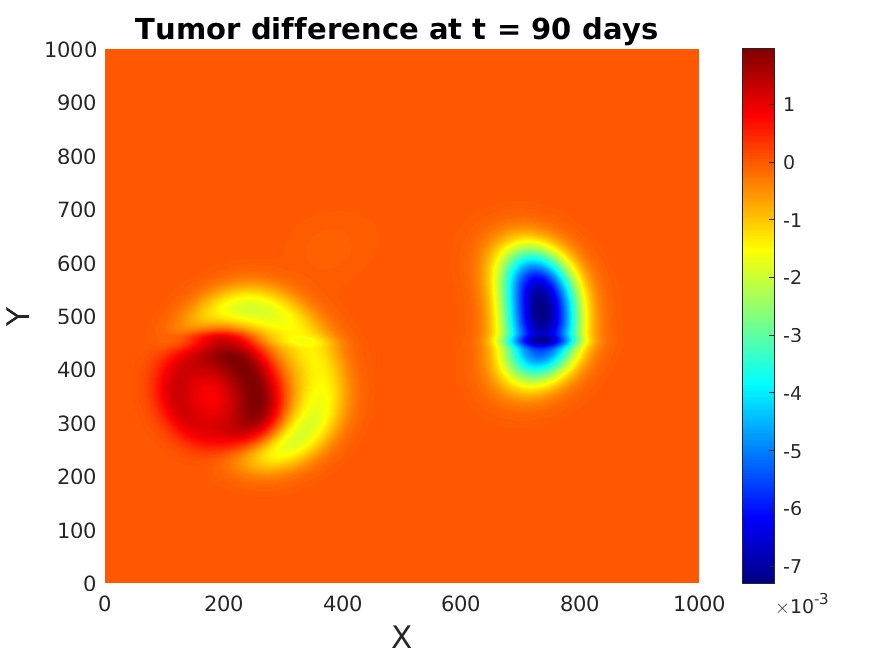}}\\
		\raisebox{1.2cm}{\rotatebox[origin=t]{90}{300 days}}{\includegraphics[width=1\linewidth, height = 3cm]{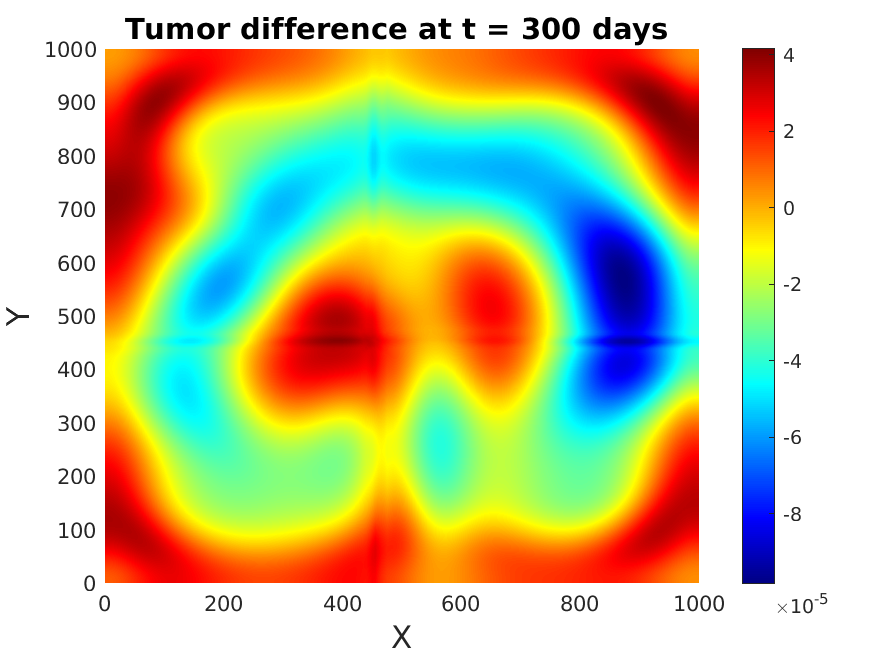}}\\
		\raisebox{1.2cm}{\rotatebox[origin=t]{90}{330 days}}{\includegraphics[width=1\linewidth, height = 3cm]{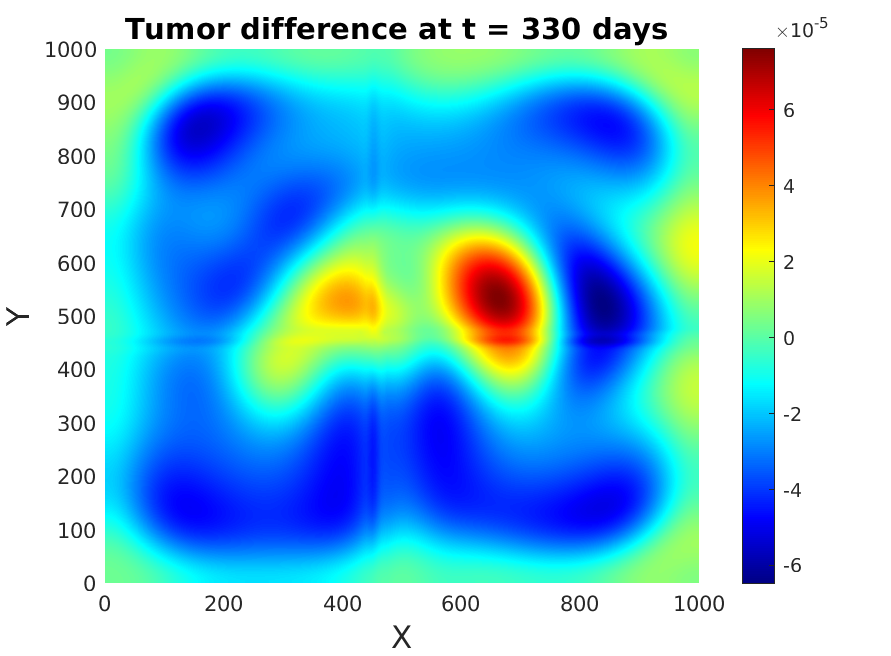}}\\
		\raisebox{1.2cm}{\rotatebox[origin=t]{90}{390 days}}{\includegraphics[width=1\linewidth, height = 3cm]{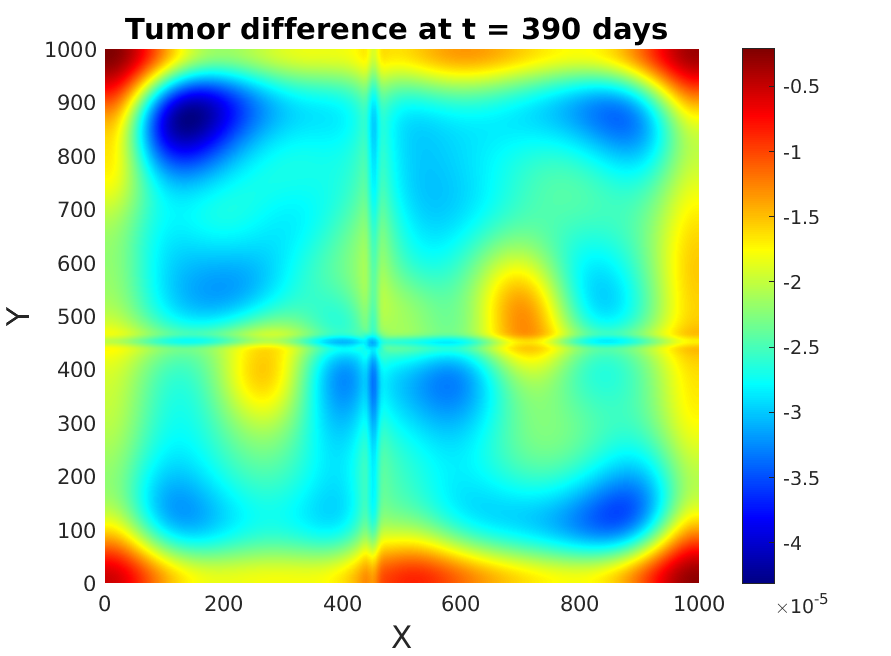}}\\
		\raisebox{1.2cm}{\rotatebox[origin=t]{90}{420 days}}{\includegraphics[width=1\linewidth, height = 3cm]{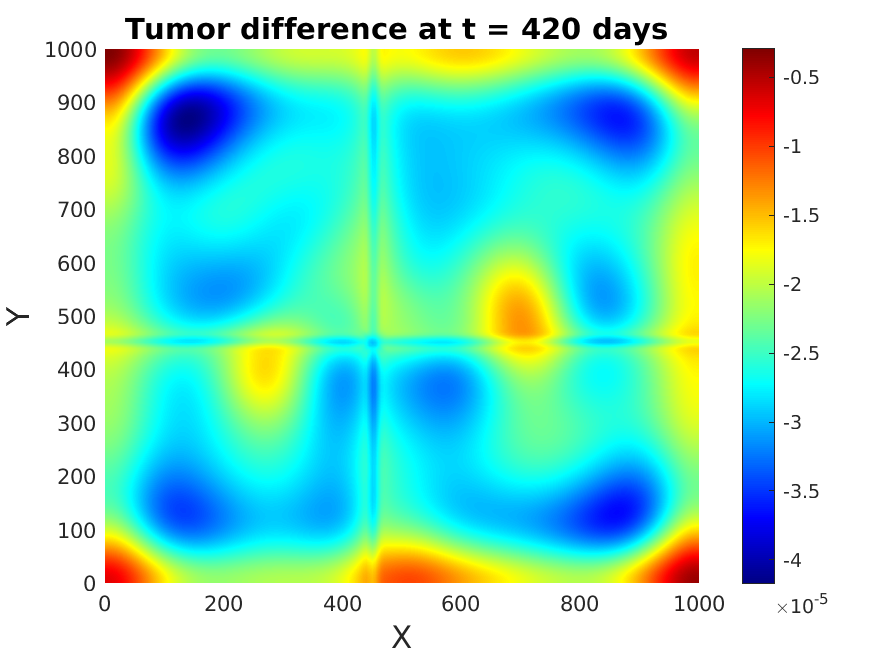}}%
		\subcaption{Tumor difference}
	\end{minipage}%
	\hspace{0.2cm}
	\begin{minipage}[t]{.24\linewidth}
		{\includegraphics[width=1\linewidth, height = 3cm]{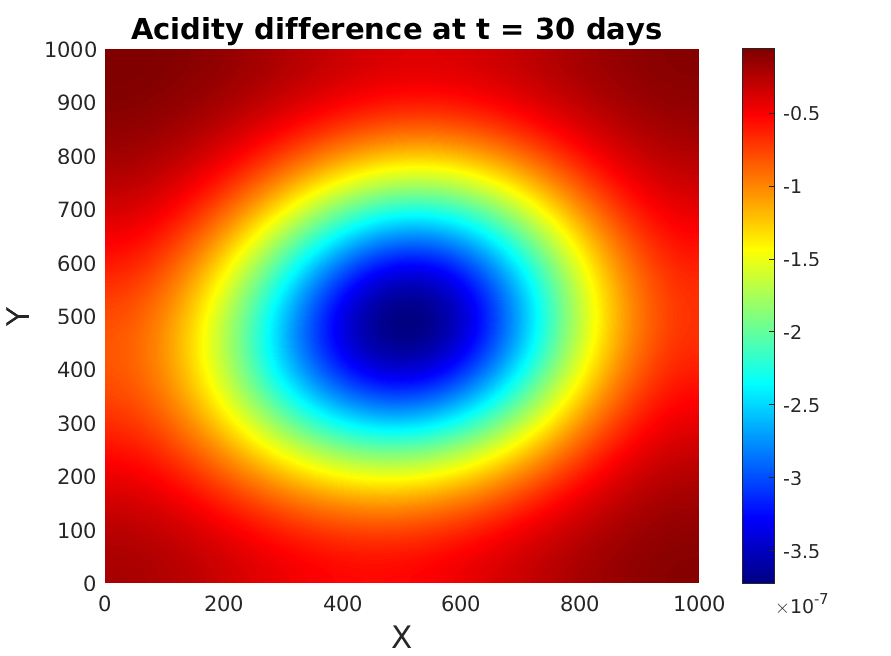}}\\
		{\includegraphics[width=1\linewidth, height = 3cm]{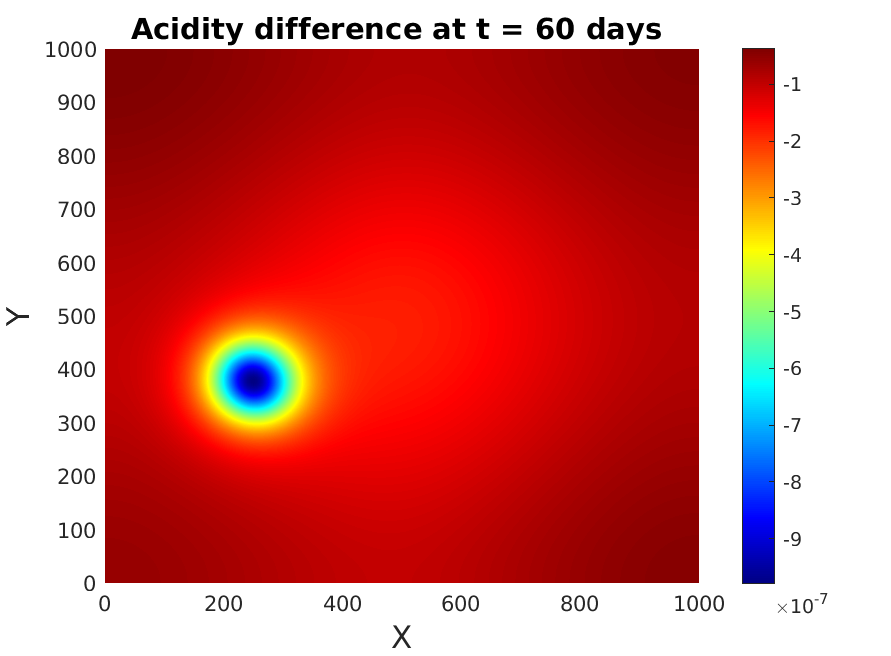}}\\
		{\includegraphics[width=1\linewidth, height = 3cm]{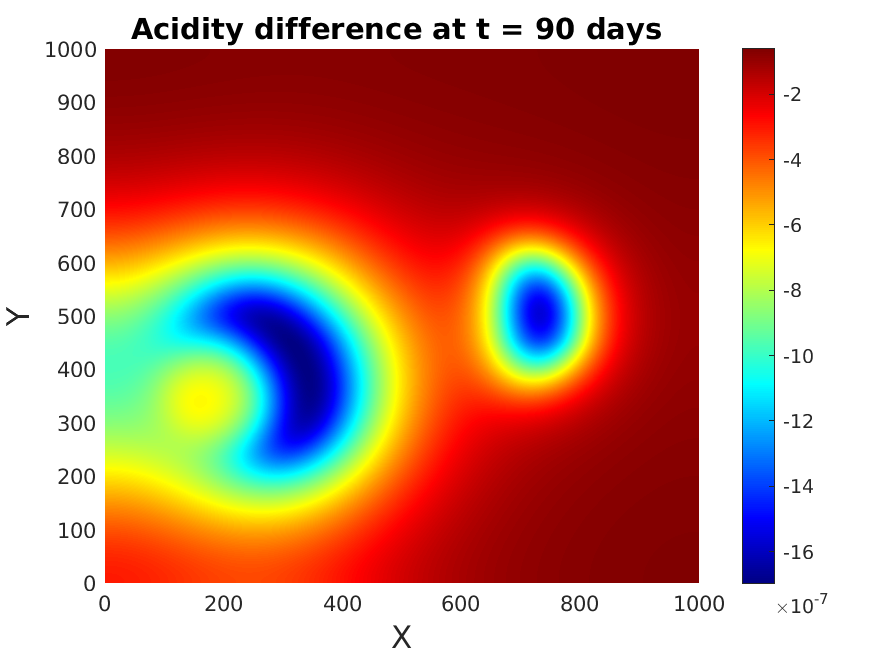}}\\
		{\includegraphics[width=1\linewidth, height = 3cm]{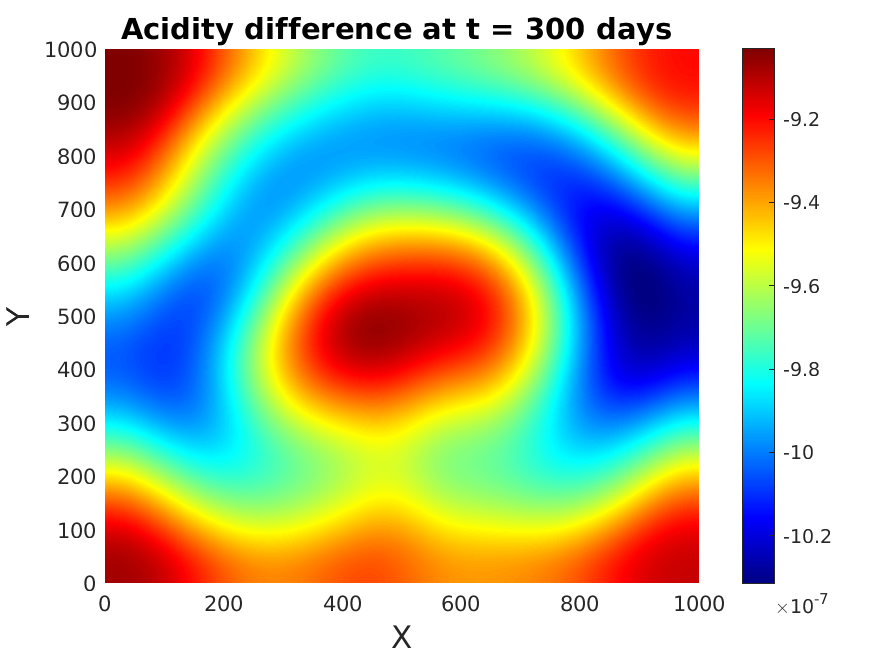}}\\
		{\includegraphics[width=1\linewidth, height = 3cm]{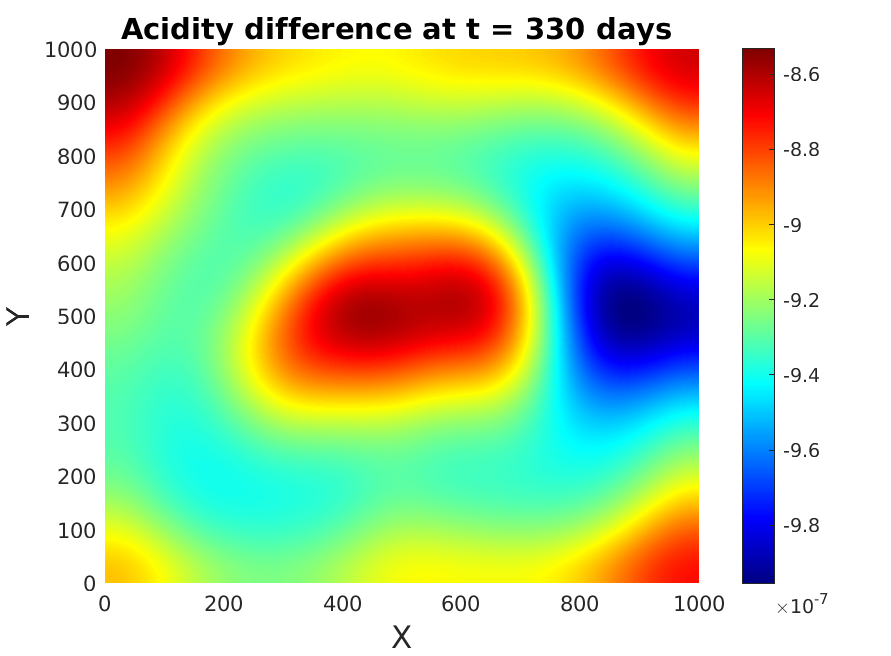}}\\
		{\includegraphics[width=1\linewidth, height = 3cm]{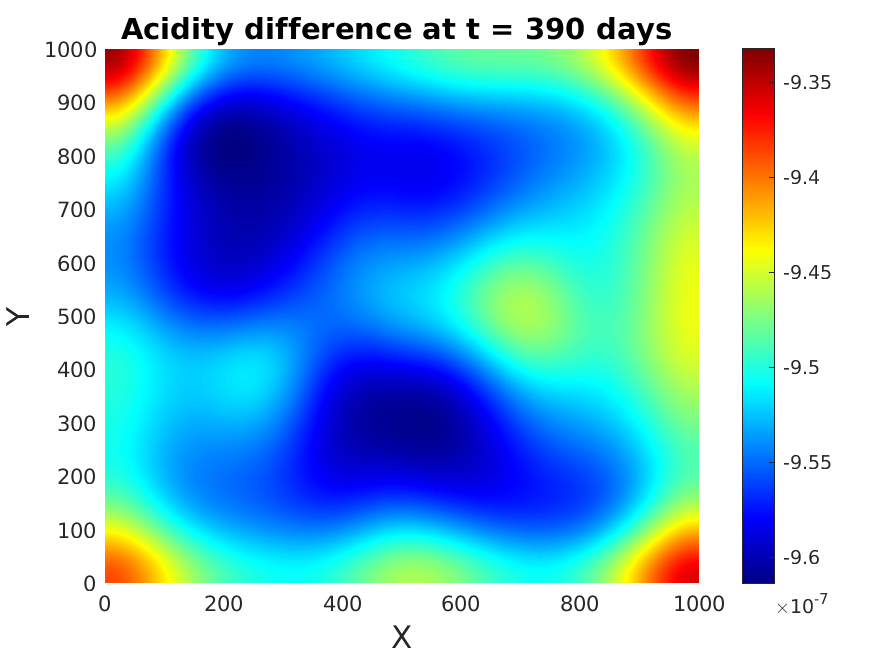}}\\
		{\includegraphics[width=1\linewidth, height = 3cm]{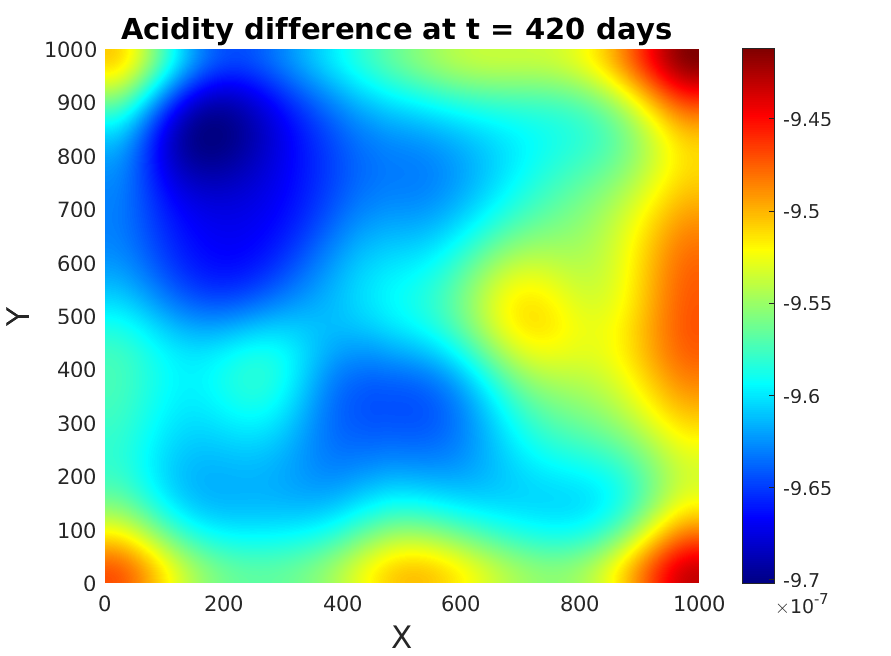}}%
		\subcaption{Acidity difference}
	\end{minipage}%
	\hspace{0.01cm}
	\begin{minipage}[t]{.24\linewidth}
		{\includegraphics[width=1\linewidth, height = 3cm]{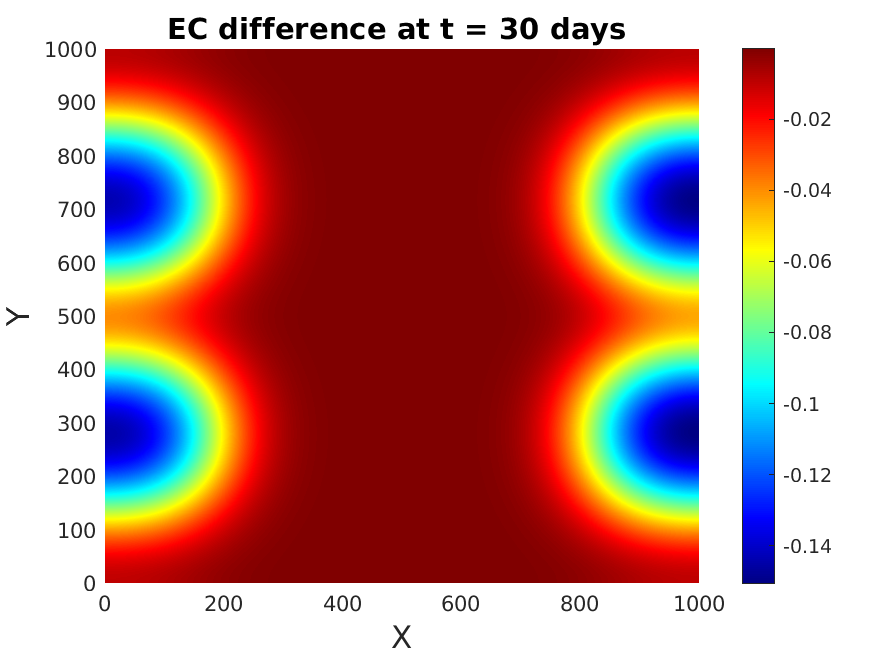}}\\
		{\includegraphics[width=1\linewidth, height = 3cm]{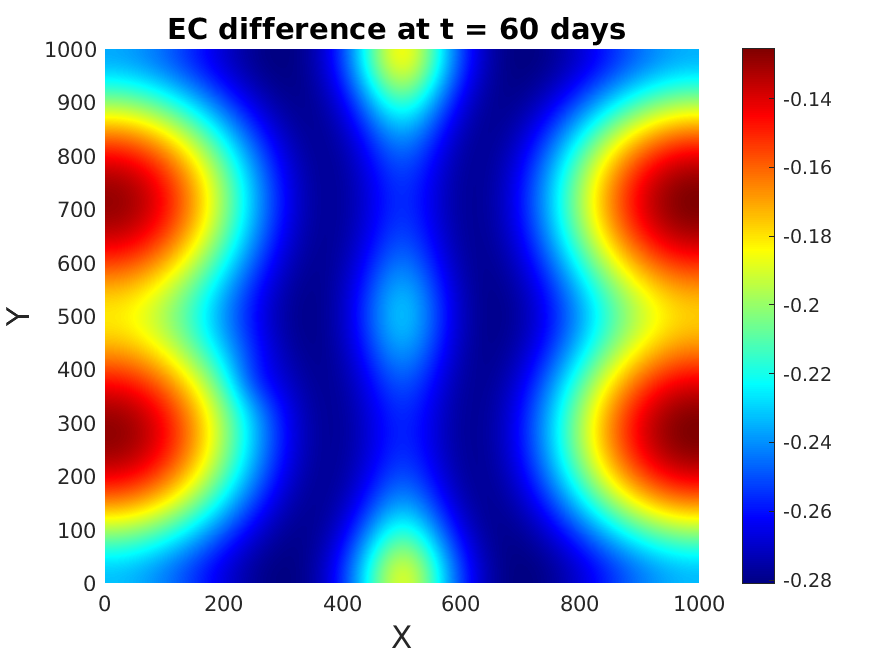}}\\
		{\includegraphics[width=1\linewidth, height = 3cm]{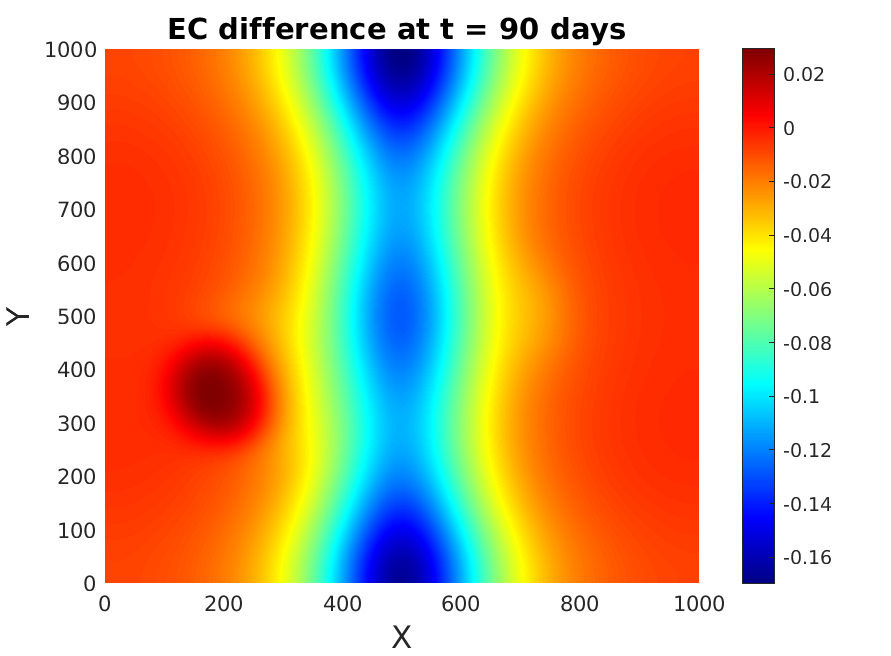}}\\
		{\includegraphics[width=1\linewidth, height = 3cm]{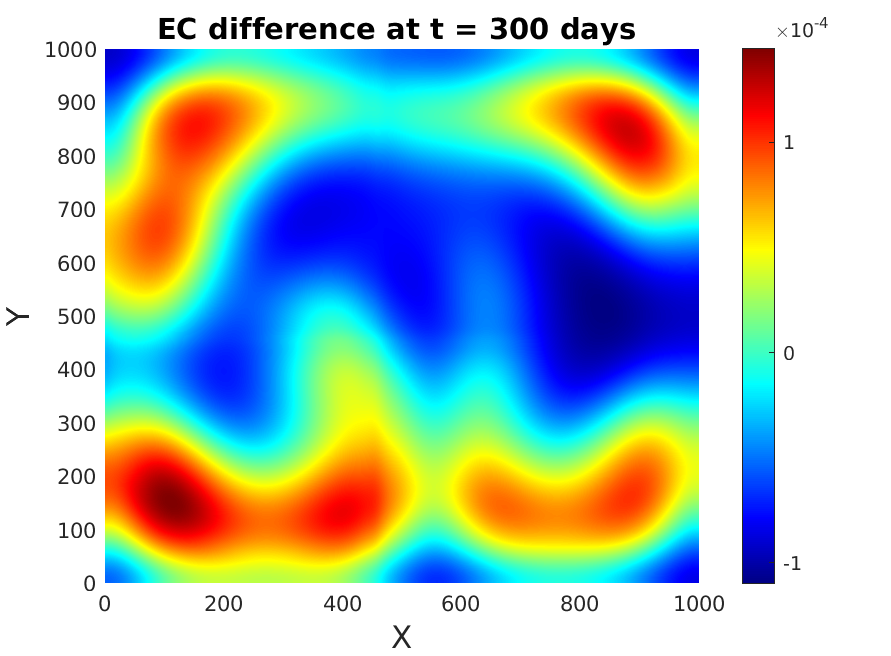}}\\
		{\includegraphics[width=1\linewidth, height = 3cm]{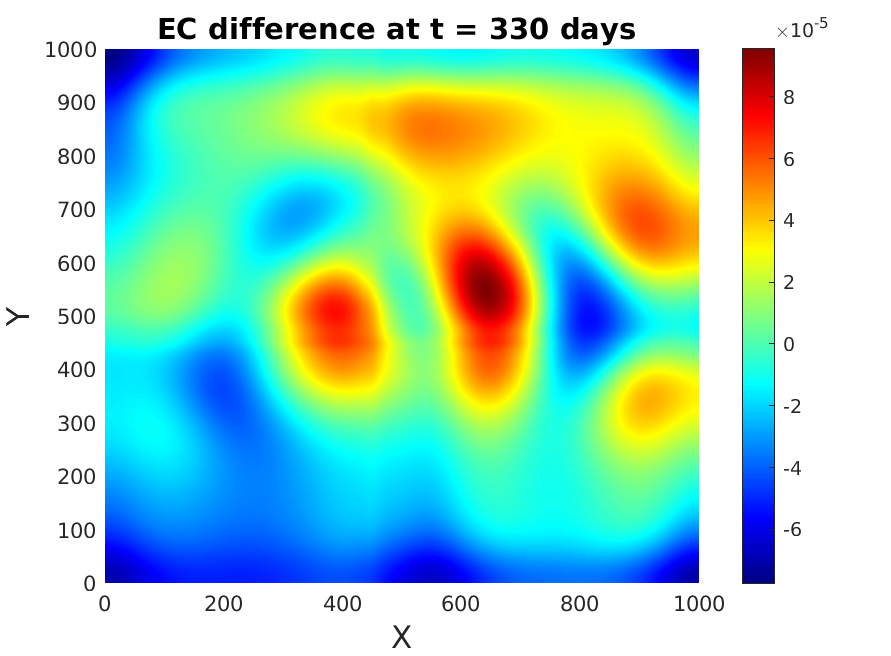}}\\
		{\includegraphics[width=1\linewidth, height = 3cm]{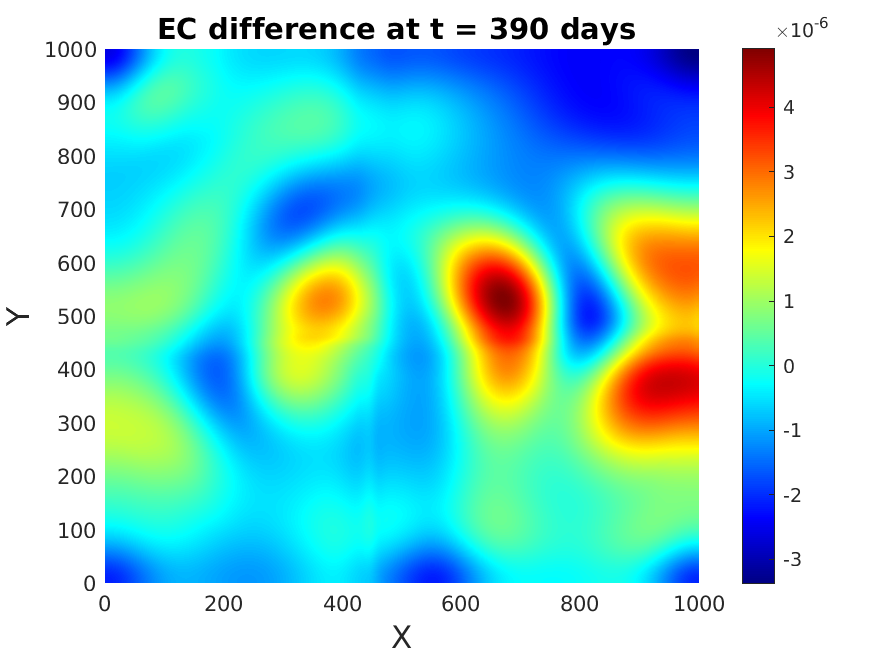}}\\
		{\includegraphics[width=1\linewidth, height = 3cm]{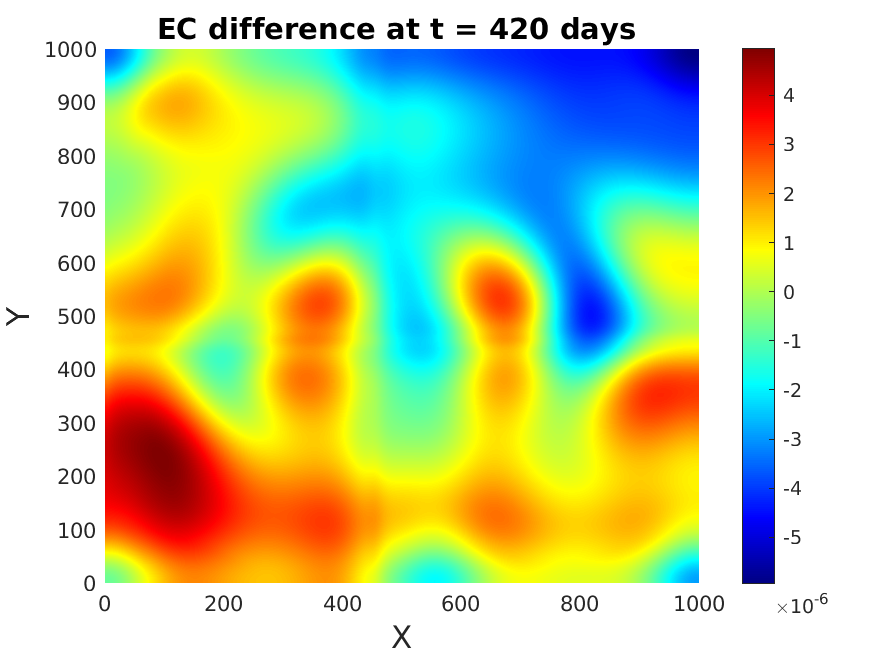}}%
		\subcaption{EC difference}
	\end{minipage}%
	\hspace{0.01cm}
	\begin{minipage}[t]{.24\linewidth}
		{\includegraphics[width=1\linewidth, height = 3cm]{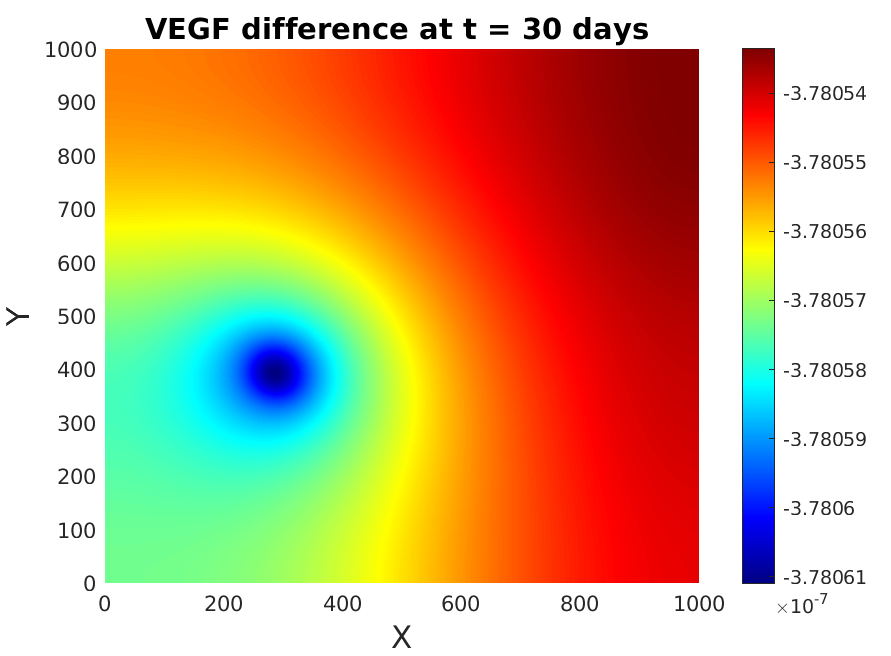}}\\
		{\includegraphics[width=1\linewidth, height = 3cm]{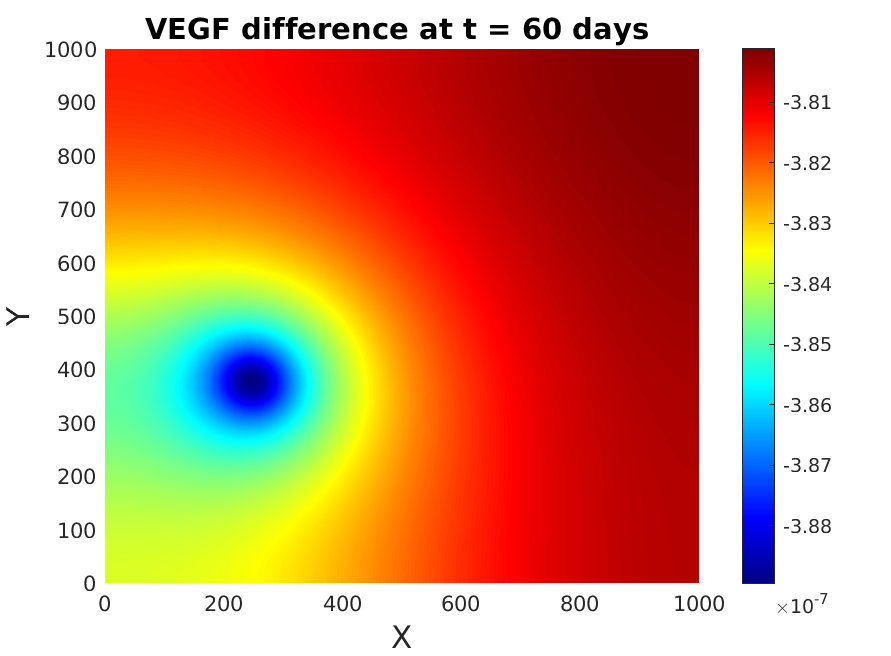}}\\
		{\includegraphics[width=1\linewidth, height = 3cm]{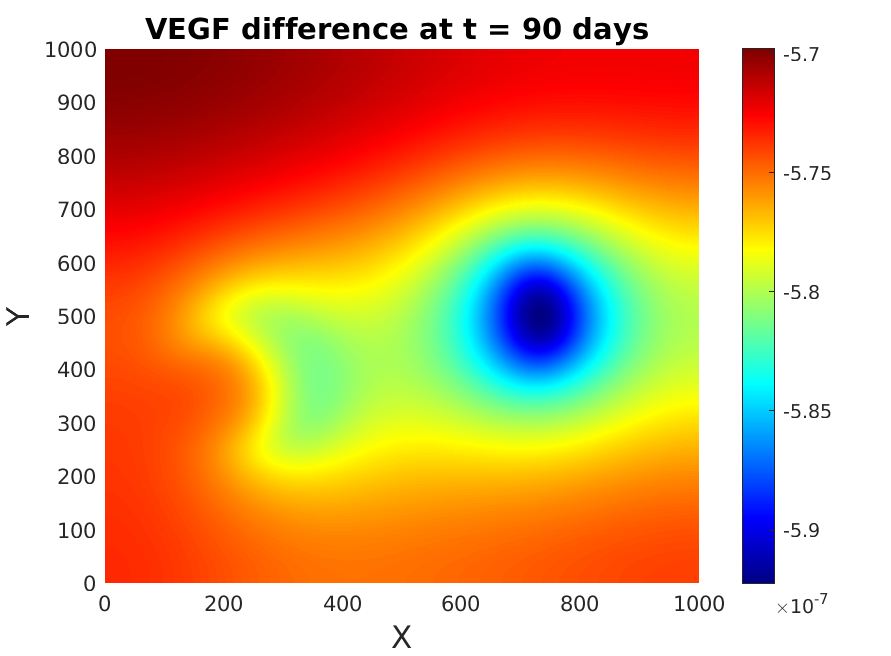}}\\
		{\includegraphics[width=1\linewidth, height = 3cm]{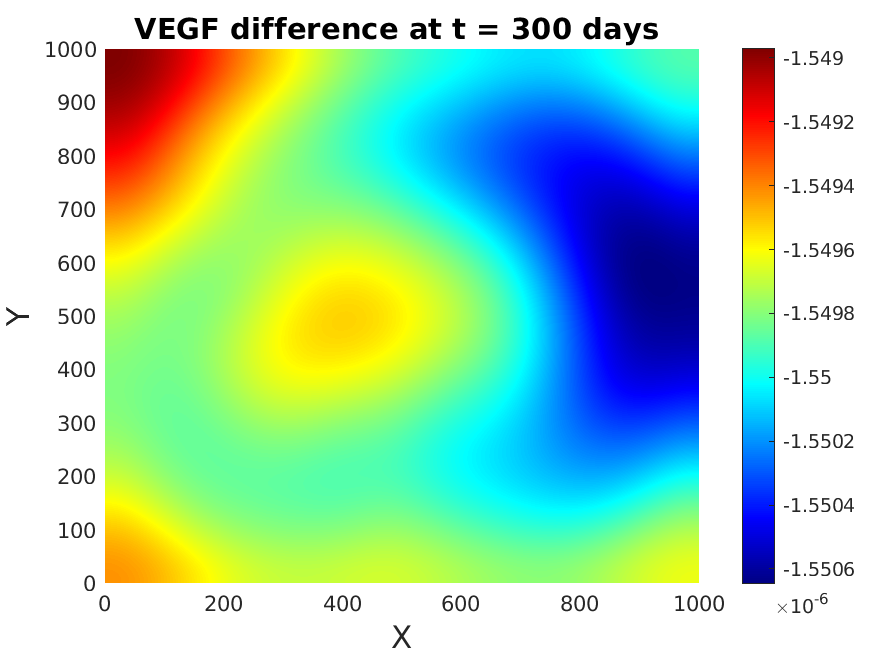}}\\
		{\includegraphics[width=1\linewidth, height = 3cm]{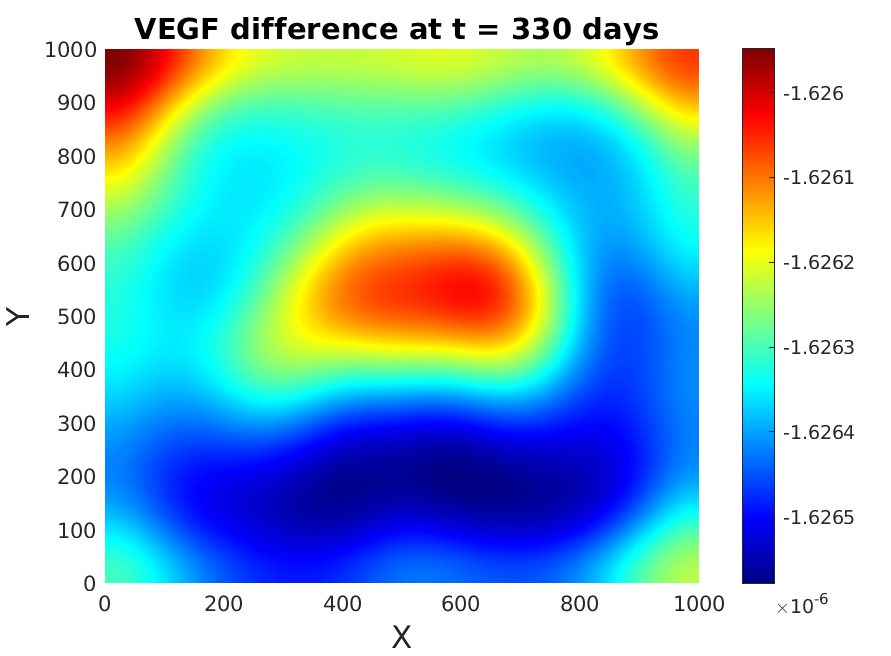}}\\
		{\includegraphics[width=1\linewidth, height = 3cm]{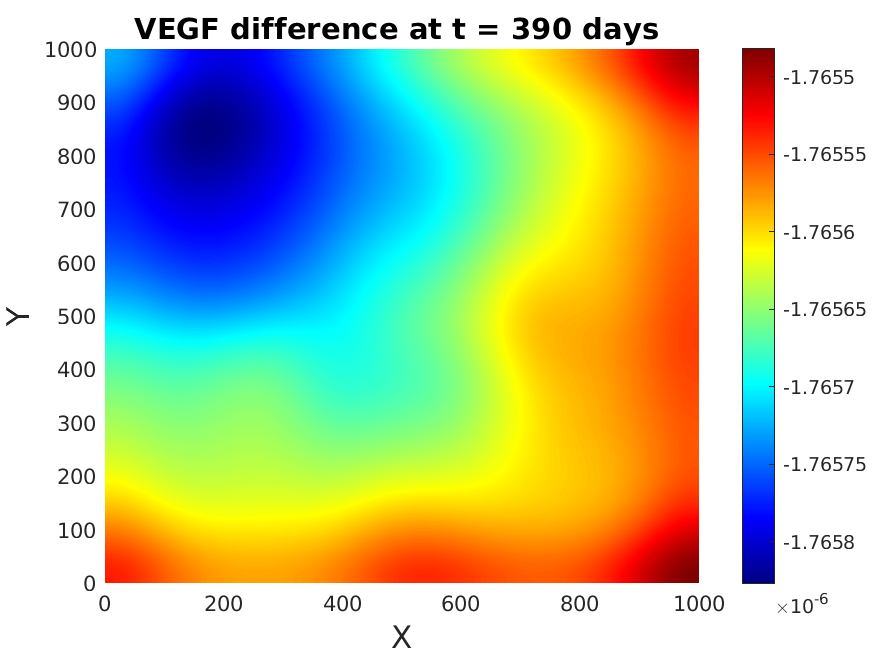}}\\
		{\includegraphics[width=1\linewidth, height = 3cm]{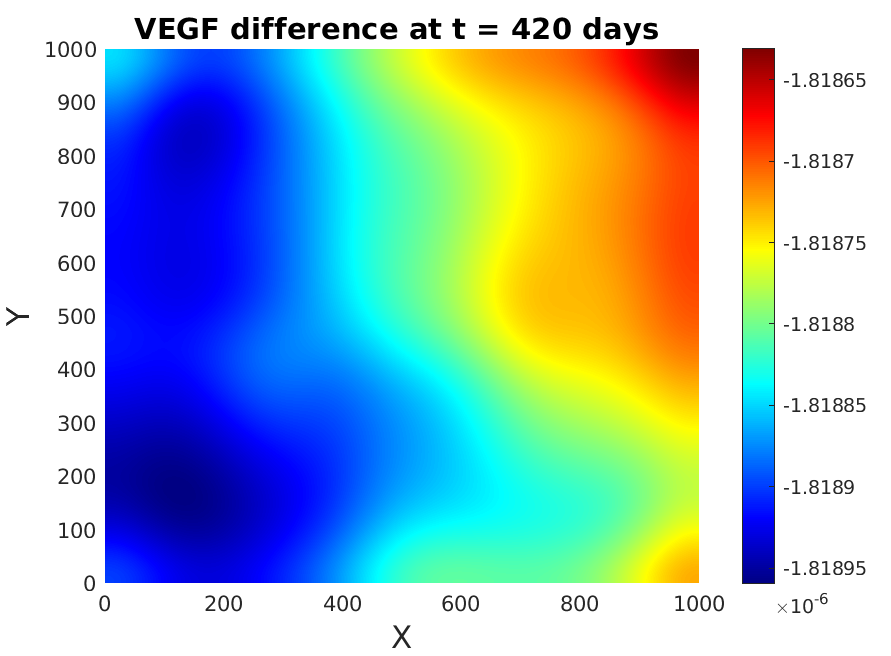}}%
		\subcaption{VEGF difference}
	\end{minipage}%
	\caption{\textbf{Experiment 4}: effect of flux-limited motility terms.   Differences between tumor density and proton concentration computed with \eqref{eq:macro-nondim} (in the parameter setting of \textbf{Experiment 2}) and those obtained for a model with glioma motility terms only involving myopic diffusion and pH-taxis without flux saturation (i.e. with glioma dynamics as in \eqref{eq:version-nofluxlim}).}\label{fig:diff_full-without_sd_ scenario2}
\end{figure}

\paragraph{Experiment 4} Effects of flux limitations. We compare the full model \eqref{eq:macro-nondim} with its counterpart involving in the motility terms only myopic diffusion and pH-taxis without flux saturation, i.e. characterizing glioma density dynamics by
\begin{align}\label{eq:version-nofluxlim}
M_t = \nabla \nabla : (\mathbb{D}_T M) + \alpha \nabla \cdot \left(\gamma_1\left( D_s \mathbb{I}_N - \mathbb{D}_T \right) M\nabla h \right) + M \left(1-M \right)\left(1-h\right)+\mu_c \frac{MW}{1+W},	
\end{align}
the rest of the equations in \eqref{eq:macro-nondim} remaining unchanged. The plots in Figure \ref{fig:diff_full-without_sd_ scenario2} illustrate the difference between solution components obtained with these two settings, in the above order. At a relatively early stage (30 days), with \eqref{eq:macro-nondim-a} there are more glioma cells at the tumor edge, with an obvious preference for moving away from the more acidic, central area. The proton concentration is higher towards the domain margins, and with time passing (plots at 60 days) it becomes consistently larger all over the domain in the model version using \eqref{eq:version-nofluxlim}, in which diffusion is dominating over taxis. In that setting the tumor cell density gets larger, too, which also applies to EC density; although \eqref{eq:macro-nondim} features (slightly) more glioma diffusion, however the flux-limited pH-taxis prevents a faster moving away from acidic regions. Instead, \eqref{eq:version-nofluxlim} ensures a wider spread of glioma, favored by the larger magnitude of the pH-tactic cell fluxes, which lets the tumor cells move faster and therewith produce the attracting VEGF at more locations. After a while (about further 30 days) the previous trend begins to get reversed, as far as glioma density is concerned: the flux limitation keeps the tumor more confined, but the acidity is still diffusing fast; the expression of VEGF is enhanced, which subsequently leads to hyperplasia, tumor enhancement, and revival of tiny glioma populations at more distant sites (in right half of the simulation domain). At the respective 'cancer spots' the cells stay more confined in the flux-limited case, while the model variant with \eqref{eq:version-nofluxlim} allows the cells to migrate faster (green-yellow margins in tumor difference plot at 90 days). The tumor cells accordingly produce acid and VEGF, attracting even more ECs at their main sites. Much later (300 days) \eqref{eq:macro-nondim} predicts larger densities of glioma cells in the center of the domain and near the boundary (the former because of the protons being increasingly removed by arriving ECs, and the latter due to the mutual proliferation support of the two cell types).
The setting with \eqref{eq:version-nofluxlim} leads to pseudopalisades with a larger diameter, being preserved for a longer while. Still later (390 days and more) the differences become smaller and smaller in magnitude, while the areas with more or less tumor cells become mixed up, with the pseudopalisades disappearing and giving way to more or less heterogeneous patterns where the glioma cells and ECs are spread over the whole domain.
\noindent
The simulations (observe in particular the plots at 390 and 420 days) also suggest that the motion of ECs is dominated by diffusion (and its limitation near tumor cells) rather than chemotaxis towards VEGF.\\[-2ex]

\begin{figure}[!htbp]
\centering
\begin{subfigure}[stb]{0.3\textwidth}
	{\includegraphics[width=1\linewidth, height = 3.3cm]{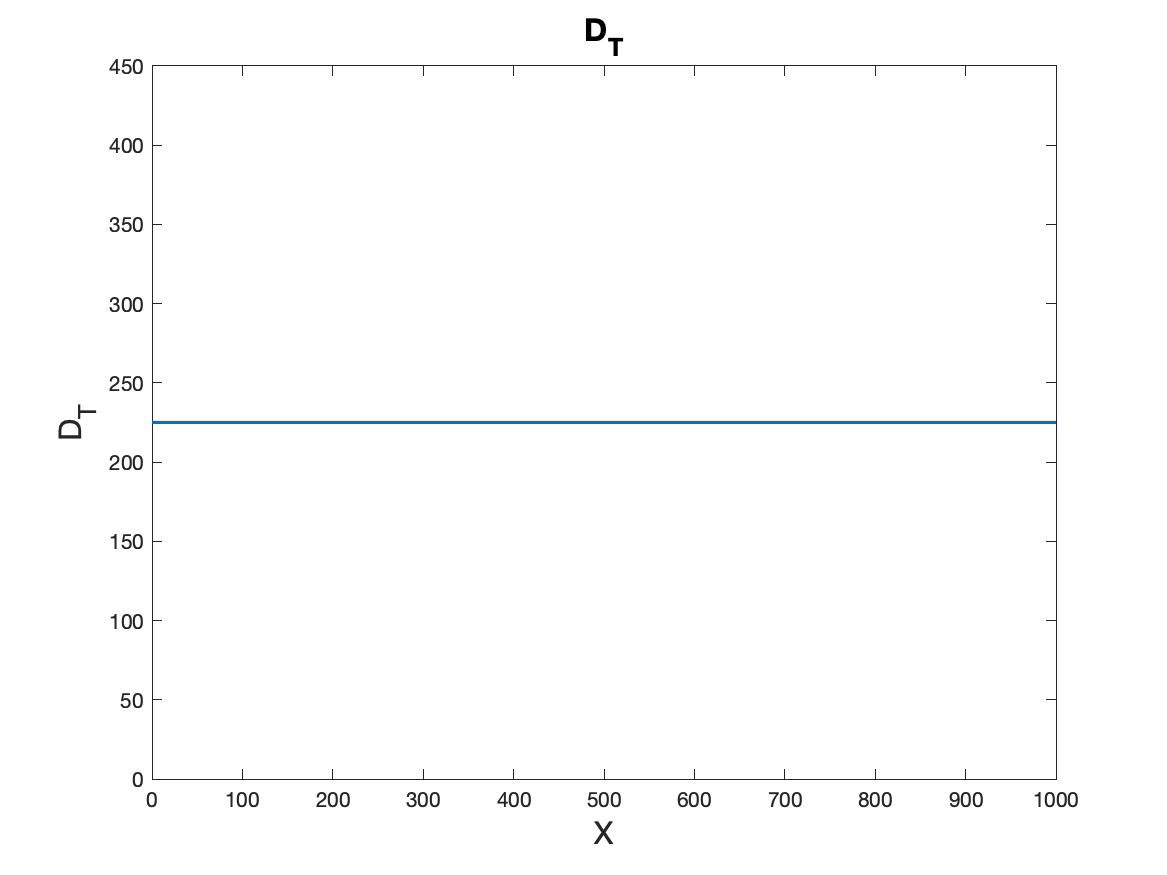}
	}
\end{subfigure}\qquad \qquad 
\begin{subfigure}[stb]{0.3\textwidth}
{\includegraphics[width=1\linewidth, height = 3.35cm]{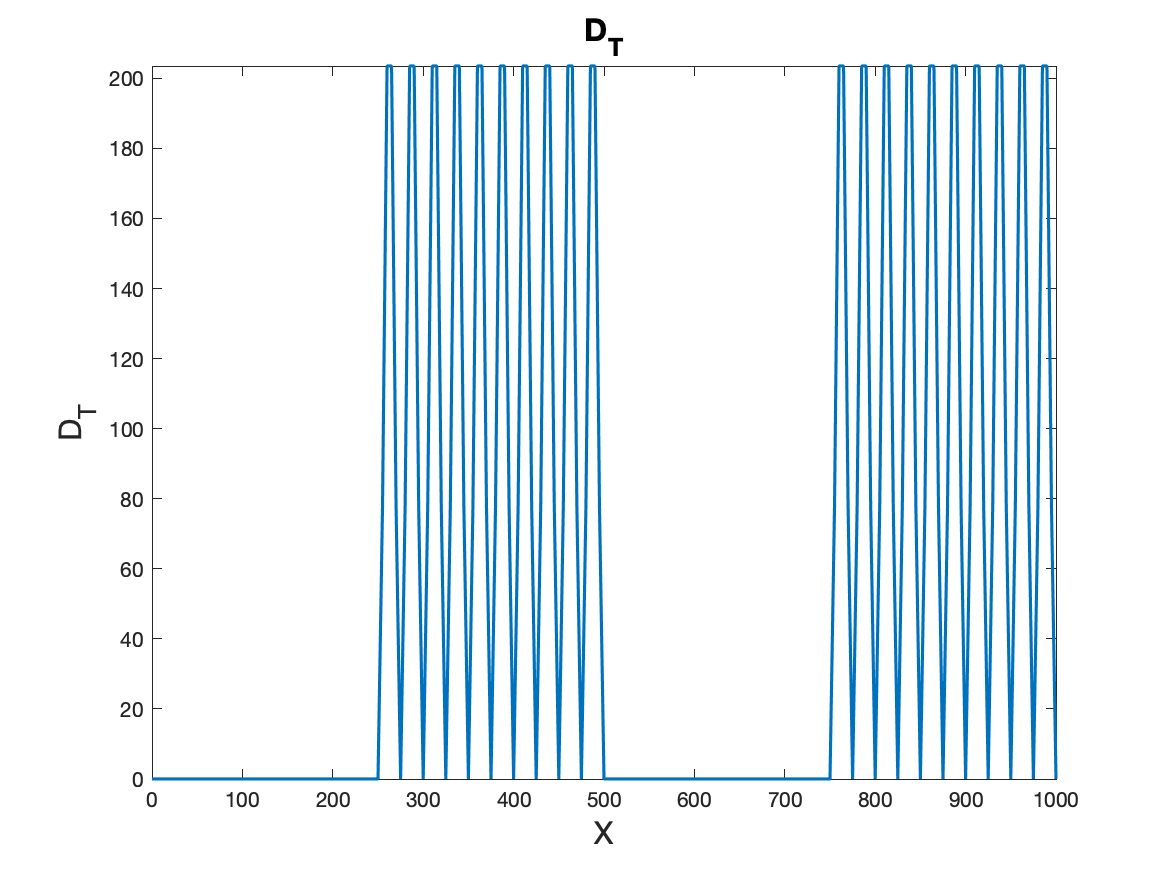}
}
\end{subfigure}\\
\begin{subfigure}[stb]{0.3\textwidth}
	{\includegraphics[width=1\linewidth, height = 3.3cm]{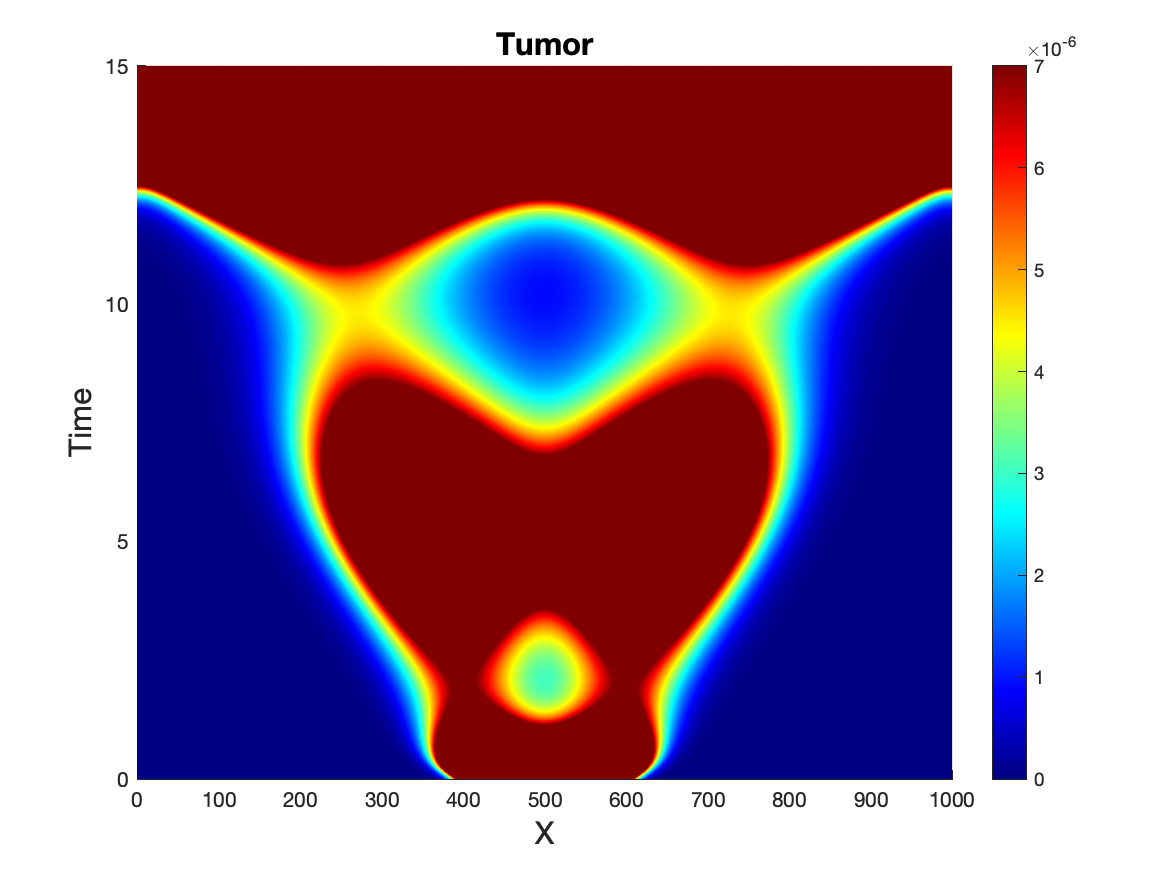}
	}
\end{subfigure}\qquad \qquad
\begin{subfigure}[stb]{0.3\textwidth}
	{\includegraphics[width=1\linewidth, height = 3.3cm]{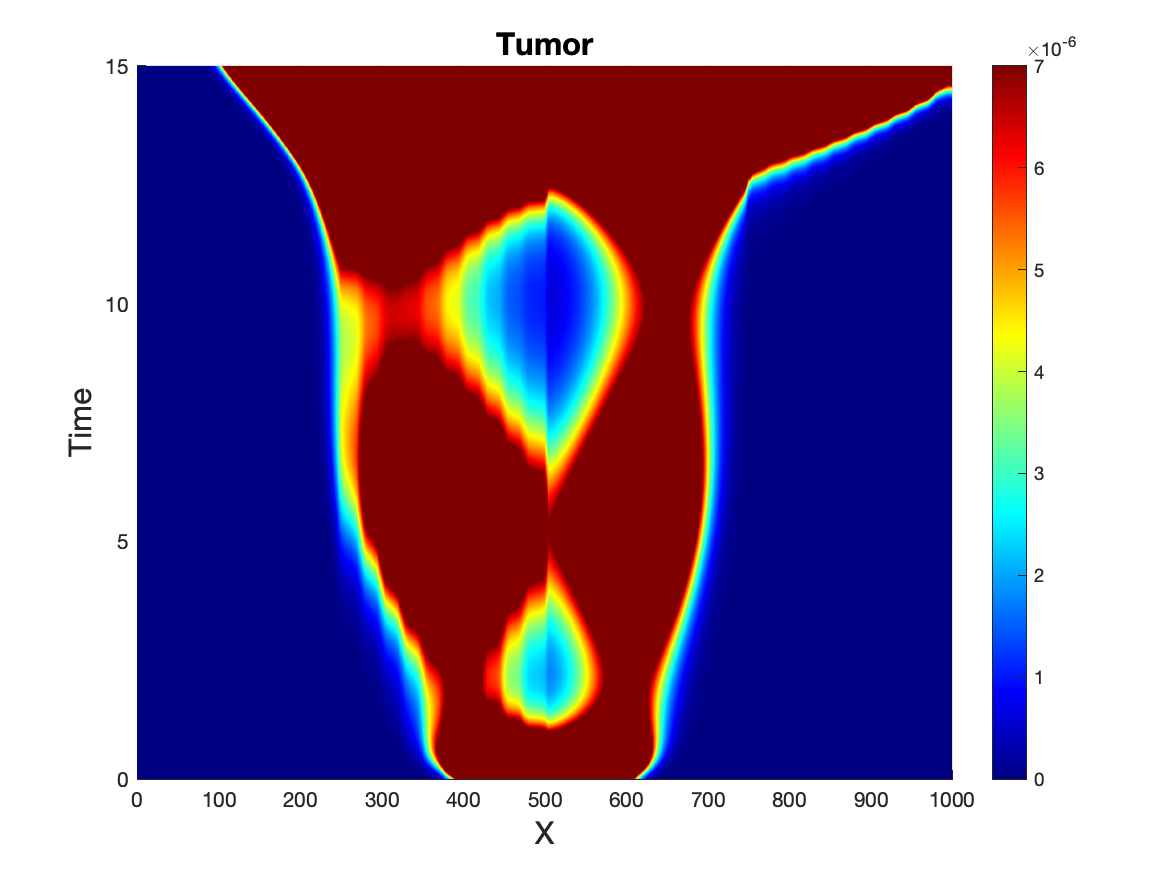}
	}
\end{subfigure}\\
\begin{subfigure}[stb]{0.3\textwidth}
	{\includegraphics[width=1\linewidth, height = 3.3cm]{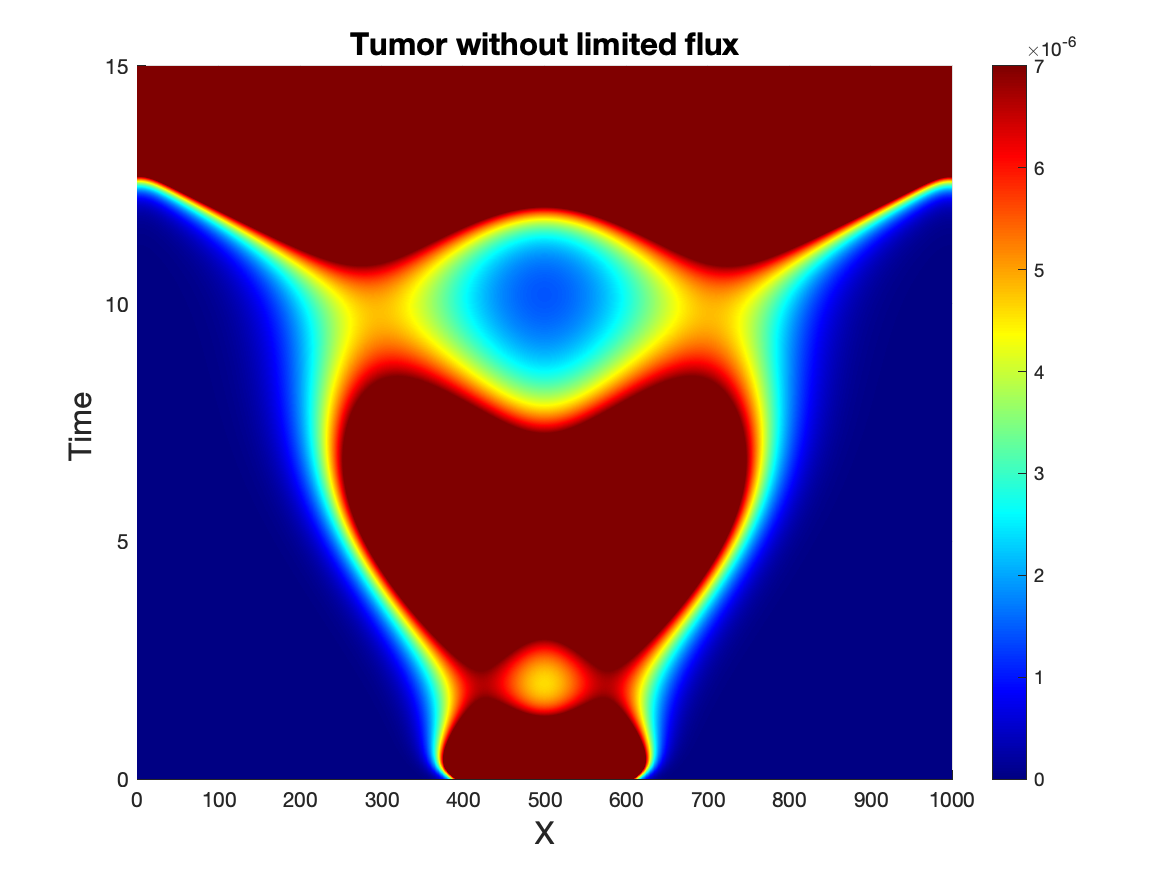}
	}
\end{subfigure}\qquad \qquad
\begin{subfigure}[stb]{0.3\textwidth}
	{\includegraphics[width=1\linewidth, height = 3.3cm]{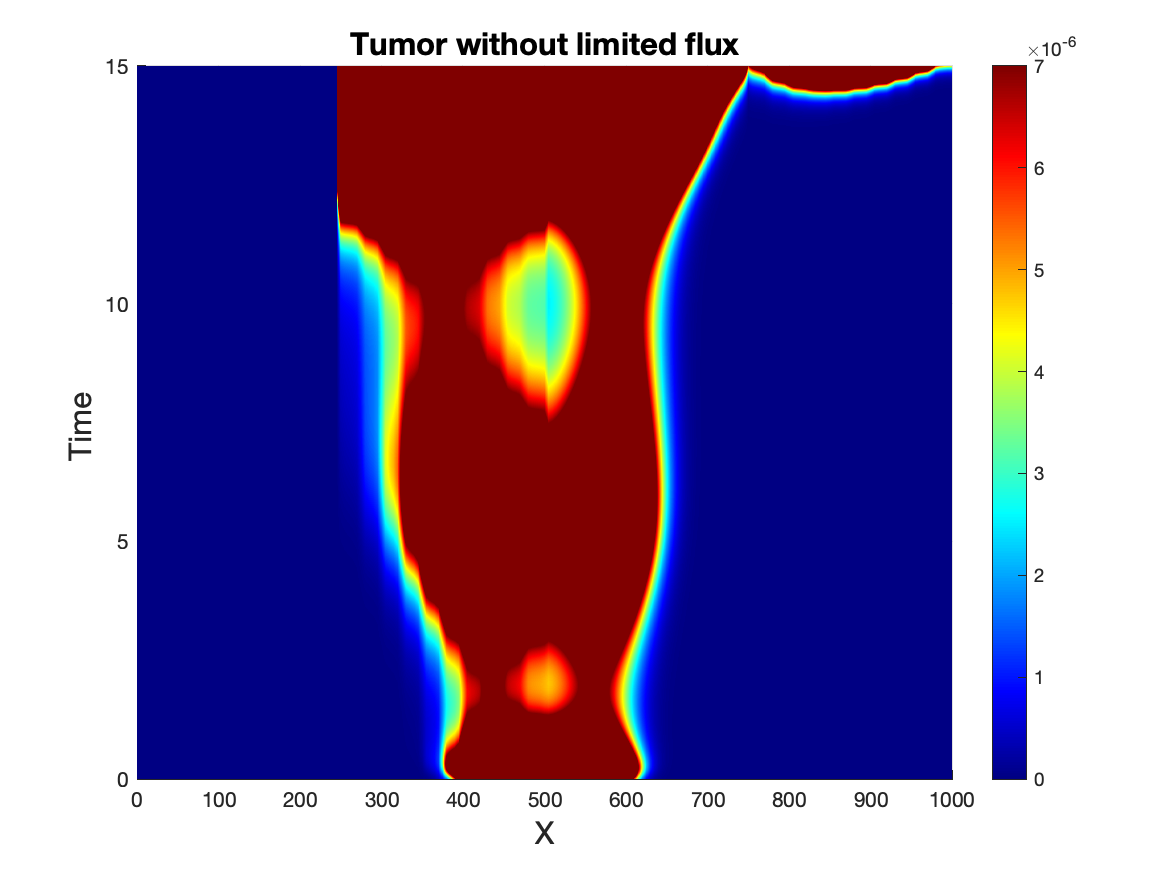}
	}
\end{subfigure}\\
\begin{subfigure}[stb]{0.3\textwidth}
	{\includegraphics[width=1\linewidth, height = 3.3cm]{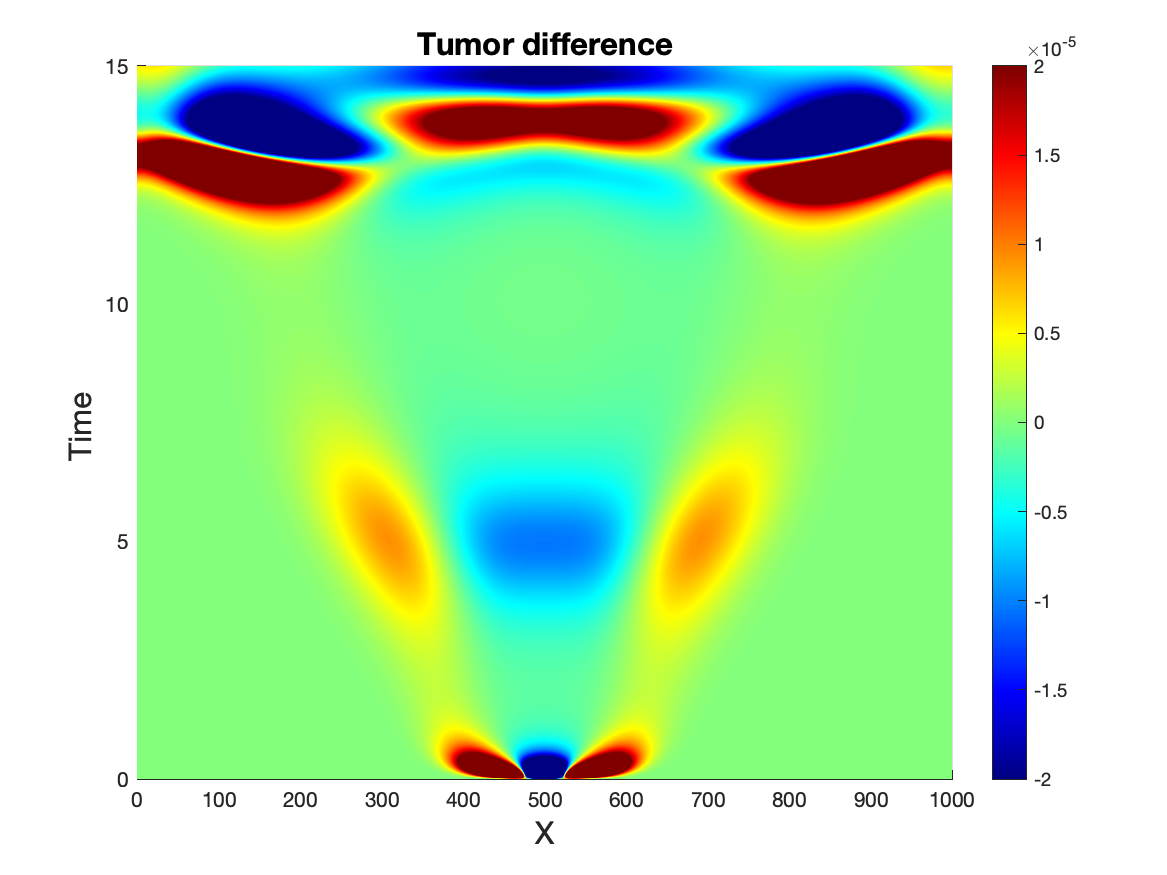}
	}
\end{subfigure}\qquad \qquad
\begin{subfigure}[stb]{0.3\textwidth}
	{\includegraphics[width=1\linewidth, height = 3.3cm]{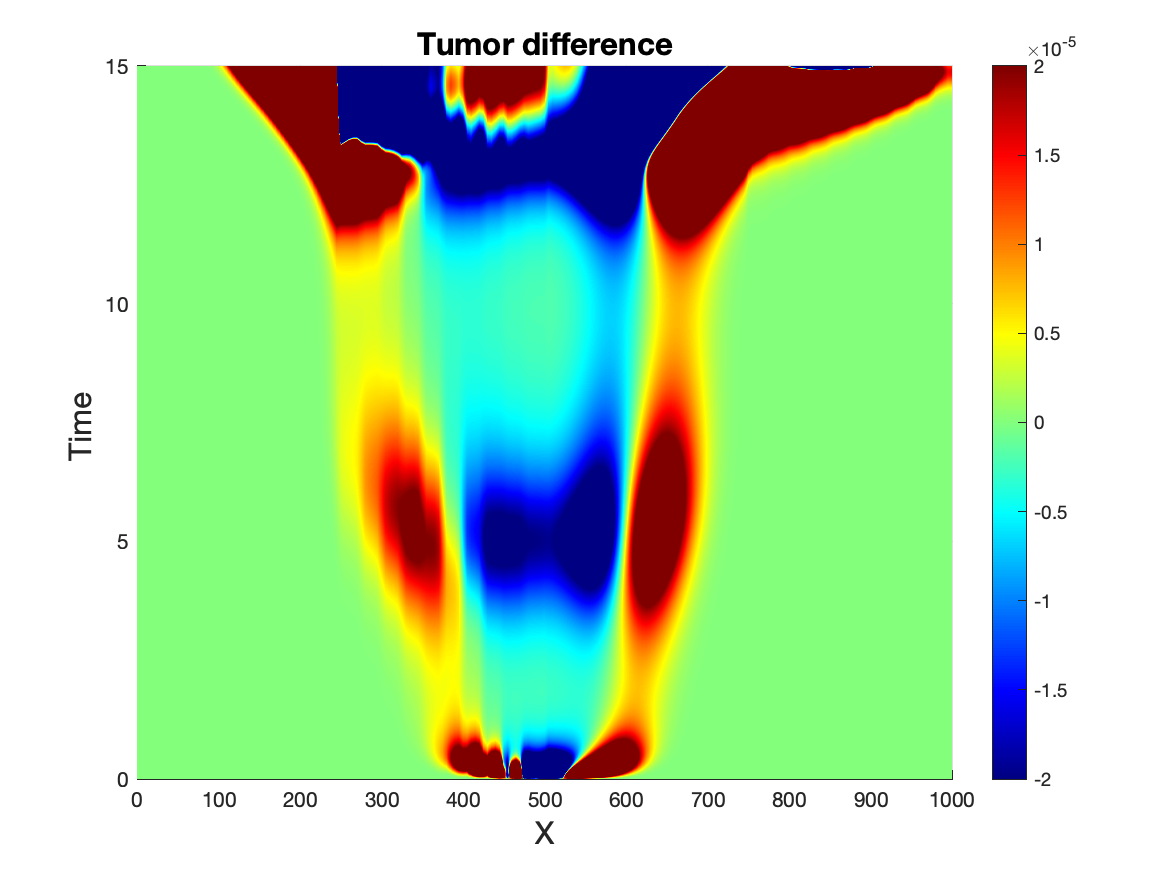}
	}
\end{subfigure}\\
\begin{subfigure}[stb]{0.3\textwidth}
	{\includegraphics[width=1\linewidth, height = 3.3cm]{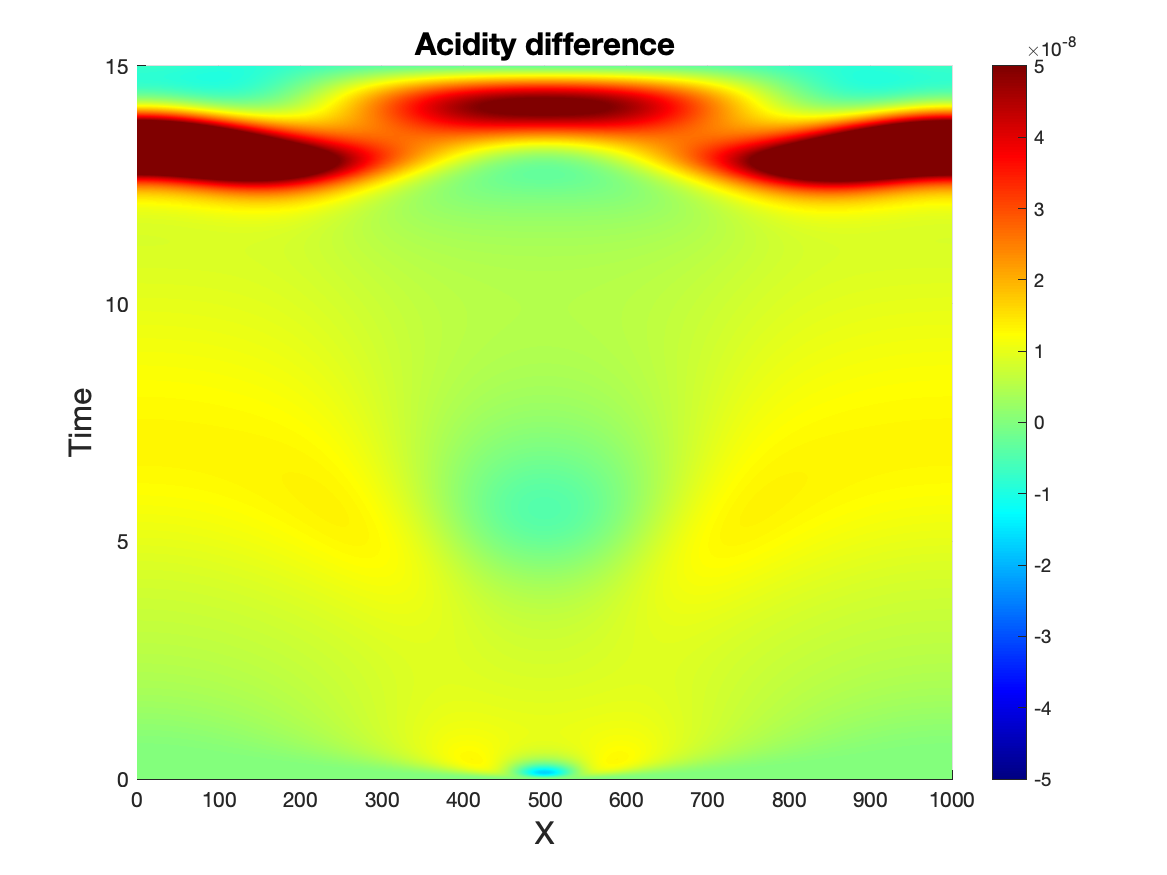}
	}
\end{subfigure}\qquad \qquad
\begin{subfigure}[stb]{0.3\textwidth}
	{\includegraphics[width=1\linewidth, height = 3.3cm]{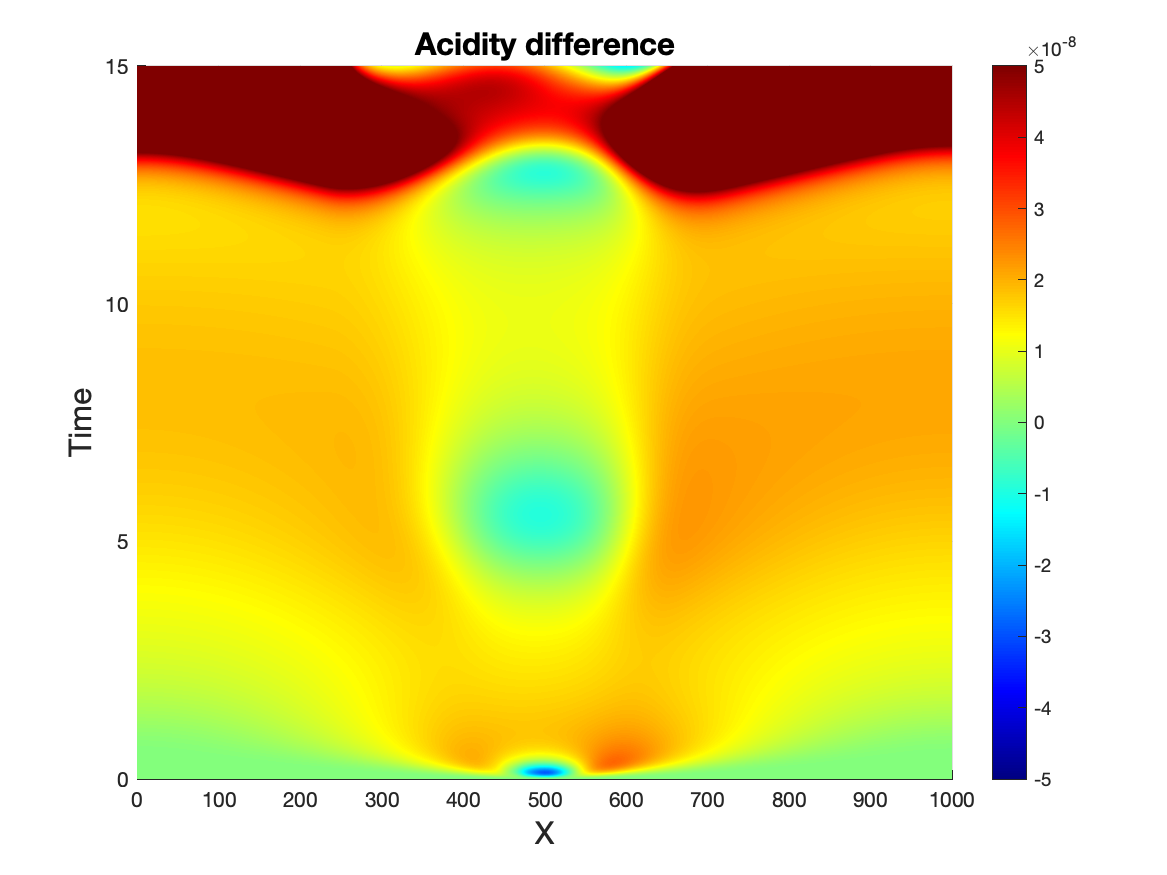}
	}
\end{subfigure}\\
\begin{subfigure}[stb]{0.3\textwidth}
	{\includegraphics[width=1\linewidth, height = 3.3cm]{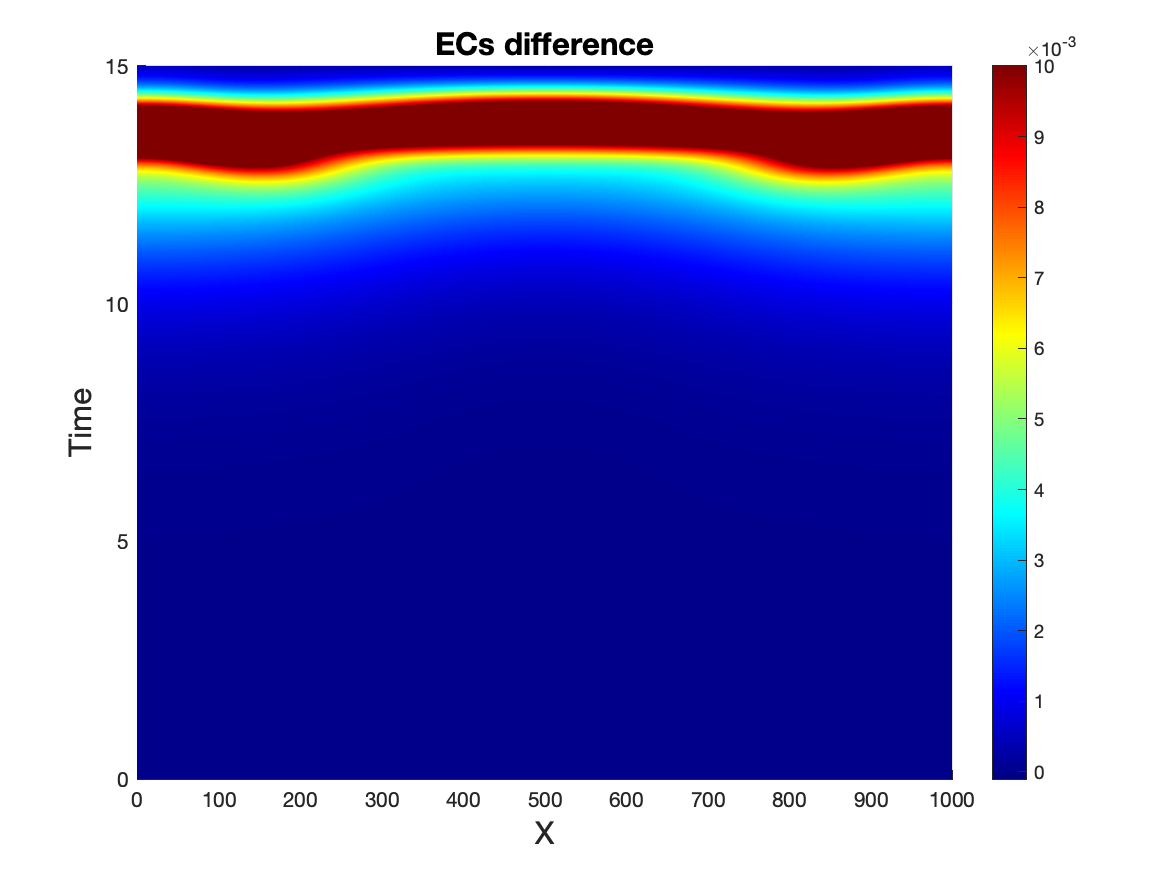}
	}
\end{subfigure}\qquad \qquad
\begin{subfigure}[stb]{0.3\textwidth}
	{\includegraphics[width=1\linewidth, height = 3.3cm]{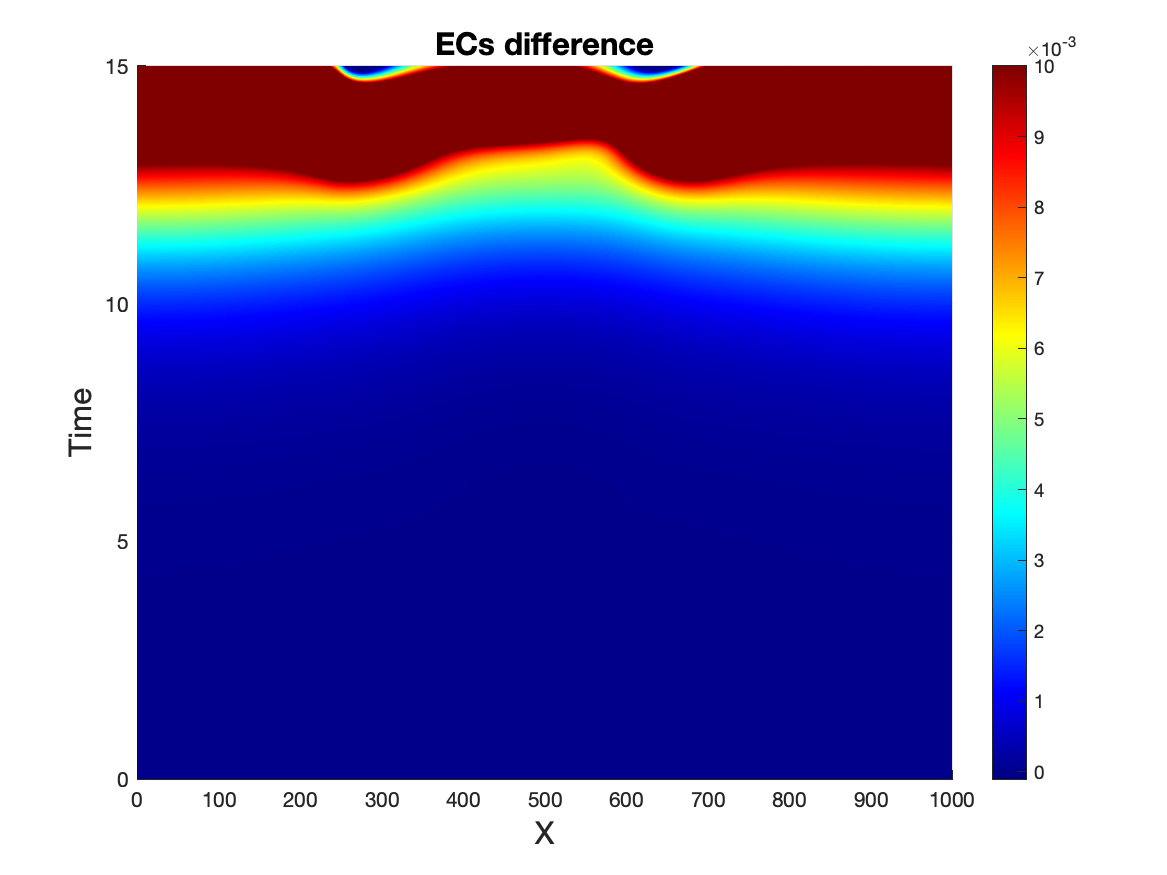}
	}
\end{subfigure}
\caption{1D patterns for two different choices of the glioma diffusion coefficient $D_T$. 2nd row: tumor patterns for model \eqref{eq:macro-nondim}. 3rd row: tumor patterns when \eqref{eq:macro-nondim-a} is replaced by \eqref{eq:version-nofluxlim}. 4th row: differences between respective patterns in rows 2 and 3. 5th and 6th rows: differences between acidity and EC densities computed with model \eqref{eq:macro-nondim} and with \eqref{eq:version-nofluxlim} replacing \eqref{eq:macro-nondim-a}. }\label{fig:patterns-1D-comp-Exp4}
\end{figure}

\noindent
Figure \ref{fig:patterns-1D-comp-Exp4} illustrates 1D tumor patterns obtained with model \eqref{eq:macro-nondim} and with \eqref{eq:version-nofluxlim} replacing \eqref{eq:macro-nondim-a}. Two different choices of the glioma diffusion coefficient $D_T(x)$ (the same as in Figure \ref{fig:patterns-1D-Exp3}) are considered.  Model \eqref{eq:macro-nondim} enables glioma diffusion even in the region where $D_T$ becomes zero, due to the flux-limited self-diffusion, hence the tumor cells can invade a wider area. In both model versions, as a consequence of VEGF availability, more ECs gather in the areas with larger glioma density, which in turn favors proton removal and tumor proliferation, followed again by glioma diffusion. For the strongly degenerating $D_T$ the setting without flux limitation leads, as in the case investigated in Figure \ref{fig:patterns-1D-Exp3}, to higher cell accumulations near the interface with highest diffusivity difference, while (as long as $M$ is bounded) the cell flux involving $\phi (h,M)$ in \eqref{eq:macro-nondim} is bounded even if the gradients of $h$ and $M$ are not; thus this model version should be less prone to singular structure formation. It also allows for sharper fronts of the cell mass, as can be seen in the right plot of the second row of Figure \ref{fig:patterns-1D-comp-Exp4}.

\section{Discussion}\label{sec:discussions}

The vast majority of continuous models for cell migration involve reaction-diffusion(-taxis) PDEs on the macroscopic scale, irrespective to whether those equations were obtained by a direct formulation or a more or less formal deduction from lower scales. Thereby, the diffusion is in some (most often constant) proportion to the density gradient, which is associated to infinite speed of propagation - an unrealistic feature, accompanied by smearing the sharp profile of the cell mass advancing into its surroundings; possible overcrowding effects cannot be properly captured either. These and other issues motivated the introduction of models with flux saturation also in this context. We refer e.g., to \cite{BBNS2010,perthame2018flux} for formal and respectively rigorous derivations of chemotaxis models from KTEs, and to \cite{CTS20,DKSS20,Kim2009} for settings specifically relating to glioma invasion. Among the latter, \cite{DKSS20} contains a deduction of macroscopic equations with flux-saturated diffusion, haptotaxis, and chemotaxis; the passage from the mesoscopic KTAP framework to the RDT system is different from the one performed here, yet still formal. Very recently, \cite{ZS-transp} started from a KTE formulation similar to that in \cite{DKSS20}, hence also related to the one in the present work, and obtained by yet another upscaling method (with rigorous convergences) an RDT with myopic diffusion for the zeroth order approximation of the solution. That PDE did not involve flux-limitation, but the equation deduced for the first order correction did. \\[-2ex]

\noindent
The model proposed here is still a bottom-up approach connecting subcellular and mesoscopic dynamics with the macroscopic evolution of cell populations in response to cues from the tumor microenvironment, leading to patterns which are qualitatively similar to those observed in histological imaging of glioblastoma. The setting is an extension of the previous one introduced in \cite{kumar2020multiscale}, since it accommodates the description of tumor vascularization in response to hypoxia. The main novelty here is the way of obtaining flux-limited pH-taxis and (supplementary) self-diffusion, namely upon considering on the single cell scale the joint effect of cell stress, chemical gradients, and population pressure to describe forces acting on the cells, cf. \eqref{eq:micro}, \eqref{eq:bigS}. The approach is akin to that employed in \cite{chauviere2007modeling,corbin2020modeling,DKSS20}, however it involves several modifications, as \cite{chauviere2007modeling,corbin2020modeling} did not contain any flux limitations, while \cite{DKSS20} does, but models differently the single cell dynamics (allowing, among others, for non-constant speeds) and does not take into account cell reorientations (by way of a turning operator) in response to local tissue anisotropy and nonlocal orientation distribution of fibers and cells. Our method also differs from \cite{BBNS2010,perthame2018flux}, where the flux limitation stems from the choice of signal response function involved in the
turning operator and depending on the directional derivative of the tactic signals (chemoattractants). We included in \eqref{eq:bigS} only effects of pH-taxis and population pressure, but in modeling glioma migration through a highly anisotropic tissue one could also involve the influence of its gradients (via macroscopic tissue density, as in \cite{DKSS20} or mesoscopic spatial and orientational distribution of tissue fibers, as in \cite{corbin2020modeling}). There are two 'sources' of diffusion in our model: the myopic diffusion comes from the description of cell reorientations via the turning operator, while the self-diffusion term with flux limitation originates from Newton's second law in \eqref{eq:micro}, \eqref{eq:bigS}. The latter diffusion part has the effect of eluding diffusion degeneration when the tensor $\mathbb D_T$ nullifies as a consequence of the mesoscopic tissue distribution $q(\xbf ,\hat \vbf )$ vanishing locally. (Strongly) degenerating myopic diffusion, alone or in combination with taxis have been related to high cell aggregates and possible singularity formation even in lower dimensions \cite{HILLEN2012,winkler-surulescu,Winkler2018}, while flux limitation could alleviate such behavior, at least under certain conditions, see e.g., \cite{BelWinkler20172,BelWinkler20171}. This tendency was also observed  in  Figure \ref{fig:patterns-1D-Exp3}, in the case with strongly degenerating $D_T$. Figure \ref{fig:patterns-1D-comp-Exp4} showed a sharper tumor cell profile at the interface with large diffusivity drop, while the model without flux limitation generated a more uniform pattern with a tendency of smearing the interface and faster filling the areas of lower cell density. We have to stress, nevertheless, that our deduction of macroscopic PDEs is merely formal; as mentioned above, a rigorous one was done in \cite{ZS-transp}, but for a simplified model and involving only limitation(s) of signal gradient(s) of the form $\frac{\nabla S}{1+|\nabla S|}$ instead of our choice encoded in $\phi (h,M)$.\\[-2ex]

\noindent
The obtained macroscopic model is able to reproduce, at least qualitatively, the histological patterns observed in patient biopsies. Indeed, the typical pseudopalisades due to severe hypoxia are formed even when (incipient) angiogenesis is present, followed by a disruption of the 'garlands' as consequence of acidity removal by vasculature (motile ECs) within the pseudopalisades and therewith associated repopulation with tumor cells of the formally acidic sites, which was not or very weakly noticed in simulations of  the previous model \cite{kumar2020multiscale}.  The flux limitation does not seem to be necessary for the latter to happen, however it does influence the duration of pattern formation and maintenance, and (to a lesser extent) their appearance. We also noticed (cf. Figures \ref{fig:scenario1} and \ref{fig:scenario2}) that the weights assigned to (flux-limited) pH-taxis and self-diffusion affect the patterns in terms of shape, duration, and persistence. \\[-2ex]

\noindent
The macroscopic system \eqref{eq:macro-nondim} features two types of taxis: a repellent, flux-limited pH-taxis of glioma cells moving in the opposite direction of the proton concentration gradient $\nabla h$, and chemotaxis of endothelial cells following the gradient of VEGF. As such, the setting does not fall into any class of models with multiple taxis as reviewed in \cite{kolbe2020modeling}, since each cell population is performing only one type of taxis, but it raises mathematical questions which are at least partially related to those models. The proton production by tumor cells gives the pH-taxis a direct character, while VEGF, the chemotactic cue of ECs, is produced by glioma cells in presence of acidity, while glioma proliferation is favorably influenced by ECs. The latter ensure uptake of protons as well as VEGF, hence exert an indirect and also a direct influence on their tactic signal. The mathematical analysis of system \eqref{eq:macro-nondim} 
with appropriate initial and boundary conditions is far from standard; this is not only due to the myopic diffusion and the nonlinearities arising from the source terms and from couplings, but especially to the flux-limited motility terms in the glioma reaction-diffusion-taxis PDE. Results about well-posedness and long time behavior of solutions to such systems are unknown. We merely provided a couple of  examples of simulation-based, 1D pattern assessments in Figures \ref{fig:patterns-1D-Exp3} and \ref{fig:patterns-1D-comp-Exp4} - with no pretension of completeness or rigor, but only for the purpose of giving a flavor of the patterns arising from such model.  Beyond that, a traveling wave analysis, particularly of glioma and EC dynamics, would be of interest, both theoretically and from the application viewpoint.

\section*{Acknowledgement}

PK acknowledges funding by DAAD in form of a PhD scholarship. CS was funded by BMBF in the project \textit{GlioMaTh} 05M2016.

\newcommand{\noopsort}[1]{}
\addcontentsline{toc}{section}{References}
\bibliographystyle{my_plain}
\bibliography{paper}

\end{document}